\documentclass[paper]{JHEP3}
\pdfoutput=1
\usepackage{amssymb}
\usepackage{amssymb}
\usepackage{cite}
\usepackage{epsfig}
\def\nc{\newcommand}
\def\phm{\phantom{-}}
\nc{\half}{\frac{1}{2}}
\nc{\shalf}{\ensuremath{\textstyle \frac{1}{2}}}

\nc{\deldag}{\mathbin{\partial\mkern-10.5mu\big/}}
\nc{\deldagss}{\mathbin{\partial\mkern-10.5mu/}}
\nc{\kdag}{\mathbin{k\mkern-10mu\big/}}
\nc{\skdag}{\mathbin{k\mkern-10mu/}}
\nc{\udag}{\mathbin{u\mkern-10mu\big/}}
\nc{\kdagss}{\mathbin{k\mkern-10mu/}}
\nc{\Pdag}{\mathbin{P\mkern-10mu\big/}}
\nc{\pp}{{\scriptscriptstyle ||}}
\nc{\stwo}{{\scriptscriptstyle 2}}
\nc{\pham}{{\phantom{-}}}

\def\lsim{\mathrel{\raise.3ex\hbox{$<$\kern-.75em\lower1ex\hbox{$\sim$}}}}
\def\gsim{\mathrel{\raise.3ex\hbox{$>$\kern-.75em\lower1ex\hbox{$\sim$}}}}

\def\Slashnew#1{#1\kern-0.55em\raise.05ex\hbox{/}}
\def\slashnew#1{#1\kern-0.5em\raise.05ex\hbox{{$\scriptstyle /$}}}
\def\sfrac#1#2{{\textstyle\frac#1#2}}

\def\ie{{\em i.e. }}
\def\wrt{{\em w.r.t. }}
\def\eg{{\em e.g. }}

\nc{\beq} {\begin{equation}}
\nc{\eeq} {\end{equation}}
\nc{\beqa}{\begin{eqnarray}}
\nc{\eeqa}{\end{eqnarray}}

\def\Slashnew#1{#1\kern-0.55em\raise.05ex\hbox{/}}
\def\slashnew#1{#1\kern-0.5em\raise.05ex\hbox{{$\scriptstyle /$}}}
\def\emph#1{{\em #1}}
\def\hepph#1{hep-ph/#1}
\def\hepth#1{hep-th/#1}
\def\sfrac#1#2{\ensuremath{\textstyle \frac{#1}{#2}}}

\title{Coherent quantum Boltzmann equations from cQPA}

\author{Matti Herranen$^a$,
        Kimmo Kainulainen$^{b,c}$ and
        Pyry Matti Rahkila$^{b,c}$ 
        \\
		\\ $^a$Institut f\"ur Theoretische Teilchenphysik und Kosmologie,
		\\ RWTH Aachen University, D--52056 Aachen, Germany
        \\ $^{b}$ Department of Physics, P.O.Box 35 (YFL), 
        \\ FIN-40014 University of Jyv\"askyl\"a, Finland, 
        \\ $^{c}$ Helsinki Institute of Physics, P.O.~Box 64, 
  	    \\ FIN-00014 University of Helsinki, Finland.\\
        \\e-mail: \email{herranen@physik.rwth-aachen.de,   
                         kimmo.kainulainen@jyu.fi, 
                         pyry.rahkila@jyu.fi}}

\abstract{We reformulate and extend our recently introduced quantum kinetic theory for  interacting fermion and scalar fields. Our formalism is based on the coherent quasiparticle approximation (cQPA) where nonlocal coherence information is encoded in new spectral solutions at off-shell momenta. We derive explicit forms for the cQPA propagators in the homogeneous background and show that the collision integrals involving the new coherence propagators need to be resummed to all orders in gradient expansion. We perform this resummation and derive generalized momentum space Feynman rules including coherent propagators and modified vertex rules for a Yukawa interaction. As a result we are able to set up self-consistent quantum Boltzmann equations for both fermion and scalar fields. We present several examples of diagrammatic calculations and numerical applications including a simple toy model for coherent baryogenesis.}

\preprint{TTK-10-34}

\preprint{}
\begin{document}

\section{Introduction}
Many interesting problems in particle physics and cosmology involve coherent, relativistic quantum systems in out-of-equilibrium conditions, examples including the electroweak baryogenesis~\cite{BG,ClassForce,ClassForceomat,SemiClassSK,PSW}, leptogenesis~\cite{leptogenesis,flavour_lepto,resonant_lepto,CTP_lepto} and the out-of equilibrium particle production~\cite{particle_prod}. While it may be straightforward to give a mathematical formulation for a given problem, for example using the Schwinger-Keldysh method, it is often hard to find an approximation scheme that is accurate, but simple enough to be used in practical applications. We have recently developed a new transport theory for relativistic quantum systems including nonlocal coherence~\cite{HKR1,HKR2,HKR3}. Our formalism is based on the observation~\cite{HKR1} that in systems with certain space-time symmetries the nonlocal coherence information is encoded in new spectral shell solutions for the dynamical 2-point functions at particular off-shell momenta. The extended spectral structure of the two-point functions, the coherent quasiparticle approximation (cQPA)~\cite{Glasgow,Thesis_Matti}, was first discovered in~\cite{HKR1} and a complete transport theory with collision terms for fermions was presented in~\cite{HKR2} and for scalar fields in~\cite{HKR3}. 

In refs.~\cite{HKR1,HKR2} the fermionic transport theory was formulated using a convenient, but somewhat less familiar spin-projected 2-component notation. Later, in refs.~\cite{Glasgow,Thesis_Matti} it was noted that the collision integrals involving the coherence functions need to be resummed to all orders of the  gradient expansion. In this paper we reformulate our formalism for fermions using a more familiar 4-component Dirac notation, accompanied by a much more convenient parametrization of the coherence-shell solutions. We will also show that the resummation issue is much more generic than what was considered in~\cite{Glasgow,Thesis_Matti}: a resummation is necessary for all external and internal lines involving coherence propagators in the diagrammatic expansion of all collision integrals appearing in the transport equations. Here we will show how this resummation can be done such that all effects are absorbed into a redefinition of the effective coherence propagators. Using these results we develop an extension of the momentum space Feynman rules for diagrammatic evaluation of self-energies and eventually collision terms including quantum coherence effects. We will give a full discussion of single flavour fermion and real scalar fields, interacting via a Yukawa type of interaction. In addition to our extension of the usual Feynman rules to the coherent case, the main results of this paper include a derivation of simple, generic quantum Boltzmann equations (qBE) for fermions and scalars. Fermionic equations couple the usual mass-shell distribution functions to new coherence functions with compact expressions for the collision integrals as traces over Dirac-matrix projections of generic self-energy functions. The scalar field qBE's on the other hand are cast into a form of a finite set of moment equations, again with a well defined collision integral that can be evaluated diagrammatically by use of our cQPA Feynman rules. 

We use our generalized Feynman rules to compute several perturbative self-energy diagrams in the Yukawa theory, including the full coherent fermion and scalar field self-energies at the one loop level. We also study a 2-loop example which bears resemblance to a more involved calculation with flavour mixing fields~\cite{FHKR}, necessary for a resonant leptogenesis problem. These examples illustrate the ease of use of our formalism for diagrammatic evaluation of collision integrals involving the nonlocal quantum coherence effects. We also compute explicit expressions for the collision integrals following from the Yukawa interaction at the one-loop level and interpret these results in terms of the mass-shell and coherence-shell contributions in scattering (decay and inverse decay) processes. As a numerical application of our formalism, we solve the cQPA phase space distribution functions for a fermion with a complex, homogeneous time varying mass parameter evolving smoothly from one phase to another, in presence of decohering interactions. These results illustrate the effect of collisions on the production and evolution of the coherence and particle numbers in the presence of interactions. We also build a toy model for coherent baryogenesis (see eg. ref.~\cite{coherent_bg}) by assuming that the coherently evolving fermion $\psi$ decays into the ordinary standard model quarks. We compute the temporal evolution of the seed quark asymmetry and the ensuing baryon asymmetry and show that such a scenario, when embedded to a more complete particle physics model, could be an attractive mechanism for baryogenesis. In this paper we only consider the spatially homogeneous and isotropic problems. 

This paper is organized as follows. In section~\ref{sec:cQPA} we briefly review the derivation of the fermionic cQPA shell structure, and the equations of motion that follow from using the cQPA structure as an ansatz in the full dynamical Kadanoff-Baym equations. We also introduce a more convenient parametrization of the coherence propagators using the familiar 4-dimensional Dirac notation with projection operators. In section~\ref{sec:fermicollision} we present the resummation method giving rise to a consistent expansion of the fermion collision integrals. In section \ref{sec:scalar} we repeat the analysis for scalar fields, and in section~\ref{sect:effective-feynman} we derive the generalized Feynman rules for the computation of the resummed self-energies for a Yukawa-theory. In section \ref{sect:self-energies} we compute examples of the self-energy diagrams and collision integrals and in section~\ref{sec:applications} we apply the formalism to a toy-model of a coherent baryogenesis. Finally, section~\ref{sec:discussion} contains our conclusions and outlook.

\section{cQPA for Fermions}
\label{sec:cQPA}
We will be using the Schwinger-Keldysh approach~\cite{SK-formalism} to non-equilibrium quantum field theory (for a review see ref.~\cite{CalHu08}). The basic objects of interest for us are the dynamical Wightman functions\footnote{Note that our sign-convention for $S^<$ is opposite to the usual one, see eg. ref.~\cite{PSW}.} $S^{<,>}$:
\begin{equation}
iS^<(u,v) = \langle \bar\psi(v)\psi(u) \rangle \quad {\rm and} \quad iS^>(u,v) = \langle \psi(u) \bar\psi(v)\rangle\,,
\label{basicfunctions}
\end{equation}
where $\langle\ldots\rangle \equiv {\rm Tr}\{\hat\rho\ldots\}$ denote the expectation values \wrt some unknown quantum density operator $\hat\rho$. 
The time-ordered (Feynman) and anti-time-ordered (anti-Feynman) 2-point functions are then related to the Wightman functions by
\def\phm{\phantom -}
\begin{eqnarray}
  S_F(u,v) &\equiv&   \theta(u_0-v_0) S^>(u,v)
                                               - \theta(v_0-u_0) S^<(u,v)
   \nonumber\\
  S_{\bar F}(u,v) &\equiv&  \theta(v_0-u_0) S^>(u,v)
                                               - \theta(u_0-v_0)S^<(u,v)\,.
\label{SFs}
\end{eqnarray}
The basic idea in the SK-approach is to write down a (Schwinger-Dyson) hierarchy of equations for all Greens functions in the theory, thus bypassing the need to define $\hat\rho$ at all. The main approximation is a truncation of this hierarchy to a closed set of equations for a restricted set of correlators. In practice the truncation is often done already at the level of the 2-point functions, assuming that the self-energy functions in the SK-equations can be evaluated perturbatively. In this case only the two Wightman functions (\ref{basicfunctions}) are needed get the closure\footnote{Given the Wightman functions, the spectral function is found from ${\cal A}=\frac{i}{2}(S^>+S^<)$.  After this the spectral relation $S_H(u,v)=-i\,{\rm sgn}(u^0-v^0){\cal A}(u,v)$ fixes the retarded and advanced correlators $S_{r,a} \equiv S_H \pm i{\cal A}$, which completes the maximal set of independent 2-point functions.}. It is convenient to separate the internal and external scales by performing Wigner transformations to the 2-point functions:
\begin{equation}
S(k,x) \equiv \int {\rm d}^{\,4} r \, e^{ik\cdot r} S(x + \sfrac{r}{2},x-\sfrac{r}{2}) \,,
\label{wigner1}
\end{equation}
where $x\equiv (u+v)/2$ is the average coordinate, and $k$ is the
internal momentum conjugate to the relative coordinate $r\equiv
u-v$. It is straightforward to show~\cite{PSW,HKR2} that in the mixed representation the correlators $S^{<,>}$ satisfy the Kadanoff-Baym equations~\cite{KadBay62}:
\begin{equation}
(\kdag + \frac{i}{2} \deldag_x - \hat m_0
      - i\hat m_5 \gamma^5) S^{<,>}
 -  e^{-i\Diamond}\{ \Sigma_H \}\{ S^{<,>} \}
 -  e^{-i\Diamond}\{ \Sigma^{<,>} \}\{ S_H \}
 = \pm {\cal C}_{\rm coll},
\label{DynEqMix}
\end{equation}
where ``$+$" refers to ``$<$" and ``$-$" to ``$>$" . The collision term is given by
\begin{equation}
{\cal C}_{\rm coll} = \frac{1}{2}e^{-i\Diamond}
                            \left( \{\Sigma^>\}\{S^<\} -
                                   \{\Sigma^<\}\{S^>\}\right)\,,
\label{collintegral}
\end{equation}
where the $\Diamond$-operator is the following generalization of the
Poisson brackets:
\begin{equation}
\Diamond\{f\}\{g\} = \frac{1}{2}\left[
                  \partial_x f \cdot \partial_k g
                - \partial_k f \cdot \partial_x g \right]\,.
\label{diamond}
\end{equation}
We are interested in problems where a complex mass function can be spatially or temporally varying $m \equiv m_R(x) + im_I(x)$, whereby the nontrivial mass operators $\hat m_0$ and $\hat m_5$ appear in Eq.~(\ref{DynEqMix}). Their action on $S$-functions is defined as:
\begin{equation}
\hat m_{\rm 0,5} S(k,x) \equiv m_{R,I}(x) e^{-\frac{i}{2}
      \partial_x^m \cdot \partial_k^S} S(k,x)\,.
\label{massoperators}
\end{equation}
The self-energies $\Sigma$ are some complicated functionals of the correlators $S^{<,>}$, and they can be computed using some explicit truncation scheme, such as the 2PI effective action in loop expansion \cite{2PI,CalHu08}. From Eq.~(\ref{DynEqMix}) we immediately see that the spectral function ${\cal A}=\sfrac{i}{2}(S^>+S^<)$ obeys an equation identical to Eq.~(\ref{DynEqMix}), but with a vanishing collision term. In addition the spectral function needs to satisfy the sum-rule:
\begin{equation}
 \int \frac{{\rm d}k_0}{\pi}
                 {\cal A}(k,x) \gamma^0 = 1\,,
\label{sumrule2}
\end{equation}
which follows for example from the equal time anti-commutation relations of the fermionic fields. We will later see that this constraint is strong enough to completely fix the spectral function in our approximation scheme.
Equations (\ref{DynEqMix}) for $S^{<}$ and ${\cal A}$, together with the sum-rule (\ref{sumrule2}) form a complete set of equations when the interactions (a scheme to compute $\Sigma$) and the mass profiles are specified. In practice, these equations are too hard to be solved in their full generality, and several approximations are needed to find a solvable set of equations.

\subsection{Approximations}
\label{sec:approximations}
The usual kinetic approach to non-equilibrium quantum field theory \cite{kinetic,PSW,CalHu08} considers weak interactions and slowly varying classical backgrounds. In this case the excitation spectrum of the system can be reasonably approximated by a singular shell structure and the KB-equations (\ref{DynEqMix}) reduce to Boltzmann equations for the on-shell excitations called quasiparticles. Moreover, the standard approach typically makes the assumption that the system is, as a consequence of being close to a thermal equilibrium, nearly translation invariant. This implies that the phase space singularity structure is also essentially the same as in thermal equilibrium. The key idea leading to the {\em coherent} quasiparticle approximation~\cite{HKR1,HKR2,HKR3,Glasgow,Thesis_Matti} scheme is to relinquish the assumption of nearly translation invariance. The ensuing new solutions which describe the nonlocal quantum coherence are typically oscillatory in the space and time coordinates in quantum scales $x \sim k^{-1}$, which explicitly violates the translation invariance.  To find the cQPA we make the following approximations to Eqs.~(\ref{DynEqMix}):

\begin{itemize}
\item[1)]
We neglect the terms $\propto S_H$ on the LHS of the KB-equations (\ref{DynEqMix}). 
\item[2)]
We neglect the terms $\propto \Sigma_H$ on the LHS of the KB-equations (\ref{DynEqMix}).
\item[3)]
We consider the spatially homogeneous and isotropic case with $\nabla S^<=0$.
\end{itemize}

\noindent 
The first approximation above is made also in the standard kinetic theory and it should be reasonable at least in the limit of weak interactions. It has been shown that for a scalar field close to thermal equilibrium the neglected terms are of higher order in the interaction rate $\Gamma$ than the dominant contribution from the collision term \cite{PSW}. From the practical point of view, these terms must be neglected in order to find a singular phase space structure for the spectral function, even in the usual Boltzmann approach. 
The second approximation is made only for simplicity; the $\Sigma_H$ corrections would merely give rise to modified dispersion relations for the quasiparticles, but not change the generic structure of the theory. These terms could be incorporated later on. The third assumption on the other hand is crucial. The simple spectral structures for the coherence solutions, which is the heart of the cQPA, arise only in systems with particular space-time symmetries, including spatially homogeneous and isotropic, and stationary planar symmetric cases~\cite{HKR1}. This paper is devoted to the former of these symmetries. 

With the approximations 1-3 the KB-equations (\ref{DynEqMix}) can be reduced and decomposed into the following Hermitian (H) and anti-Hermitian (AH) parts:
\begin{eqnarray}
{\rm (H)}:\quad 2k_0 \bar S^< &=& \hat H \bar S^< + \bar S^< \hat H^\dagger + i\gamma^0 ({\cal C}_{\rm coll} - {\cal C}_{\rm coll}^\dagger)\gamma^0
\label{Hermitian22}
\\ 
{\rm (AH)}:\quad i \partial_t \bar S^< &=& \hat H \bar S^< - \bar S^< \hat H^\dagger + i\gamma^0 ({\cal C}_{\rm coll} + {\cal C}_{\rm coll}^\dagger)\gamma^0\,,
\label{AntiHermitian22}
\end{eqnarray}
where we defined a hermitian Wightman function $\bar S^< \equiv iS^<\gamma^0$, and the operator
\begin{equation}
\hat H \equiv {\bf k} \cdot {\alpha} 
             + \gamma^0 \hat m_0  + i \gamma^0\gamma^5 \hat m_5
\label{Hamiltonian}
\end{equation}
can be interpreted as a local free field Hamiltonian. Because of the spatial homogeneity the mass functions $m_{R,I}(t)$ depend now only on time. Note that in the collisionless limit (${\cal C}_{\rm coll}\rightarrow 0$) and in the zeroth order in gradients ($\hat m_{\rm 0,5} \rightarrow m_{R,I}$) only the AH-equation contains an explicit time derivative of $S^<$, while the hermitian equation becomes a purely algebraic matrix equation. For this reason Eq.~(\ref{AntiHermitian22}) is called a {\em ``kinetic equation}'', and it describes the time evolution of the Wightman function $S^<$. The role of the hermitian {\em ``constraint equation}'' (\ref{Hermitian22}) on the other hand is to restrict the phase space structure of $S^<$.

Even after approximations 1-3 the resulting equations (\ref{Hermitian22}-\ref{AntiHermitian22}) remain formidable to solve. Note in particular that they still involve the mass operators containing terms which are of arbitrarily high order in gradients. In order to find a tractable set of equations we will make the following two additional approximations:

\begin{itemize}
\item[4)]
We will expand the non-dynamical KB-equation (\ref{Hermitian22}) 
to the zeroth order in the scattering width $\Gamma = \frac{i}{2}(\Sigma^> + \Sigma^<)$ and the mass gradients $\partial_t m$. 
This gives rise to the singular cQPA phase space structure.
\item[5)]
We insert the singular cQPA structure as an {\em ansatz} to the dynamical KB-equation~(\ref{AntiHermitian22}), and expand also this equation to the lowest nontrivial order in $\Gamma$ and $\partial_t m$.
\end{itemize}
As we shall find out in the subsection~\ref{sec:shell} below, the cQPA phase space structure contains two independent mass-shell distribution functions $f_{mh\pm}$ and two coherence distribution functions $f_{ch\pm}$ for each helicity $h$. When this shell structure is fed into the dynamical equation~(\ref{AntiHermitian22}) and the equation is integrated over $k_0$, it reduces to the sought after quantum Boltzmann equations for the on-shell functions $f_\alpha$. We shall derive these equations in the subsection~\ref{sec:eom}. However, because the coherence-shell functions are rapidly oscillating in time, a special care is needed when expanding the collision terms in these equations. A novel resummation scheme that gives a consistent expansion of the collision integrals to leading order in $\Gamma$ and $\partial_t m$ will be introduced in section~\ref{sec:fermicollision}.

\subsection{cQPA phase space structure}
\label{sec:shell}

The most general fermionic 2-point function $S(k,t)$ consistent with spatial homogeneity and isotropy consists of the terms with products of $\gamma^0$, ${\bf k} \cdot {\gamma}$ and $\gamma^5$ only, and thus spans 8 of 16 independent components of the full Dirac algebra. Furthermore, we observe that the helicity operator: $\hat h = ({\bf k} \cdot \gamma^0 \vec \gamma \gamma^5)/|{\bf k}|$, commutes with the Hamiltonian (\ref{Hamiltonian}), implying that different helicity projections of $S^<$ do not mix in a collisionless theory, \ie helicity is a good quantum number. For this reason, it is convenient to decompose the 8 independent components of the 2-point function $S^<(k,t)$ in terms of helicity projectors: $P_h = \frac12(1 + h \hat h)$, $h = \pm 1$ to get:  
\begin{equation}
i S^<(k,t)  = \frac12 \sum_{h=\pm 1} P_h \big[\gamma^0 g_{h0} + h \mathbf{\hat k} \cdot \gamma\,g_{h3} + g_{h1} + i\gamma^5 g_{h2}\big]\,,
\label{bloch_prop}
\end{equation}
where $g_{h\alpha}(k,t)$ are the 8 real scalar components. Using the the chiral (Weyl) representation of the Dirac algebra: $\gamma^0 = \sigma^1 \otimes 1$, $\vec \gamma = i\sigma^2 \otimes \vec \sigma$ and $\gamma^5 = -\sigma^3 \otimes 1$, the helicity block-diagonal Wightman function in Eq.~(\ref{bloch_prop}) can be written as
\begin{equation}
 \bar S^<(k,t) = \sum_{h=\pm 1} g_h^<(k,t) \otimes
 \frac12(1 + h \hat {\bf k} \cdot \sigma)\,,
\label{chiral_basis}
\end{equation}
where the Bloch representation of chiral part is just $g^<_h \equiv \frac12 \left(g_{h0}^< +  g_{hi}^< \sigma^i \right)$.

We now proceed to step 4 in our approximation scheme and analyze the constraint equation (\ref{Hermitian22}) in the zeroth order in $\Gamma \sim \Sigma^{<,>}$ (neglecting the collision term) and $\partial_t m$. Inserting this decomposition into the constraint equation (\ref{Hermitian22}) reduces it to two independent homogeneous matrix equations for the helicities $h=\pm 1$ (note that the index ordering is here defined as $\alpha = 0,3,1,2$):
\begin{equation}
\sum_\beta (B_h)_{\alpha\beta}\,g_{h\beta}^< = 0 
\,,\qquad B_h = \left( \begin{array}{cccc}
   k_0  &  h|{\bf k}|  &  -m_R  &  m_I \\
  h|{\bf k}| &  k_0   &  0     &  0   \\
  -m_R &  0     &  k_0   &  0   \\
  m_I  &  0     &  0     &  k_0 
  \end{array} \right) \,.
\label{bloch_constraint}
\end{equation}
Equation (\ref{bloch_constraint}) may have nontrivial solutions only if 
\begin{equation}
\det(B_h) = \left(k^2 - |m|^2 \right)k_0^2 = 0\,.
\label{constraint_det}
\end{equation}
This constraint gives rise to a {\em singular shell structure}, because the solutions need to be proportional to either $\delta(k_0^2 - {\bf k}^2 - |m|^2)$ or $\delta(k_0)$. The full matrix structure of the corresponding singular solutions can then be worked out by setting $k_0 \neq 0$ and $k_0=0$ in Eq.~(\ref{bloch_constraint}), respectively, so that the combined solution is eventually found to be~\cite{HKR1,HKR2}:
\begin{eqnarray} iS^<(k,t) &=& \sum_h P_h \bigg[2\pi\,{\rm sgn}(k_0)
   (\kdag + m_R -i \gamma^5 m_I ) f_{mh\,{\rm sgn}(k_0)}^< \delta(k^2-|m|^2) 
\nonumber\\
&& \quad + \pi\Big[\hat {\bf k} \cdot  \gamma 
\Big(\frac{m_R}{|{\bf k}|}f_{h1}^< + \frac{m_I}{|{\bf k}|}f_{h2}^<\Big) + f_{h1}^< -i \gamma^5 f_{h2}^<  \Big]\delta(k_0)\bigg]\,.
\label{full_wightman_OLD}
\end{eqnarray}
Here the first line is the standard one-particle mass-shell solution with dispersion relations $k_0 = \pm \omega_{\bf k} \equiv \pm ({\bf k}^2 + |m|^2)^{1/2}$, while the second line is a new coherence solution living at shell $k_0=0$. The real mass-shell distribution functions $f_{mh\pm}^<({\bf k},t)$ can be related to the phase space densities for particles and antiparticles via the Feynman-St{\"u}ckelberg interpretation\footnote{We associate the negative energy states with the CP- rather than C conjugates to the positive energy states: $\bar n_{{\bf k}h} \equiv (n_{{\bf k}h})^{\rm CP}$.}: 
\beq
n_{{\bf k}h}(t) = f_{mh+}^<({\bf k},t) 
\qquad
\bar n_{{\bf k}h}(t) = 1 - f^<_{mh-}(-{\bf k},t)
\label{Fey-Stuck}
\eeq
and the new (real) $k_0=0$-shell distribution functions $f_{h1,2}^<({\bf k},t)$ describe nonlocal quantum coherence between these modes. The representation (\ref{full_wightman_OLD}) for $S^<$ was derived in refs.~\cite{HKR1,HKR2}. For our purposes here it is actually more convenient to reparametrize the $k_0=0$ coherence-shell functions as follows:
\begin{equation}
f_{ch\pm}^< \equiv \frac{\omega_{\bf k}^2}{2(\omega_{\bf k}^2-m_R^2)}\bigg[f_{h1}^< \mp \frac{ih}{\omega_{\bf k}} \Big(|{\bf k}| f_{h2}^< + m_{I}\Big(\frac{m_R}{|{\bf k}|}f_{h1}^< + \frac{m_I}{|{\bf k}|}f_{h2}^< \Big)\Big)\bigg]\,.
\end{equation}
Using these new functions the Wightman function Eq.~(\ref{full_wightman_OLD}) 
becomes 
\begin{eqnarray}
iS^<(k,t) &=& 2\pi \sum_{h\pm} \pm \frac{1}{2\omega_{\bf k}} P_h 
 (\kdag_{\pm} + m_R -i \gamma^5 m_I)
\nonumber\\
&& \qquad\qquad\quad\;\;\,\times
 \Big[f_{mh\pm}^< \,\delta(k_0 \mp \omega_{\bf k}) 
+ \Big(\gamma^0 \mp \frac{m_R}{\omega_{\bf k}}\Big) f^<_{ch\pm}\,\delta(k_0)\Big]\,,
\label{full_wightman}
\end{eqnarray}
where we denote $k^{\mu}_{\pm} = (\pm\omega_{\bf k},{\bf k})$. The complete singular phase space structure (\ref{full_wightman}) is the coherent quasiparticle approximation for $S^<$. \\

\noindent {\em A note on conventions.} The shell functions appearing in Eq.~(\ref{full_wightman}) are labelled by three indices: $j=m,c$ referring either to mass- or coherence-shell, $h=\pm1$ referring to helicity, and $\pm$ referring to the positive and negative energy. These indices will be appearing, always in this order, in many other constructs throughout the paper. Sometimes one or the other of the indices is to be summed over, and in order to suppress the notational clutter we will always use an implied summing convention, where a non-displayed index is summed over.  That is, we use for example 
\begin{equation}
\sum_{jh\pm} Q_{jh\pm} =  \sum_{jh} Q_{jh} = \sum_j Q_j = Q \,,
\label{eq:notaatio}
\end{equation}
where $Q$ can be for example $S^{<,>}$, $f^{<,>}$ or any other construct defined using them. 

Note that the time dependence of the on-shell functions $f_\alpha$ is not defined yet; in the next section we will use the cQPA  structure as an ansatz in the dynamical equation (\ref{AntiHermitian22}) to derive the generalized quantum Boltzmann equations for $f_\alpha$. First however, we have to determine the cQPA propagators also for the function $S^>$, the spectral function and the pole propagators.

\subsubsection{Pole functions and thermal limit}

In the collisionless limit $S^>$ and the spectral function ${\cal A}=\frac{i}{2}(S^>+S^<)$ satisfy the same constraint and kinetic equations as $S^<$. The spectral solution must then be identical to Eq.~(\ref{full_wightman}) with four yet undefined on-shell functions $f^{\cal A}_{h\alpha}$ for both helicities. However, the spectral function must in addition satisfy the sum rule (\ref{sumrule2}), which completely fixes these functions: $f^{\cal A}_{mh\pm} = \frac12$ and $f^{\cal A}_{ch\pm} = 0$~\cite{HKR1}. The spectral function then reduces to the familiar form:
\begin{equation}
{\cal A}(k,t) = \pi {\rm sgn}(k_0) (\kdag +  m_R - i \gamma^5 m_I)
           \delta(k^2-|m|^2) \,.
\label{full_spectral}
\end{equation}
The retarded and advanced propagators are directly related to the spectral function: $S_{r,a}(u,v) = \mp2i\,\theta(\pm(u_0-v_0)){\cal A}(u,v)$, so that they are also given by their standard expressions:
\begin{equation}
  S_{r,a}(k,t) = \frac{\kdag +  m_R - i \gamma^5 m_I}
                      {k^2-|m|^2 \pm i{\rm sgn}(k_0) \epsilon}\,.
\label{eq:standardpole}
\end{equation}
Next, the relation ${\cal A} = \frac{i}{2}(S^>+S^<)$ together with Eqs.~(\ref{full_wightman}) and (\ref{full_spectral}) completely specify $S^>$; it has an expression identical to Eq.~(\ref{full_spectral}), where the on shell functions are given by
\begin{equation}
f_{mh\pm}^> = 1 - f_{mh\pm}^< \qquad {\rm and} \qquad f_{ch\pm}^> = -f_{ch\pm}^<\,.
\label{lessgreater}
\end{equation}
Thus, the four on-shell functions $f^<_{\alpha}$ are the only dynamical variables in the cQPA; all other 2-point functions can be expressed in terms of these. For this reason we only need to consider the dynamical equations for $S^<$ and the on-shell functions $f^<_{\alpha}$. 

Let us stress that the new coherence solutions appear only in the dynamical functions $S^{<,>}$ in the cQPA, but not in the spectral function ${\cal A}$ or in the pole propagators $S_{r,a}$. This is quite natural, as it merely expresses the fact that there cannot be coherence without the interfering mass-shell excitations; coherence is a purely dynamical phenomenon. Of course the Feynman and anti-Feynman propagators {\em do} have well defined coherence contributions that can be computed from their respective definitions: 
\begin{eqnarray} 
S_F &=& S_r - S^<
\nonumber \\
S_{\bar F} &=& -S_a - S^< \,.
\label{eq:standardfeynman}
\end{eqnarray} 
To conclude this section we note that in thermal equilibrium limit the full translation invariance kills the coherence-shell contributions also from the Wightman functions $S^{<,>}$~\cite{HKR1}. Moreover, the thermal density operator implies Kubo-Martin-Schwinger (KMS) relation \cite{KMS}: $S^>_{\rm eq}(k_0) = e^{\beta k_0}S^<_{\rm eq}(k_0)$, which fixes the remaining mass-shell functions to Fermi-Dirac distribution: $f_{mh\pm}^< \rightarrow f_{\rm eq}(\pm \omega_{\bf k}) \equiv 1/(e^{\pm \beta \omega_{\bf k}} + 1)$.
This reduces $S^<$ to the standard thermal form \cite{LeBellac00}:
\begin{equation}
iS^<_{\rm eq} = 
      2\pi\,{\rm sgn}(k_0) (\kdag +  m_R - i\gamma^5 m_I) \,f_{\rm
        eq}(k_0) \,\delta(k^2-|m|^2)\,.
\label{thermal_full}
\end{equation}
The thermal limits of the other standard propagators can then be found by using the general relations given above. 

\subsection{Quantum Boltzmann equations}
\label{sec:eom}
As the final step in our approximation scheme, we insert the spectral cQPA-solution (\ref{full_wightman}) as an ansatz into the kinetic equation (\ref{AntiHermitian22}). This will lead to a closed set of equations of motion for the dynamical variables $f_{h\alpha}^<$. Let us first define a useful shorthand notation:
\begin{equation}
  iS^< \equiv 2\pi \sum_{h\pm} \big( {\cal S}^<_{mh\pm}\delta(k_0 \mp \omega_{\bf k}) 
                       + {\cal S}^<_{ch\pm}\delta(k_0) \big) \,,
\label{specsol2}
\end{equation}
\vskip-0.2truecm
\noindent
where the spin-matrix functions ${\cal S}^<_{jh\pm}$ are easily read off from Eq.~(\ref{full_wightman}). Inserting this form to the kinetic equation (\ref{AntiHermitian22}) and integrating over $k_0$ we find the master equation:
\begin{equation}
\partial_t \bar{\cal S}^<  = -i [H, \bar{\cal S}^<] + \gamma^0 \langle{\cal C}_{\rm coll} + {\cal C}_{\rm coll}^\dagger\rangle\gamma^0\,,
\label{eom_prop}
\end{equation}
where we denoted $\langle\ldots\rangle \equiv \int \frac{{\rm d} k_0}{2\pi}\,(\ldots)$ and, in accordance with the notation (\ref{specsol2}):
\beq
\langle \bar S^< \rangle = \bar {\cal S}^< = \sum_{h\pm} (\bar {\cal S}^<_{mh\pm} + \bar {\cal S}^<_{ch\pm})Ê\,,
\eeq
\vskip-0.2truecm \noindent
where $\bar {\cal S}^<_{jh\pm} \equiv {\cal S}^<_{jh\pm}\gamma^0$. Note that all $k_0$-derivatives in the operator $\hat H$ have vanished in the integration process as total derivatives, reducing $\hat H$ to the usual local Hamiltonian:
\begin{equation}
H = {\bf k} \cdot \alpha + \gamma^0 m_R + i\gamma^0\gamma^5 m_I \,.
\label{localhamiltonian}
\end{equation}
The master equation (\ref{eom_prop}) is in fact the most compact form of the quantum transport equations in the cQPA scheme. One can see that the lowest moment of the correlation function, $\langle \bar S^< \rangle = \bar {\cal S}^<$, corresponds to a density matrix in the Dirac indices, and what remains to be done is to express the collision term in Eq.~(\ref{eom_prop}) in terms of the degrees of freedom included in $\bar {\cal S}^<$. This is of course not possible in general, but it can be done in the context of the spectral cQPA ansatz. In the case of fermions it is convenient to go further than just computing the collision matrices however, and rewrite Eq.~(\ref{eom_prop}) as a set of scalar Boltzmann equations for the on-shell functions $f_\alpha$, in exact analogy to the ordinary Boltzmann equation.

We begin by writing the integrated correlation function $\bar {\cal S}^<$ as (we will drop the superscript $<$ on $f_\alpha$ when there is no risk of confusion):
\begin{equation}
\bar {\cal S}^< 
= \sum_{h\pm} P_h P_{{\bf k}\pm} \Big[f_{mh\pm} 
                               + {\cal R}_{{\bf k}\pm} f_{ch\pm}\Big] \,,
\label{int_wightman}
\end{equation}
\vskip-0.1truecm \noindent
where
\begin{equation}
P_{{\bf k}\pm} \equiv \frac12\Big(1 \pm \frac{H}{\omega_{\bf k}}\Big) \quad 
{\rm and} \quad {\cal R}_{{\bf k}\pm} \equiv \gamma^0 \mp \frac{m_R}{\omega_{\bf k}}\,.
\end{equation}
Here $P_{{\bf k}\pm}$ are positive and negative energy projection operators. Indeed, using the the identity $H^2 = \omega_{\bf k}^2$ one can easily show that $H P_{{\bf k}\pm} = \pm\omega_{\bf k} P_{{\bf k}\pm}$ and then
\begin{equation}
  P_{{\bf k} +}P_{{\bf k} -} = P_{{\bf k} -}P_{{\bf k} +} = 0 \quad {\rm and}Ê\quad
  P_{{\bf k}\pm}^2 = P_{{\bf k}\pm}\,.
\label{proj_identity1}
\end{equation}
Moreover, the coherence matrix ${\cal R}_{{\bf k}\pm}$ and energy projectors $P_{{\bf k}\pm}$ obey the relations:
\begin{eqnarray}
 P_{{\bf k}\pm}{\cal R}_{{\bf k}\pm} = {\cal R}_{{\bf k}\mp} P_{{\bf k}\mp} = P_{{\bf k}\pm}\gamma^0 P_{{\bf k}\mp}\,,
\label{PR_identity}
\end{eqnarray}
where the last equality follows from the anticommutator $\{ H,\gamma^0\} = 2 m_R$.
The matrix ${\cal R}_{{\bf k}\pm}$ is not a projector, but combined with the energy projector it acts as an effective projector onto the coherence solution under the trace operation, such that the on-shell functions $f_\alpha$ can be solved from Eq.~(\ref{int_wightman}) tracing over Dirac algebra:
\begin{eqnarray}f_{mh\pm} &=& {\rm Tr}\Big[P_{{\bf k}\pm} P_h \, \bar {\cal S}^< \Big]
\nonumber\\
f_{ch\pm} &=&\xi_{\bf k}  {\rm Tr}\Big[{\cal R}_{{\bf k}\pm} P_{{\bf k}\pm} P_h \, \bar {\cal S}^< \Big]\,,
\label{on-shell_proj}
\end{eqnarray}
where $\xi_{\bf k} \equiv \omega_{\bf k}^2/(\omega_{\bf k}^2 - m_R^2)$. The equations of motion for the on-shell functions $f_{h\alpha}$ can now be obtained from the master equation (\ref{eom_prop}) by forming the appropriate traces, or alternatively by taking time-derivatives of Eqs.~(\ref{on-shell_proj}) and using the kinetic equation (\ref{eom_prop}) for $\partial_t \bar{\cal S}^<$, the identities (\ref{proj_identity1}-\ref{PR_identity}) and finally computing the traces over Dirac algebra. After some algebra one finds the sought after {\em quantum Boltzmann equations}:
\begin{eqnarray}
  \partial_t f_{mh\pm} &=& \pm \frac12 \big( \Phi_{{\bf k}h+}^\prime f_{ch+} + \Phi^\prime_{{\bf k}h-} f_{ch-}\big) + {\cal C}_{mh\pm}[f_\alpha]
\label{eom_on-shell1}
\\
  \partial_t f_{ch\pm} &=& \mp i 2\omega_{\bf k} f_{ch\pm} +  \xi_{\bf k}\Phi^\prime_{{\bf k}h\mp} \Big[\frac{m_R}{\omega_{\bf k}} f_{ch\pm} - \frac12 \Delta f_{mh}\Big] + {\cal C}_{ch\pm}[f_\alpha]\,,
\label{eom_on-shell2}
\end{eqnarray}
where $\Delta f_{mh} \equiv f_{mh+} - f_{mh-}$ and all dependence on the mass gradients is encoded in 
\begin{equation}
\Phi^\prime_{{\bf k} h\pm} \equiv \Big(\frac{m_R}{\omega_{\bf k}}\Big)^\prime 
              \pm ih \frac{|{\bf k}| m_I^\prime}{\omega_{\bf k}^2}\,.
\end{equation}
After using cyclicity of trace the collision integrals ${\cal C}_{jh\pm}$ can be written as:
\begin{eqnarray}
{\cal C}_{mh\pm}[f_\alpha] &=& {\rm Tr}\Big[\gamma^0 \langle{\cal C}_{\rm coll} + {\cal C}_{\rm coll}^\dagger\rangle\gamma^0 P_{{\bf k}\pm} P_h\Big]\,,
\label{coll_integrals1}
\\
{\cal C}_{ch\pm}[f_\alpha] &=& \xi_{\bf k} {\rm Tr}\Big[\gamma^0 \langle{\cal C}_{\rm coll} + {\cal C}_{\rm coll}^\dagger\rangle\gamma^0{\cal R}_{{\bf k}\pm} P_{{\bf k}\pm}P_h\Big]\,.
\label{coll_integrals2}
\end{eqnarray}
Eqs.~(\ref{eom_on-shell1}-\ref{eom_on-shell2}) together with the collision integrals (\ref{coll_integrals1}-\ref{coll_integrals2}) form a closed set of equations of motion for the on-shell functions $f_{\alpha}$, once the self-energy functionals $\Sigma^{<,>}$ appearing in the collision integrals are specified. Beyond the introduction of the convenient 4-dimensional notation, the main contribution of this paper is to show how the collision terms (\ref{coll_integrals1}-\ref{coll_integrals2}) can be consistently computed to a given (lowest) order in gradients and coupling constants. As we shall see, this is a nontrivial task due to the need for resummation of the coherence propagators, which eventually leads to a new set of momentum space Feynman rules.

\subsection{Resummed fermion collision term}
\label{sec:fermicollision}

The basic quantity appearing in the projected collision terms (\ref{coll_integrals1}-\ref{coll_integrals2}) is the zeroth moment of the collision term $\langle {\cal C}_{\rm coll} \rangle$. We shall now carefully explain how this quantity can be expanded consistently to the leading order in the interaction width and the gradient of the mass function, denoted by ${\cal O}^1 \equiv {\cal O}(\Gamma, \partial_t m)$. Let us begin with a straightforward, but limited in validity, derivation using the mixed space representation:
\begin{equation}
\langle {\cal C}_{\rm coll} \rangle 
= \frac{1}{2}\int \frac{{\rm d} k_0}{2\pi}\,e^{-i\Diamond}
                             \big( \{\Sigma^>(k,t)\}\{S^<(k,t)\} -
                                    \{\Sigma^<(k,t)\}\{S^>(k,t)\}\big) \,.
\label{EffColl1}
\end{equation}
Naively, one would simply truncate the $\Diamond$-expansion to the lowest order, as was done in refs.~\cite{HKR1,HKR2,HKR3}. However, it was later realized that because of their oscillatory behaviour the coherence-shell functions contribute to the leading order term at all orders of the $\Diamond$-expansion~\cite{Glasgow,Thesis_Matti}. We can see this quite easily when we observe that to order ${\cal O}^1$ the equations of motion (\ref{eom_on-shell1}-\ref{eom_on-shell2}) imply that $\partial_t f_{mh\pm} = 0$ and  $\partial_t f_{ch\pm} = \mp \,2 i\omega_{\bf k} f_{ch\pm}$. This holds true also for the full mass-shell and coherence-shell Wightman functions (in the distribution sense):
\begin{equation}
\partial_t S_{m\pm}^< = {\cal O}^1\,,\qquad\quad
\partial_t S_{c\pm}^< = \, \mp \, 2i\omega_{\bf k} S_{c\pm}^< 
  + {\cal O}^1\,,
\label{zeroth_eq0}
\end{equation}
where $S^< \equiv \sum_\pm ( S^<_{m\pm} + S^<_{c\pm})$ in accordance with the  notation (\ref{eq:notaatio}). If we make the additional {\em assumption} that $\partial_t\Sigma \sim {\cal O}^2$, it is easy to show that $(-i\Diamond)^p \{\Sigma^>(k,t)\}\{S^<(k,t)\} 
= ( (\pm \omega_{\bf k} \partial_{k_0})^p \Sigma^>) S_{c\pm}^<  + {\cal O}^2$ when $p\ge1$. That is, we can use the naive expansion for the mass-shell part, but for the coherence-shell part we have to replace the $-i\Diamond$-operator by $\pm \omega_{\bf k}\partial_{k_0}$ acting only on the self-energy function $\Sigma^{>}$:
\begin{equation}
\langle {\cal C}_{\rm coll} \rangle 
\approx
\frac{1}{2}\sum_{\pm}  \int \frac{{\rm d} k_0}{2\pi}\, 
     \Big( \Sigma^> S^<_{m\pm} 
         + (e^{\pm\omega_{\bf k}\partial_{k_0}}\Sigma^>)S^<_{c\pm} 
          -  (\, > \,\,  \leftrightarrow \, \,  < \,) \Big) 
          \,,
\label{effectiveC2} 
\end{equation}
where we have dropped all terms of order ${\cal O}^2$ or higher.
We can now see that the exponential gradient operator in Eq.~(\ref{effectiveC2}) precisely translates the self-energy functions associated with the coherence solutions $S_{c\pm}$ from the actual $k_0=0$ shell to the corresponding mass shells. This feature will persist also in the more accurate treatment where the smallness of  $\partial_t\Sigma$ is not assumed.

\subsubsection{Consistent expansion and $\Sigma_{\rm eff}$}
\label{sec:collisionold}

Equation (\ref{effectiveC2}) consistently accounts for the {\em resummation} of all zeroth order gradient corrections from an external coherence-shell propagator in the collision integral. This introduces a correction to the naive gradient expansion of Eq.~(\ref{EffColl1}). It is easy to see that the same issue is encountered with the coherence propagators that appear in the perturbative expansion of the self-energy $\Sigma$. Indeed, whenever the time derivative acts on an internal coherence propagator in $\Sigma$, it returns as a leading term that propagator multiplied by $\pm 2i\omega_q$, where $\omega_q$ is the energy associated with the propagator. Such terms do not contain any gradient suppression and the assumption $\partial_t\Sigma \sim {\cal O}^2$ made below Eq.~(\ref{zeroth_eq0}) fails. Thus, we need a resummation scheme also for the coherence contributions inside $\Sigma$, before a consistent gradient expansion for $\langle {\cal C}_{\rm coll} \rangle $ can be established. We shall now formulate a general method of expanding $\langle {\cal C}_{\rm coll} \rangle$ consistently to the leading order. To this end, it is convenient to rewrite Eq.~(\ref{EffColl1}) in terms of the (partially transformed) coordinate space functions: 
\begin{equation}
\langle {\cal C}_{\rm coll} \rangle
= \frac{1}{2}\int {\rm d} w_0 \left[\Sigma^>(t,w_0,{\bf k})S^<(w_0,t,{\bf k}) - \Sigma^<(t,w_0,{\bf k})S^>(w_0,t,{\bf k})\right]\,. 
\label{EffColl2}
\end{equation}
Note that this expression has only one external time $t$; forcing equal times exactly corresponds to integrating over the energy in the mixed representation. Because we are only interested in the spatially homogeneous case, {\em all} gradients vanish in the expression (\ref{EffColl2}). Clearly, for the equations of motion (\ref{eom_prop}) to close, we need to be able to express $\langle {\cal C}_{\rm coll} \rangle$ entirely in terms of functions ${\cal S}_\alpha^<({\bf k},t)$. We begin by writing the propagators in the two-time representation:
\begin{eqnarray}S(w_0,w_0',{\bf k}) 
  &\equiv& \int {\rm d}^{\,3}({\bf w} - {\bf w}') \, 
  e^{-i{\bf k}\cdot({\bf w} - {\bf w}')} S(w,w')
\nonumber\\ 
&=& \int \frac{{\rm d} k_0}{2\pi}\, e^{-i k_0(w_0 - w_0')}S\big(k_0,{\bf k}, \frac{w_0+w_0'}{2}\big) \,.
\label{partfourier}
\end{eqnarray}
Inserting the spectral solution (\ref{specsol2}) to the definition (\ref{partfourier}) we get the two-time representation of our cQPA-Wightman functions ($t\equiv (u_0+v_0)/2$):
\begin{equation}
iS^{<,>}(u_0,v_0,{\bf k}) = \sum_\pm\left[e^{\mp i \omega_{\bf k}(u_0 - v_0)} 
{\cal S}_{m\pm}^{<,>}({\bf k},t) + {\cal S}_{c\pm}^{<,>}({\bf k},t) \right]\,.
\label{uv_prop}
\end{equation}
Note that $iS^{<}(t,t,{\bf k}) = {\cal S}^{<}({\bf k},t)$. We clearly need to expand the functions $S^{<,>}(w_0,t,{\bf k})$ appearing explicitly in Eq.~(\ref{EffColl2}). However, the perturbative expressions for the self-energy functions $\Sigma^{<,>}(t,w_0,{\bf k})$ in general involve further integrations over internal vertices $w_0'$, $w_0''$, and so we will need to expand correlators $S(w_0,w_0',{\bf k})$ with {\em arbitrary} time-coordinates $w_0$ and $w_0'$ with respect to the ``external" time $t=(u_0+v_0)/2$. This is easily done by Taylor expanding correlators around $t$: 
\begin{equation}
S^{<,>}\big(k_0,{\bf k}, \frac{w_0+w_0'}{2}\big)= \sum_{n=0}^\infty \frac{1}{n!}
    \Big(\frac{w_0+w_0'}{2}-t\Big)^n \partial_t^n S^{<,>}(k_0,{\bf k},t) \,.
\end{equation}
We can determine the derivative terms in the above equation by recursively using the zeroth order equations (\ref{zeroth_eq0}) for $n\ge 1$: 
\begin{equation}
\partial_t^n S_{m\pm}^< = {\cal O}^1\,,\qquad\quad
\partial_t^n S_{c\pm}^< = (\mp i2\omega_{\bf k})^n S_{c\pm}^< + {\cal O}^1\,.  
\label{zeroth_eq_higher}
\end{equation}
This implies that the Taylor expansion of the mass-shell functions is trivial to the lowest order in mass gradients, and that the coherence solutions get multiplied by a simple phase factor. Inserting these results to Eq.~(\ref{partfourier}) one finds:
\begin{equation}
iS^{<,>}(w_0,w_0',{\bf k}) \approx \sum_\pm\left[e^{\mp i \omega_{\bf k}(w_0 - w_0')} 
{\cal S}_{m\pm}^{<,>}({\bf k},t) + e^{\mp i \omega_{\bf k}(w_0 + w_0' - 2t)}
{\cal S}_{c\pm}^{<,>}({\bf k},t) \right] \,,
\label{eff_prop} 
\end{equation}
where we dropped terms of order ${\cal O}^1$. As a trivial consistency check we see that this expansion reduces to Eq.~(\ref{uv_prop}) as $w_0 \to u_0$ and $w_0' \to v_0$. Physically the propagator (\ref{eff_prop}) is an approximation which takes into account the rapid temporal variations due to the oscillations of the coherence functions, but neglects the corrections due to temporal variations in the background fields. In particular equation (\ref{eff_prop}) implies that $iS^{<,>}(w_0,t,{\bf k}) =  \sum_\pm \exp(\mp i\omega_{\bf k}(w_0-t)){\cal S}^{<,>}_\pm$, so that applying it to Eq.~(\ref{EffColl2}), we find
\begin{equation}
\langle {\cal C}_{\rm coll} \rangle
= -\frac{i}{2}  \sum_\pm \Big( 
     \Sigma^>_{\rm eff}(\pm \omega_{\bf k}) {\cal S}^<_\pm
   - \Sigma^<_{\rm eff}(\pm \omega_{\bf k}) {\cal S}^>_\pm
   \Big) \, ,
\label{effectiveC3}
\end{equation}
where
\begin{equation}
\Sigma^{<,>}_{\rm eff}(k_0,{\bf k},t) 
  \equiv
  \int {\rm d} w_0 e^{ i k_0  (t-w_0)}\Sigma^{<,>}(t,w_0,{\bf k})\,.
\label{effsigma}
\end{equation}
The result (\ref{effectiveC3}) is similar to the equation (\ref{effectiveC2}), although in the new derivation we did not assume the smallness of $\partial_t\Sigma$. This is as expected, as both equations account for the resummation of the external coherence propagator. However, the correct self-energy function $\Sigma_{\rm eff}$ in Eq.~(\ref{effsigma}) is not in general equivalent with the projected mixed representation function $\Sigma$ appearing in Eq.~(\ref{effectiveC2}).  Yet it is interesting to see that in the end, after the resummations have been carried out, Eq.~(\ref{effectiveC3}) can be written in an integral form {\em formally} equivalent to the naive lowest order gradient expansion of the equation~(\ref{EffColl1}):
\begin{equation}
\langle {\cal C}_{\rm coll} \rangle
\equiv \frac{1}{2} \int \frac{{\rm d}k_0}{2\pi }\big( 
     \Sigma^>_{\rm eff}(k,t) S^<_{\rm eff}(k,t)
   - \Sigma^<_{\rm eff}(k,t) S^>_{\rm eff}(k,t)
   \big) \,.
\label{effectiveC3b}
\end{equation}
The crucial point is that the original cQPA $S^{<,>}$-functions have been replaced by the effective ones:
\begin{equation}
iS^{<,>}_{\rm eff}(k,t) \equiv 2\pi \sum_{\pm} 
 \big( {\cal S}^{<,>}_{m\pm}({\bf k},t) + {\cal S}^{<,>}_{c\pm}({\bf k},t)\big) \,\delta(k_0 \mp \omega_{\bf k}) 
 \equiv iS^{<,>}_{m}  + iS^{<,>}_{c,\rm eff} \,. 
\label{effectiveS1}
\end{equation}
\vskip-0.2truecm\noindent
The resummed propagators $iS^{<,>}_{\rm eff}$ will also appear in perturbative expansions of the effective self-energies $\Sigma^{<,>}_{\rm eff}$.  In section \ref{sect:effective-feynman} we will return to this issue and derive a set of extended momentum space Feynman rules that can be used to formulate perturbative expansions and compute arbitrary self-energy diagrams in the cQPA framework.

At the first sight it might look surprising that the resummations effectively pushed the coherence shells back to the mass shells in the effective propagators appearing in the collision term. On the second thought this is of course perfectly natural: despite the fact that the $k_0=0$ solutions {\em are} real (they have been constructed explicitly in exactly solvable free theories elsewhere~\cite{HKR3,Joni}), they still physically correspond to coherence information between the on-shell states. What we are observing is our theory being consistent with the fact that it would make no sense to have a mere coherence information colliding with something; to have an interaction there always need to be a colliding particle present. When the coherence contributions are pushed to mass-shells their role is sensible: to modulate the collision frequencies of the mass-shell states depending on their state of coherence.


\subsubsection{Collision integrals for the on-shell functions}
\label{sec:cisforshellfs}

Given the generic result Eq.~(\ref{effectiveC3}) we can write the collision terms in Eqs.~(\ref{coll_integrals1}-\ref{coll_integrals2}) explicitly as traces over the effective self-energy functions  (\ref{effsigma}) multiplied by simple propagator matrices. To this end it is convenient to introduce some further notation, rewriting the coefficient matrices ${\cal S}_{mh\pm}$ and ${\cal S}_{ch\pm}$ as: 
\begin{eqnarray}
  {\cal S}_{mh\pm}^<({\bf k},t) 
   &\equiv& \frac{1}{2\omega_{\bf k}}{\cal K}_{mh\pm}({\bf k}) f_{mh\pm}({\bf k}) \,,
\nonumber\\
  {\cal S}_{ch\pm}^<({\bf k},t) 
   &\equiv& \frac{1}{2\omega_{\bf k}}{\cal K}_{ch\pm}({\bf k}) f_{ch\pm}({\bf k}) \,,
\label{4spectral2Cov}
\end{eqnarray}
where: 
\begin{eqnarray}
 {\cal K}_{mh\pm}({\bf k}) \equiv (2\omega_{\bf k})P_{{\bf k}\pm} P_h({\bf k})\gamma^0 = \pm (\kdag_{\pm} +  m_R - i\gamma^5 m_I)P_h({\bf k}) 
\label{PR_identity3}
\end{eqnarray}
and 
\begin{eqnarray}
 {\cal K}_{ch\pm}({\bf k}) \equiv  (2\omega_{\bf k})P_{{\bf k}\pm}{\cal R}_{{\bf k}\pm}P_h({\bf k})\gamma^0 = \frac{1}{2\omega_{\bf k}}{\cal K}_{mh\pm}({\bf k}){\cal K}_{mh\mp}({\bf k})  \,.
\label{PR_identity4}
\end{eqnarray}
To derive the last form for ${\cal K}_{ch\pm}$ we used the identity (\ref{PR_identity}). Note that the operators ${\cal K}_{mh\pm}$ and ${\cal K}_{ch\pm}$ have the natural interpretations in terms of the spinor products. First, it is easy to show that\footnote{If one interprets our negative energy states as positive energy antiparticles by replacing ${\bf k} \rightarrow -{\bf k}$ for the negative energy solutions one finds the familiar result: $\sum_h{\cal K}_{mh\pm} = \skdag \pm m$, where $m \equiv m_R- i\gamma^5 m_I$.} 
\beq
{\cal K}_{mh\pm} = u_\pm(k,h)\bar u_\pm(k,h) \,,
\eeq
where $u_\pm(k,h)$ are the free positive and negative energy spinors normalized such that $u_\pm^\dagger u_\pm= 2\omega_{\bf k}$. From Eq.~(\ref{PR_identity4}) one then immediately finds that the coherence propagators are proportional to off-diagonal spinor products:
\beq
{\cal K}_{ch\pm} = A_{h\pm} u_\pm(k,h)\bar u_\mp(k,h) \,,
\eeq
where the normalization factor
\beq
A_{h\pm} = \frac{1}{2\omega_{\bf k}} \bar u_\mp(k,h) u_\pm(k,h) 
=  \frac{h|\bf k|}{\omega_{\bf k}} 
 \pm i h\Big( 1- \frac{h|\bf k|}{\omega_{\bf k}} \Big) \frac{m_Im_R}{|m|^2} \,.
\eeq
This observation gives rise to a natural normalization for the coherence-shell functions $f_{ch\pm}$; the {\em canonically normalized} coherence functions clearly are:
\begin{figure}
\centering
\includegraphics[width=0.6\textwidth]{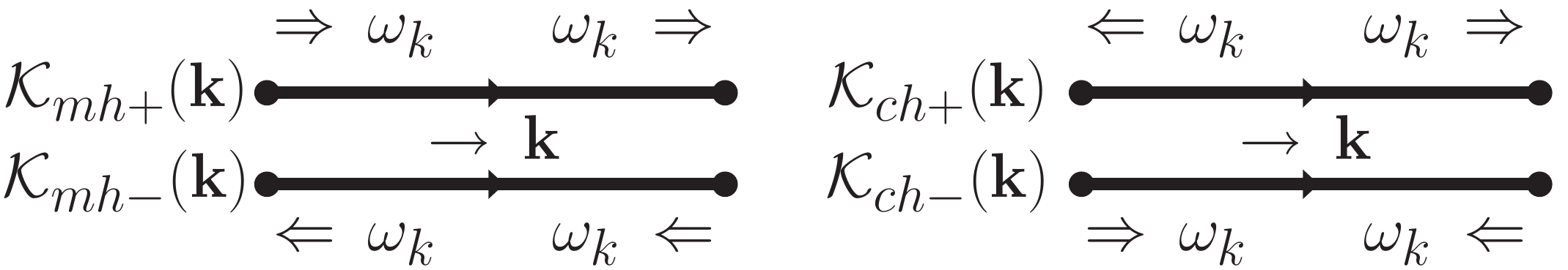}
    \caption{Graphical representation of Dirac structures of different on-shell
     propagators. Three momentum and fermion number flow from left to right.
     Arrows in fermion lines represent the fermion number flow and double arrows the direction of the energy flow.}
    \label{fig:K:s}
\end{figure}
\beq
\tilde f_{ch\pm} = A_{h\pm} f_{ch\pm} \,.
\label{eq:canonicalnorm}
\eeq
While it will be convenient to use our initial definition of the coherence functions in practical calculations, we will always present our results using the canonical normalization.

The fact that ${\cal K}_{ch\pm}$ can be expressed as products of covariant projectors onto opposite energies allows a nice pictorial representation of the propagators shown in figure~\ref{fig:K:s}: the mass-shell propagator ${\cal K}_{mh+}$ corresponds to the usual positive energy states moving along the direction of momentum $\bf k$ and ${\cal K}_{mh-}$ to negative energy states that move against the direction of $\bf k$ (that is, their physical momentum is $-{\bf k}$). The coherence propagators on the other hand, can be viewed as transporting states simultaneously in from both directions (${\cal K}_{ch-}$) or out to both directions (${\cal K}_{ch+}$), both along and against the direction of $\bf k$. From practical point of view however, it is important that the matrices ${\cal K}_{mh\pm}$ and ${\cal K}_{ch\pm}$ form an orthogonal system with respect of Dirac traces. Indeed, using first the definitions Eq.~(\ref{4spectral2Cov}) the collision integrals~(\ref{coll_integrals1}-\ref{coll_integrals2}) can be written as:
\begin{eqnarray}
{\cal C}_{mh\pm}[f_\alpha] 
&=&  \frac{1}{2\omega_{\bf k}}{\rm Tr}\Big[ \langle{\cal C}_{\rm coll} + {\cal C}_{\rm coll}^\dagger\rangle\gamma^0{\cal K}_{mh\pm} \Big] \,,\label{Cov-coll_integrals1} \nonumber\\ 
{\cal C}_{ch\pm}[f_\alpha]
&=& \frac{\xi_{\bf k}}{2\omega_{\bf k}}{\rm Tr}\Big[\langle{\cal C}_{\rm coll} + {\cal C}_{\rm coll}^\dagger\rangle\gamma^0 {\cal K}_{ch\mp}\Big] \,,
\label{Cov-coll_integrals2}
\end{eqnarray}
where $\xi_{\bf k} \equiv \omega_{\bf k}^2/(\omega_{\bf k}^2-m_R^2)$. Now using the orthogonality properties of ${\cal K}_{j h\pm}$ and the generic expression for the collision term Eq.~(\ref{effectiveC3}) we find
\begin{eqnarray}
{\cal C}_{mh\pm}[f_\alpha] &=& 
   - \Re \Big( \big[\; \Sigma^>_{mh\pm} f^<_{mh\pm} +             
                 \Sigma^>_{ch\mp} f^<_{ch\mp} \; \big]
   \; - \; \big[ >  \,\,  \leftrightarrow \, \, < \big] \Big)   
\nonumber \\
{\cal C}_{ch\pm}[f_\alpha] &=& 
   - \frac{1}{2} \Big( \big[\; \Sigma^{>}_{mh\pm} f^<_{ch\pm}  
   + ( \xi_{\bf k} \Sigma^>_{ch\pm})^* f^<_{mh\pm}   
  \; + \; (\pm \leftrightarrow \mp)^* \; \big]
  \; - \; \big[ >  \,\,  \leftrightarrow \, \,  < \big] \Big) \,,
\label{coll_types1}
\end{eqnarray}
where the functions $\Sigma^{<,>}_{jh\pm}$ are the following traces of the self-energy functions $\Sigma^{<,>}_{\rm eff}$:
\begin{eqnarray}
   \Sigma^{<,>}_{mh\pm} &=& \frac{1}{2\omega_{\bf k}}{\rm Tr}\Big[i\Sigma^{<,>}_{\rm eff}(\pm \omega_{\bf k}){\cal K}_{mh\pm}({\bf k}) \Big]\,,
\nonumber \\
   \Sigma^{<,>}_{ch\pm} &=& \frac{1}{2\omega_{\bf k}}{\rm Tr}\Big[i\Sigma^{<,>}_{\rm eff}(\pm \omega_{\bf k}){\cal K}_{ch\mp}({\bf k}) \Big]\,.
\label{coll_types2}
\end{eqnarray}
All other contributions from the collision integrals $\langle {\cal C}_{\rm coll} \rangle$ get annihilated by the projection operators inside the traces.
Note that the coherence-shell self-energy functions $\Sigma^{<,>}_{ch\pm}$ get projected with an ``inverted" spin structure ${\cal K}_{ch\mp}$ under the trace. If the coherence functions are set to be identically zero, and the source terms are neglected in the flow terms, equations~(\ref{eom_on-shell1}) 
with the collision integrals~(\ref{coll_types1}) reduce to the usual Boltzmann equation for the mass-shell excitations. Observe however, that any departure from equilibrium, even in the absence of the source terms, would  lead to creation of coherence through $\xi_{\bf k}$-terms. Coherence would be created in this way for example in massive particle decays, which could be of relevance for example for leptogenesis calculations.

The quantum Boltzmann equations (\ref{eom_on-shell1}-\ref{eom_on-shell2}) together with the explicit forms of the collision integrals Eq.~(\ref{coll_types1}) constitute one of the main results of this paper. We wish to stress that they are generic equations in the cQPA framework, where arbitrary couplings of the coherently evolving system to its surroundings are encoded into the perturbative expressions of the self-energy functions (\ref{coll_types2}). As we saw in section~\ref{sec:collisionold}, to compute these functions we need special techniques that account for the resummations over the coherence propagators. We will develop the necessary calculational rules in section~\ref{sect:effective-feynman}, after we first extend the results of this section to the case of scalar fields.

To conclude this section, let us stress that despite their apparent completeness, the cQPA  quantum Boltzmann equations (\ref{eom_on-shell1}-\ref{eom_on-shell2}) correspond to an {\em approximation} scheme. Indeed, we found the spectral representation for the correlator in terms of the eight $f_\alpha$-functions by expanding and solving the collisionless constraint equations (\ref{Hermitian22}) to the zeroth order in gradients. Because these equations were purely algebraic they had a formally exact spectral solution which was in one-to-one correspondence to the lowest moment of the correlator ${\cal S}$. If we included gradient corrections into the constraint equations however, a more complicated set of distribution functions with new independent terms (for example like $\sim a^{(1)}_\alpha\partial_{k_0}\delta(k^2-m^2)$) would be needed to find closure consistently order by order. In this sense the qBE's (\ref{eom_on-shell1}-\ref{eom_on-shell2}) written in terms of $f_\alpha$'s are almost too promising; in reality the heart of the cQPA-scheme remains to be that it allows a reasonable approximation for the collision integral in the master equation (\ref{eom_prop}) for the lowest moment of the correlation function ${\cal S}$. Adding more gradients to constraint equations would lead to a more complicated ansatz, and correspondingly, to find a complete set of qBE's with the new associated shell-functions, more moments of the dynamical equations would be needed. Most physical observables can be written in terms of the lowest moment functions however, and in most physical applications the gain in the accuracy from going to higher orders in gradients would be very limited. Our restriction to the lowest moment expansion (lowest order in gradients) is, at any rate, exactly analogous to the derivation of the usual Boltzmann equation.


\section{cQPA for scalar fields}
\label{sec:scalar}

We now formulate the cQPA formalism for scalar fields using analogous approximations to the ones we introduced for fermions in section \ref{sec:approximations}. As was shown in~\cite{HKR3}, one obtains a qualitatively similar phase space structure, with the mass shells at $k_0 = \pm\omega_{\bf k}$ and coherence shells at $k_0 = 0$. However, the integration procedure leading to a closed set of Boltzmann equations is somewhat different from the fermionic case. First, since there is only one component in the scalar field correlator $\Delta^<(k,t)$, one needs to introduce a finite number of $k_0$-moments of the initial singular correlators~\cite{HKR3}. Second, as the constraint equations for scalars are not algebraic there are no formally exact spectral solutions to them. In other words, the gradient expansion is slightly more delicate for scalars than for fermions. As a result, it will be more convenient to present the scalar qBE's in terms of moments and use the spectral solutions only to compute the collision integrals. 

We begin with the Kadanoff-Baym equations for the Wightman functions 
\begin{equation}
i\Delta^<(u,v) = \langle \phi^\dagger(v)\phi(u) \rangle 
\qquad {\rm and}Ê\qquad 
i\Delta^>(u,v) = \langle \phi(u) \phi^\dagger(v)\rangle \,.
\end{equation}
In the Wigner representation, the Kadanoff-Baym equation for these functions become (see \eg ref.~\cite{PSW}):
\begin{equation}
\Big( k^2-\frac{1}{4}\partial_x^2 + i k_0 \partial_x - m^2e^{-\frac{i}{2}\overleftarrow{\partial_x}\partial_k} \Big) \Delta^{<,>}  -  e^{-i\Diamond}\{ \Pi_H \}\{ \Delta^{<,>} \}
  -  e^{-i\Diamond}\{ \Pi^{<,>} \}\{ \Delta_H \}
= \mathcal{C}_{\rm coll}\,,
\label{KB_scalar}
\end{equation}
where $\Delta_H = \Delta_F - (\Delta^> + \Delta^<)/2$ and $\Pi_H = \Pi_F - (\Pi^> + \Pi^<)/2$, while $\Delta_F$ and $\Pi_F$ denote the time-ordered (Feynman) propagator and the corresponding self-energy. The collision term can be found from  Eq.~(\ref{collintegral}) with $S \to \Delta$ and $\Sigma \to \Pi$. Proceeding through steps 1-3 in the approximations detailed in section \ref{sec:approximations} and breaking equations into hermitian and anti-Hermitian parts we find:
\begin{eqnarray}
  \Big( k^2 -\frac{1}{4}\partial_t^2 
-  \cos\big(\frac{1}{2}\partial^m_t\partial^\Delta_{k_0}\big) m^2
  \Big) i\Delta^{<,>} 
&=& -\mathcal{C}_{A} 
\label{ConstraintEq} \\ 
 \Big( k_0 \partial_t 
+ \sin\big(\frac{1}{2}\partial^m_t\partial^\Delta_{k_0}\big) m^2
 \Big)i\Delta^{<,>} 
&=& \phantom{-}\mathcal{C}_{H} \,,
\label{EvoEq}
\end{eqnarray}
where $\mathcal{C}_{H} \equiv (\mathcal{C}_{\rm coll} + \mathcal{C}_{\rm coll}^\dagger)/2$ and $\mathcal{C}_{A} \equiv (\mathcal{C}_{\rm coll} - \mathcal{C}_{\rm coll}^\dagger)/(2i)$. 

\subsection{Phase space structure}
We now proceed to approximation step 4 and analyze the KB-equations (\ref{ConstraintEq}-\ref{EvoEq}) in the zeroth order in $\Gamma$ and $\partial_t m$ to find out the singular phase space structure. That is, we initially set:
\begin{eqnarray}
  \Big( k^2 -\frac{1}{4}\partial_t^2 - m^2 \Big) i\Delta^{<,>} 
&=& 0 
\nonumber \\ 
 k_0 \partial_t i\Delta^{<,>} &=& 0 \,.
\nonumber
\end{eqnarray}
Because both of these equations contain explicit time derivatives even in zeroth order, one need to use them both to get one algebraic constraint. The appropriate approximate solution was found in~\cite{HKR3}: 
\begin{equation}
i\Delta^{<,>}(k,t) = 
    2\pi\,{\rm sgn}(k_0)f_{ms_{k_0}}^{<,>}(|{\bf k}|,t)\delta\big(k^2 -m^2\big) 
  + 2\pi\, f_c^{<,>}(|{\bf k}|,t) \delta(k_0) \,, 
\label{SpecSolHOM}
\end{equation}
with $s_{k_0} ={\rm sgn}(k_0)$, and $m=m(t)$. Following the fermionic analog, it will be convenient to define two new (dependent) coherence-shell solutions:
\begin{equation}
f_{c\pm}^{<,>} \equiv \Big(\omega_{\bf k} \pm \frac{i}{2} \partial_t \Big) f_{c}^{<,>}  \,.
\label{fcohpm}
\end{equation}
With these variables we can write Eq.~(\ref{SpecSolHOM}) as:

\begin{equation} 
 i\Delta^{<,>}(k,t) = \frac{\pi}{\omega_{\bf k}} \sum \limits_{\pm} 
 \Big( \pm f_{m\pm}^{<,>}({\bf k},t)\delta\left(k_0 \mp \omega \right) 
         + f_{c\pm}^{<,>}({\bf k},t)\delta(k_0) \Big)\,.
\label{Dist}
\end{equation}

As in the fermionic case, the KB-equation for the spectral function ${\cal A} = \frac{i}{2}(\Delta^> - \Delta^<)$ is identical to the ones for $\Delta^{<,>}$, and consequently, the solution is of the same form as Eq.~(\ref{Dist}). In addition however, the spectral function must obey the sum rule, which follows from the equal time commutation relations of the field operators $\phi$:
\beq
\int \frac{d k_0}{\pi} \big(k_0 + \frac{i}{2}\partial_t \big) {\cal A}(k,t) 
= 1\,.
\eeq
Again the spectral relation completely determines the spectral on-shell functions, setting $f^{\cal A}_{\pm} = \frac12$ and $f^{\cal A}_{c\pm} = 0$  (see ref.~\cite{HKR3}), reducing ${\cal A}$ to its standard form:
\begin{equation}
 {\cal A} = \pi\,{\rm sgn}(k_0) \delta(k^2 - m^2) \,.
\label{full_spectral_sca}
\end{equation}
Using this result with the defining relation $2i{\cal A} = \Delta^< - \Delta^>$, one can easily show that the dynamic functions $f^>$ and $f^<$ are related:
\begin{equation}
f_{m\pm}^> = 1 + f_{m\pm}^<  \,,\qquad {\rm and}Ê\qquad  
f_{c\pm}^> = f_{c\pm}^< \,.
\label{gtr_less_sca}
\end{equation}
That is, only half of the on-shell functions appearing in $\Delta^{<,>}$ are free variables. In what follows, we derive equations of motion for the on-shell functions $f_\alpha \equiv f_\alpha^<$.  Finally, let us write down the cQPA  pole propagators which are equivalent to the standard expressions:
\begin{equation}
  i\Delta_{r,a}(k,t) = \frac{i}
                      {k^2-|m|^2 \pm i{\rm sgn}(k_0) \epsilon}\,.
\label{eq:standardpolesc}
\end{equation}
From these one can find out the cQPA Feynman and the anti-Feynman propagators:
\begin{eqnarray} 
\Delta_F &=& \Delta_r + \Delta^< 
\nonumber \\
\Delta_{\bar F} &=& -\Delta_a + \Delta^< \,,
\label{eq:standardfeynmansc}
\end{eqnarray} 
where $\Delta^<$ is of course given by Eq.~(\ref{Dist}).

\subsection{Equations of motion}
We again define the cQPA transport equations by treating the functions $f_\alpha$ as free parameters, and inserting the spectral solution (\ref{Dist}) as an ansatz back into the full KB-equations (\ref{ConstraintEq}) and (\ref{EvoEq}). Because we have only one scalar function $\Delta^<$ for three unknown shell functions, we need to integrate these equations with a number of different weights functions to get a closure. To be specific, we will use the moment functions:
\begin{equation}
\rho_n(\mathbf{k},t) = \int \frac{{\rm d}k_0}{2\pi}k_0^n\,i\Delta^<(k,t)\,,
\label{moment}
\end{equation}
which we need at least three to get the closure. Taking the three lowest moments and working to the zeroth order in gradients, we find the following invertible relations between the moments and the on-shell functions $f_{\pm}^<$ and $f_{c}^<$:  
\begin{eqnarray}
\rho_0 &=& \frac{1}{2\omega_{{\bf k}}}(f_+ - f_-) + f_c 
\nonumber\\
\rho_1 &=& \frac{1}{2}(f_+ + f_-)
\nonumber\\[1mm]
\rho_2 &=& \frac{\omega_{{\bf k}}}{2}(f_+ - f_-)\,.
\label{rho-f_HOM}
\end{eqnarray}
We will clearly need three evolution equations for our three moments, and it is natural to define them as the zeroth moment of equation (\ref{ConstraintEq}) and as first and second moments of Eq.~(\ref{EvoEq}):
\begin{eqnarray}
\frac14 \partial_t^2 \rho_0 + \omega_{{\bf k}}^2 \rho_0 - \rho_2 
&=& -\left<{\cal C}_A\right>
\nonumber\\
\partial_t\rho_1 &=& \left<{\cal C}_H\right>
\nonumber\\
\partial_t\rho_2 - \frac12 \partial_t(m^2) \rho_0 
&=&  \left<k_0{\cal C}_H\right>\,.
\label{rho_Eq_Coll1}
\end{eqnarray}
It is now evident that the moment connections (\ref{rho-f_HOM}) are the key element of the cQPA approximation, as they will allow us to rewrite the collision integrals appearing in Eq.~(\ref{rho_Eq_Coll1}) in terms of the moments $\rho_{0,1,2}$ to get the closure. Unlike in the case of fermions, we do not attempt to rewrite these equations in terms of $f$-functions, although it could be done formally by differentiating the inverted equations (\ref{rho-f_HOM}) and using recursively the evolution equations (\ref{rho_Eq_Coll1}). This change of variables carries a delicate issue related to the gradient expansion, however. Remember that the moment equations (\ref{rho_Eq_Coll1}) are {\em exact} in the sense of gradient expansion, and to avoid introducing a loss of accuracy in going to $f$-variables, one should treat also the relations (\ref{rho-f_HOM}) as exact, despite the fact that they were computed only to the lowest order in gradients. 
The inverted equations would thus have a mixed set of gradient terms, coming from both the exact moment equations and from the approximate inversion process. In particular, second order gradient terms would be invoked because of the second time derivative in the equation of motion for $\rho_0$. These terms should not be neglected in the spirit of gradient expansion, however, as they arise from the defining relations (\ref{rho-f_HOM}). This issue did not arise in the fermionic case where the equations of motion include only first order time derivatives. As a result of this complication we prefer to write our qBE's in terms of the moments rather than the $f$-functions in the scalar case. 

However, also the scalar coherence solutions are rapidly oscillating and thus the scalar collision terms need to be resummed with respect to these oscillations. We can see this by solving the evolution of the coherence-shell solutions to the lowest order in gradients from equations (\ref{rho_Eq_Coll1}) in the context of the formulae (\ref{rho-f_HOM}) and the definition (\ref{fcohpm}):
\beq
\partial_t f_{c\pm} = \mp 2i\omega_{\bf k} f_{c\pm} + {\cal O}^1 \,,
\label{eq:scalarcoherence}
\eeq
where the correction term ${\cal O}^1$ has the same meaning as in the fermionic case. Indeed, equations (\ref{eq:scalarcoherence}) are identical to the zeroth order limit of fermionic qBE:s (\ref{eom_on-shell2}). The mass-shell solutions on the other hand, are constants to the lowest order: $\partial_t f_{m\pm} = {\cal O}^1$, and so, analogously to equations (\ref{zeroth_eq0}), we find the following lowest order equations for the singular shell-solutions:
\beq
\partial_t \Delta^<_{m\pm} =  {\cal O}^1\,, \qquad  	
\partial_t \Delta^<_{c\pm} = \mp 2i\omega_{\bf k} \Delta^<_{c\pm} 
+ {\cal O}^1 \,.
\label{eq:scalarcoherence2}
\eeq
\subsection{Resummed scalar collision term}
We need to express the collision integrals appearing in Eq.~(\ref{rho_Eq_Coll1})
in terms of the distribution functions $f_\pm$ and $f_{c\pm}$ (and eventually in terms of the moments using the connection Eq.~(\ref{rho-f_HOM})). The basic quantity we encounter is:
\beq
\langle{\mathcal{C}}_\alpha\rangle 
=  \int \frac{{\rm d} k_0}{2\pi}  k_0^\alpha\, 
  \frac{1}{2} e^{-i\Diamond} 
     \big( \{\Pi^>(k,t)\}\{\Delta^<(k,t)\} 
         - \{\Pi^<(k,t)\}\{\Delta^>(k,t)\} \big)\,.
\label{eq:scalarcollision}
\eeq
where $\alpha=0,1$. Again, to re-sum the oscillatory gradients of the distribution functions in the $\Diamond$-expansion, we write Eq.~(\ref{eq:scalarcollision}) in the two-time representation:
\beqa
\langle{\mathcal{C}}_\alpha\rangle 
 &=& \frac{1}{2} \int {\rm d} w_0  (-i\partial_{r_0})^\alpha
  \Big[ \Pi^>(t+\frac{r_0}{2},w_0) \Delta^<(w_0,t-\frac{r_0}{2}) 
\nonumber\\
&& \qquad\qquad\qquad\;\;\; - \Pi^<(t+\frac{r_0}{2},w_0) \Delta^>(w_0,t-\frac{r_0}{2})\Big]_{r_0 = 0} \,. 
\label{coll_int_sca}
\eeqa
For $\alpha=0$ this immediately reduces to a formula analogous to Eq.~(\ref{EffColl2}) for fermions. For $\alpha=1$ the extra $\partial_{r_0}$-derivative gives rise to an additional complication, but we still continue to search for a consistent expansion around the external time $t$ as before. We begin by writing the spectral propagator (\ref{Dist}) in the two-time representation:
\begin{eqnarray}
 i\Delta^<(w_0,w_0',\mathbf{k}) &=& \int \frac{{\rm d}k_0}{2\pi} e^{-ik_0(w_0-w_0')}\Delta^< \big(k_0,\mathbf{k},\frac{w_0+w_0'}{2}\big)
\nonumber
\\
&\approx& \frac{1}{2\omega_{\bf k}} 
 \sum \limits_{\pm}\Big[ 
  \pm e^{\mp i \omega (w_0-w_0')} f_{m\pm}(\mathbf{k},t) 
    + e^{\mp i \omega (w_0+ w_0'-2t)} f_{c\pm}(\mathbf{k},t)\Big]\,. 
\label{two-time_sca}
\end{eqnarray}
Here we Taylor expanded $\Delta^<_x\big(\mathbf{k},\frac{w_0+w_0'}{2}\big)$ around the external time variable $t$, and used the recursive zeroth order equations of motion:
\begin{equation}
\partial_t^n \Delta^<_{m\pm} =  {\cal O}^1\,, \qquad 
\partial_t^n \Delta^<_{c\pm} = (\mp 2i\omega_{\bf k})^n \Delta^<_{c\pm} + {\cal O}^1\,,
\end{equation}
exactly as in the fermionic case. Now, using the expanded propagator (\ref{two-time_sca}) in the collision integrals (\ref{coll_int_sca}) we get:
\beq
\langle {\cal C}_\alpha \rangle 
= \frac{(-1)^{\alpha+1}}{2}  \sum \limits_{\pm} \Big( \big[ \; 
  \pm i\Pi^>_{m\alpha\pm} f^<_{m\pm}  
   + i\Pi^>_{c\alpha\pm} f^<_{c\pm} \;\big] 
 \; - \; \big[ >  \,\,  \leftrightarrow \, \, < \big] \Big) \,,
\label{coll_final_sca}
\eeq
where the effective self-energies are:
\beq
\Pi^{<,>}_{m\alpha\pm} 
  = \left(\pm \omega_{\bf k} + \sfrac{i}{2}\partial_t\right)^\alpha
   \Pi^{<,>}_{{\rm eff}}(\pm \omega_{\bf k})
\quad \; {\rm and}Ê\quad \;
  \Pi^{<,>}_{c\alpha\pm} 
  = \left(\sfrac{i}{2}\partial_t\right)^\alpha
\Pi^{<,>}_{{\rm eff}}(\pm \omega_{\bf k})
\label{Sigma_eff_proj_sca}
\eeq
with 
\beq
\Pi^{<,>}_{{\rm eff}}(k_0,{\bf k},t) \equiv \frac{1}{2|k_0|} \int {\rm d} w_0 e^{i k_0(t-w_0)} \, \Pi^{<,>}(t,w_0,\mathbf{k})\,.
\label{Sigma_eff_proj}
\eeq
Note again the simplicity of these results; to get the collision integral for an arbitrary moment equation, we only need to evaluate one generic function $\Pi^{<,>}_{{\rm eff}}(\pm \omega_{\bf k})$, which is of the same form as the fermionic effective self-energy function Eq.~(\ref{effsigma}). For the cases $\alpha = 0,1$ appearing in (\ref{caleffsneeded}) we have:
$\Pi^{<,>}_{m0\pm} = \Pi^{<,>}_{c0\pm} = \Pi^{<,>}_{\rm eff\pm}$, 
$\Pi^{<,>}_{m1\pm} = \pm \omega_{\bf k} \Pi^{<,>}_{\rm eff\pm} +\sfrac{i}{2}\partial_t \Pi^{<,>}_{{\rm eff\pm}}$ and $\Pi^{<,>}_{c1\pm} = \sfrac{i}{2}\partial_t \Pi^{<,>}_{{\rm eff\pm}}$, so that
\beqa
\langle {\cal C}_0 \rangle &=& 
-\frac{1}{2}\sum \limits_{\pm} \Big(\big[\;
i\Pi^>_{\rm eff\pm} ( \,f^<_{c\pm} \pm f^<_{m\pm}  \, ) \; \big] 
 \; - \; \big[ >  \,\,  \leftrightarrow \, \, < \big] \Big)\,,
\nonumber \\
\langle {\cal C}_1 \rangle &=& 
\frac{1}{2}\sum \limits_{\pm}  
\Big( \Big[\; \omega_{\bf k} i\Pi^>_{\rm eff\pm}f^<_{m\pm} 
+\Big( \frac{i}{2}\partial_t i\Pi^>_{\rm eff\pm} \Big) 
\big(\, f^<_{c\pm} \pm f^<_{m\pm} \,\big)\;\Big]
 \; - \; \big[ >  \,\,  \leftrightarrow \, \, < \big] \Big)\,.
\label{Sigma_eff_proj_sca2}
\eeqa
The collision integrals appearing explicitly in Eqs.~(\ref{rho_Eq_Coll1}) are just real and complex parts of these expressions:
\begin{eqnarray}
\langle {\cal C}_H \rangle = \Re \langle {\cal C}_0 \rangle \,, 
\quad
\langle {\cal C}_A \rangle = \Im \langle {\cal C}_0 \rangle   
\quad {\rm and} \quad
\langle k_0{\cal C}_H \rangle = \Re \langle {\cal C}_1 \rangle .
\label{caleffsneeded}
\end{eqnarray}
Equations (\ref{Sigma_eff_proj_sca}) deserve some comments. First, the resummation of leading oscillatory terms in the $\Diamond$-expansion pushed the (zeroth moment) coherence self-energy functions onto mass-shells, analogously to what happened in the fermionic case. This again conforms with our expectation that collisions cannot be sensibly defined for pure coherence; although coherence solutions {\em do} live on $k_0=0$ in the phase space, their effect is, after resummation, only felt as a modification of the collision rates for the mass-shell states.  Second, the resummation effectively transforms the moment function $k_0^\alpha$ in equation (\ref{eq:scalarcollision}), into an operator $(k_0 + \sfrac{i}{2}\partial_t)^\alpha$ which, after projection onto mass- and coherence shells, gives rise to expressions (\ref{Sigma_eff_proj_sca}). Obviously, a naive truncation of the diamond expansion would have missed the derivative terms in (\ref{Sigma_eff_proj_sca}). Note that these gradients cannot be ignored due to the same reason that led us to perform the diamond expansions in the first place; whenever the gradient acts on any coherence function occurring in the perturbative expansion for $\Pi^{<,>}_{{\rm eff}}$, the result is not formally suppressed by mass-gradients. Thus we have to use Equations (\ref{Sigma_eff_proj_sca2}) as such in the collision integrals~(\ref{caleffsneeded}).

In practical calculations it is most convenient to express the collision integrals in terms of $f_\alpha$'s and use the inverse relations of Eq.~(\ref{rho-f_HOM}) to write them in terms of the moments $\rho_{0,1,2,}$. For the record we write the necessary inverse relations explicitly here:
\begin{eqnarray}
f_{m\pm} &=& \rho_1 \pm \frac{1}{\omega_{\bf k}}\rho_2
\nonumber\\
f_{c\pm} &=& (\omega_{\bf k} \pm \frac{i}{2}\partial_t)\rho_0 - \frac{1}{\omega_{\bf k}}\rho_2 \,.
\label{rho-f_HOMinv}
\end{eqnarray}
Note the appearance of the quantity $\partial_t\rho_0$, which appears as a ``hidden" variable in the evolution equations (\ref{rho_Eq_Coll1}).


\section{Momentum space Feynman rules}
\label{sect:effective-feynman}

We now derive generalized Feynman rules for computing the effective self-energy functions $\Sigma_{\rm eff}$ and $\Pi_{\rm eff}$ through perturbative techniques, including the coherence effects. Standard methods, such as the 2PI formalism~(see \eg \cite{PSW}), exist for diagrammatic expansion of the two-time self-energies $\Sigma^{ab}(u,v)$ and $\Pi^{ab}(u,v)$ appearing in equations (\ref{effsigma}) and (\ref{Sigma_eff_proj}), and our task is to reduce the computation of the diagrams generated by these methods into a set of momentum space Feynman rules.  We derive these rules as usual by replacing the propagators in an arbitrary diagram by our resummed propagators (\ref{eff_prop}) and (\ref{two-time_sca}) and performing all time integrations related to internal vertices. The only essential complication comes from the nontrivial phase structure associated with the coherence-shell parts in the dynamical propagators (\ref{eff_prop}) and (\ref{two-time_sca}). We can account for all phase factors by rewriting all propagators in a generic 4-dimensional representation as follows:
\begin{equation}
G(w_0,w_0',{\bf k}) = \int \frac{{\rm d}k_0}{2\pi} e^{-ik_0(w_0-c w_0') 
+ ik_0(1-c )t} G(k_0,{\bf k},t) \,.
\label{eq:effphase}
\end{equation}
Here $G$ can refer either to fermion or scalar propagators (the internal degrees of freedom, included in $S(k_0,{\bf k},t)$, are not necessary for our treatment here):  $G=S_{r,a}, S_{m\pm}^{<,>}, S_{c,{\rm eff}\pm}^{<,>}$ or $G=\Delta_{r,a}, \Delta_{m\pm}^{<,>}, \Delta_{c,{\rm eff}\pm}^{<,>}$, where $\Delta_{c,{\rm eff}\pm}^{<,>}$ are defined below in Eq.~(\ref{eq:Geffsnew}). The sign factor $c=c(k_0)$ has the ``normal" value $c= +1$ for the dynamical mass-shell propagators $G=G^{<,>}_{m}$ and for the pole propagators $G=G_{r,a}$ which do not contain any coherence solutions. However, for the resummed coherence-shell propagators $G=G_{c,\rm eff}^{<,>}$ the sign factor is negative, $c=-1$, as required by Eqs.~(\ref{eff_prop}) and (\ref{two-time_sca}). In the latter case the overall phase is ``abnormal", except  in the particular case of $w_0'=t$, where the $c$-terms cancel, and the phase factor becomes normal also for the coherence propagator. This implies that the phases associated with a given vertex in a self-energy diagram are normal in all cases but those where the vertex time (not equal to $t$) corresponds to the second time argument of at least one coherence propagator connected to the vertex. In these cases the signs of the phases coming from the corresponding coherence lines are reversed and extra phases proportional to the reference time $t$ are added. This is a general rule to be used in addition to the usual combinatorics after the interactions have been specified. 

Let us stress that the effective mixed representation propagators $G_{c,\rm eff}(k_0,{\bf k},t)$ in (\ref{eq:effphase}), corresponding to the resummed coherence propagators (\ref{eff_prop}) and (\ref{two-time_sca}), by definition  have their poles on the mass shells. Let us recall their explicit expressions: 
\begin{eqnarray}
iS^{<,>}_{\rm eff\pm} &=& iS^{<,>}_{m\pm} + iS^{<,>}_{c,\rm eff\pm}
\; = \; 2\pi \big( {\cal S}^{<,>}_{m\pm} + {\cal S}^{<,>}_{c\pm} \big) 
   \,\delta(k_0 \mp \omega_{\bf k})  
 \nonumber \\[2mm]
i\Delta^{<,>}_{\rm eff\pm} &=& 
 i\Delta^{<,>}_{m\pm} + i\Delta^{<,>}_{c,\rm eff\pm} 
\; = \;\frac{\pi}{\omega_{\bf k}}  
 \big( \pm f^{<,>}_{m\pm} + f^{<,>}_{c\pm}\big) 
 \,\delta(k_0 \mp \omega_{\bf k})  \, ,
\label{eq:Geffsnew}
\end{eqnarray}
where ${\cal S}$- matrix functions are defined in Eq.~(\ref{specsol2}). Let us remind that in terms of the complex path indexing the Wightman functions correspond to the off-diagonal propagators $S^{12}\equiv S^<$ and $S^{21}\equiv S^>$. In loop calculations we often encounter also the diagonal Feynman and anti Feynman propagators $S^{11}$ and $S^{22}$. Because the novel coherence solutions only appear in the Wightman functions, we can find the coherent $S^{ii}$-functions directly by using  Eqs.~(\ref{eq:standardfeynman}) and (\ref{eq:standardfeynmansc}) together with 
Eqs.~(\ref{eq:Geffsnew}) and the expressions for the standard pole propagators given in Eqs.~(\ref{eq:standardpole}) and (\ref{eq:standardpolesc}):
\begin{eqnarray}
S^{11}_{\rm eff} &=&  \phantom{-} S_r - S^<_{\rm eff}
\;\, = \;\, S_{F,0} - S^-_{\rm eff}
 \nonumber \\
S^{22}_{\rm eff} &=&  - S_a - S^<_{\rm eff} 
\;\, = \;\, S_{\bar F,0} - S^-_{\rm eff}
 \nonumber \\
\Delta^{11}_{\rm eff} &=& \phantom{-} \Delta_r + \Delta^<_{\rm eff}
\;\, = \;\, \Delta_{F,0} + \Delta^-_{\rm eff}
 \nonumber \\
\Delta^{22}_{\rm eff} &=& -\Delta_a + \Delta^<_{\rm eff}
\;\, = \;\, \Delta_{\bar F,0} + \Delta^-_{\rm eff} \,,
\label{eq:sigmaekaeff}
\end{eqnarray}
where $S_{F,0}$, $\Delta_{F,0}$, $S_{\bar F,0}$ and $\Delta_{\bar F,0}$ refer to the standard vacuum Feynman- and anti Feynman propagators. Moreover, the  quantities $S^-_{\rm eff}$ and $\Delta^-_{\rm eff}$ correspond to the effective Wightman functions $S^<_{\rm eff}$ and $\Delta^<_{\rm eff}$ from which the vacuum parts have been subtracted off. Indeed, one can show for example that
\begin{equation}
S_{r} =  S_{F,0} 
+ 2\pi i \theta(-k_0)(\kdag + m_R -i \gamma^5 m_I )\delta(k^2-|m|^2)\,,
\end{equation}
and when the delta-function is absorbed into $S^<_{\rm eff}$ it exactly cancels the vacuum part (included in the negative frequency mass-shell function $S^<_{m-} \propto f^<_{m-} = 1 - \bar n$) 
from $S^-_{\rm eff}$:
\begin{equation}
S^-_{\rm eff} \equiv S^<_{\rm eff} 
- 2\pi i \theta(-k_0)(\kdag + m_R -i \gamma^5 m_I )\delta(k^2-|m|^2)\,.
\end{equation}
It is straightforward to show that similar relations hold for all propagators in (\ref{eq:sigmaekaeff}).

This completes our rules for the propagator functions in the 4-dimensional mixed state representation. Before we turn to the derivation of the vertex rule, let us write down the Hermiticity properties of our propagators in the mixed representations
\begin{eqnarray}
(iG(k,t))^\dagger &=& iG(k,t)
\nonumber \\
(iG^{<,>}_{c,{\rm eff\pm}}(k,t))^\dagger &=& iG^{<,>}_{c,{\rm eff\mp}}(k,t) \,,
\label{eq:complexconj}
\end{eqnarray}
where $iG = \bar S^{<,>}_{m\pm}$, $\bar S_{r,a}$, $i\Delta^{<,>}_m$ or $i\Delta_{r,a}$ in the first line and $iG = \bar S$ or $i\Delta$ in the second. Note that taking the complex conjugate flips the $\pm \rightarrow \mp$ in the coherence propagator. Finally, it will be useful to observe that for a real scalar field propagator
\begin{eqnarray}
 \Delta_{r,a}(k,t) &=& \Delta_{a,r}(-k,t) \,,
\nonumber \\
 \Delta^>_{m,\pm}(k,t) &=& \Delta^<_{m,\pm}(-k,t)\,,
\nonumber \\
 \Delta^>_{c,{\rm eff\pm}}(k,t) &=& \Delta^<_{c,{\rm eff\pm}}(k,t) 
 \;\, \equiv \;\, \Delta_{c,{\rm eff\pm}}(k,t) \,.
\label{eq:greaterless}
\end{eqnarray}
Note in particular that there actually exists only one distinct type of coherence propagators.

\begin{figure}
  \centering
  \begin{minipage}[b]{7 cm}
    \includegraphics[width=1\textwidth]{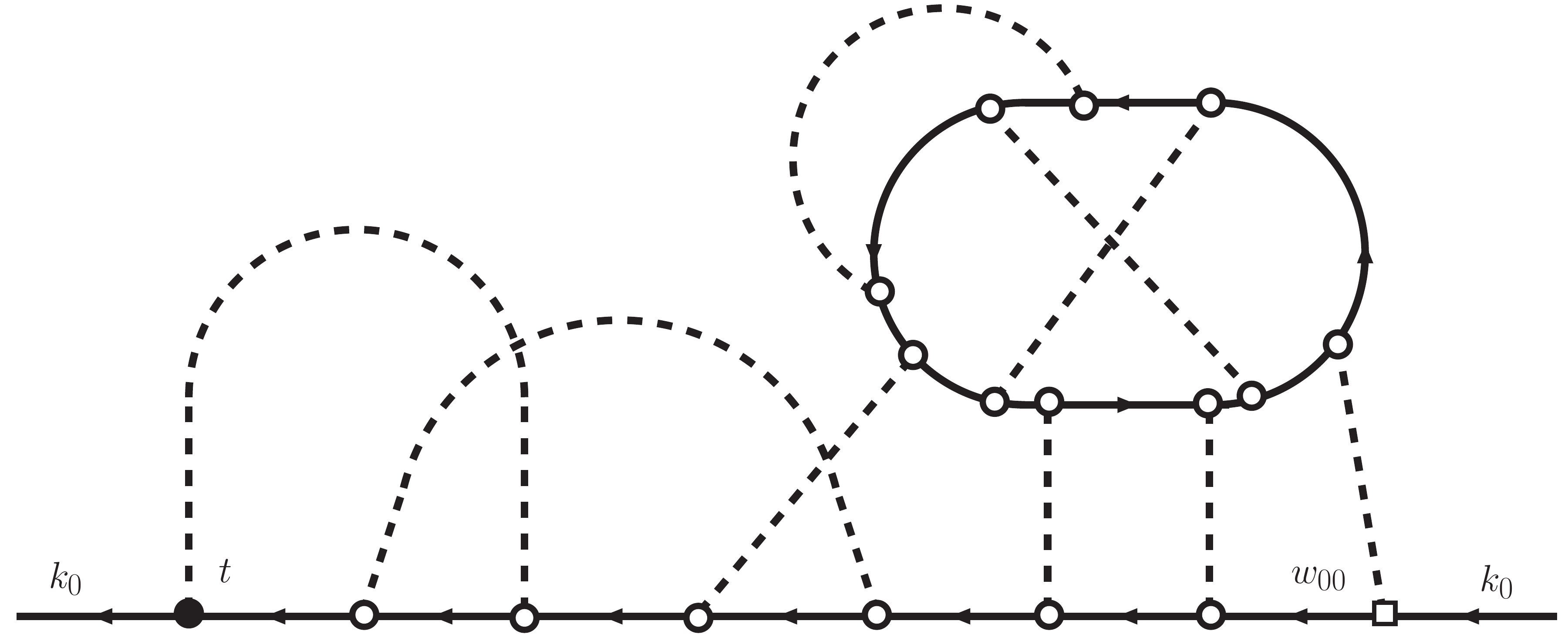}  
  \end{minipage} \hskip 1cm
  \begin{minipage}[b]{5 cm}
    \includegraphics[width=0.95\textwidth]{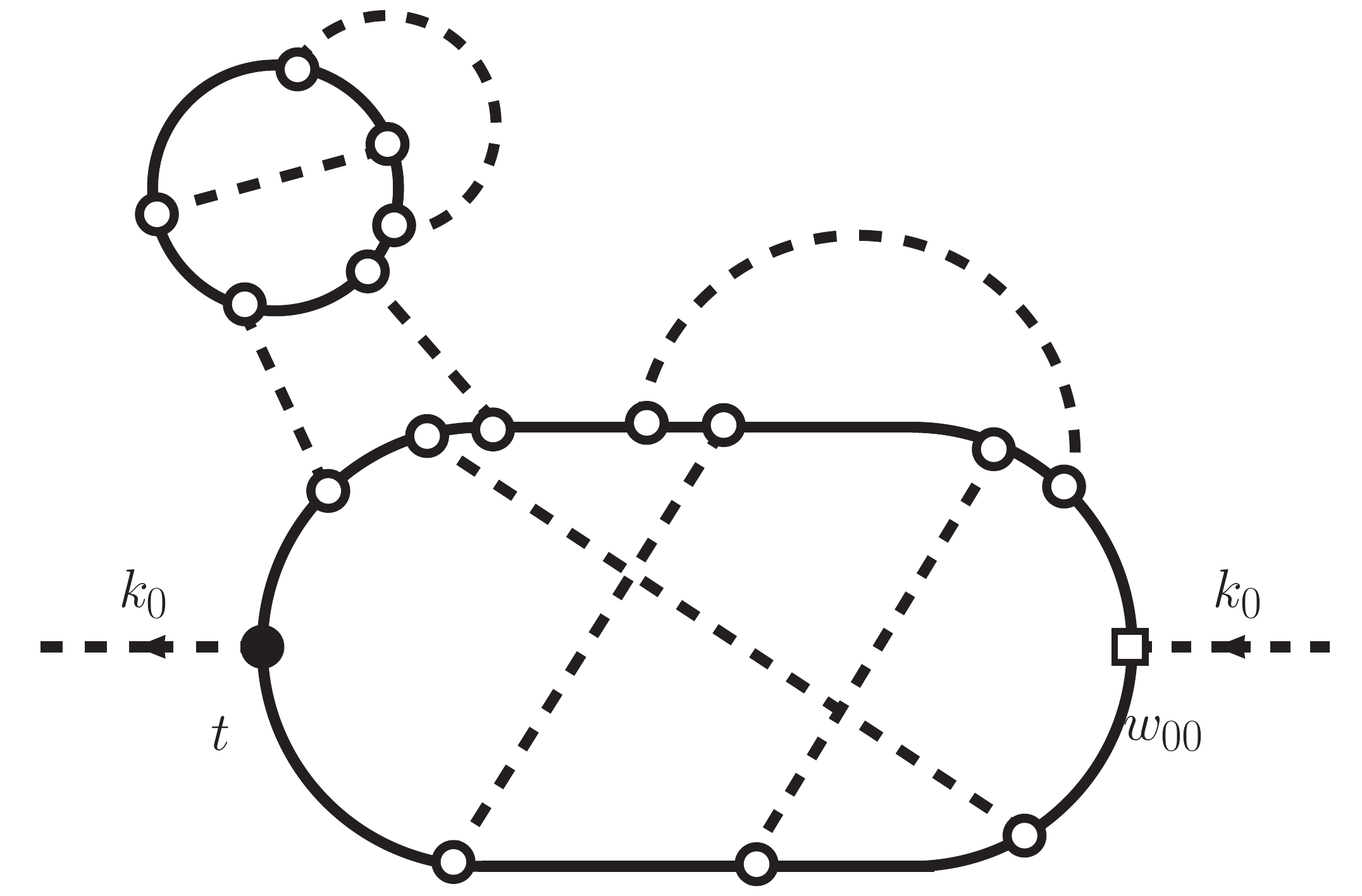}  
  \end{minipage}
  \caption{Generic diagrams contributing to the fermionic self-energy function $\Sigma_{\rm eff}$ (left) and to the scalar self-energy function $\Pi_{\rm eff}$ (right).}
  \label{fig:genericDiagrams}
\end{figure}

\subsection{Vertex rules}

In order to define vertex rules we need to specify a concrete model. We shall adopt the following Yukawa interactions between our scalar- and fermion fields:
\begin{equation} 
{\cal L}_{\rm int} = - y\; \bar \psi\, \phi \, \psi + h.c. \,
\label{interaction}
\end{equation}
Generic examples of self-energy functions generated by this interaction are shown in figure~\ref{fig:genericDiagrams}. Let us consider the fermionic self-energy diagrams first. The most general diagram has one continuous fermion line connecting the initial and final times, accompanied by an arbitrary number of closed fermion loops where all fermionic lines can be connected and associated with an arbitrary number of scalar lines. We wish to derive a generic mixed space representation for such a diagram and extract the local Feynman rules from the resulting expression. In particular we will need to show that the global phase proportional to the external time $t$, arising from our propagators~(\ref{eq:effphase}) vanishes. 

\begin{figure}
\centering
\includegraphics[width=0.55 \textwidth]{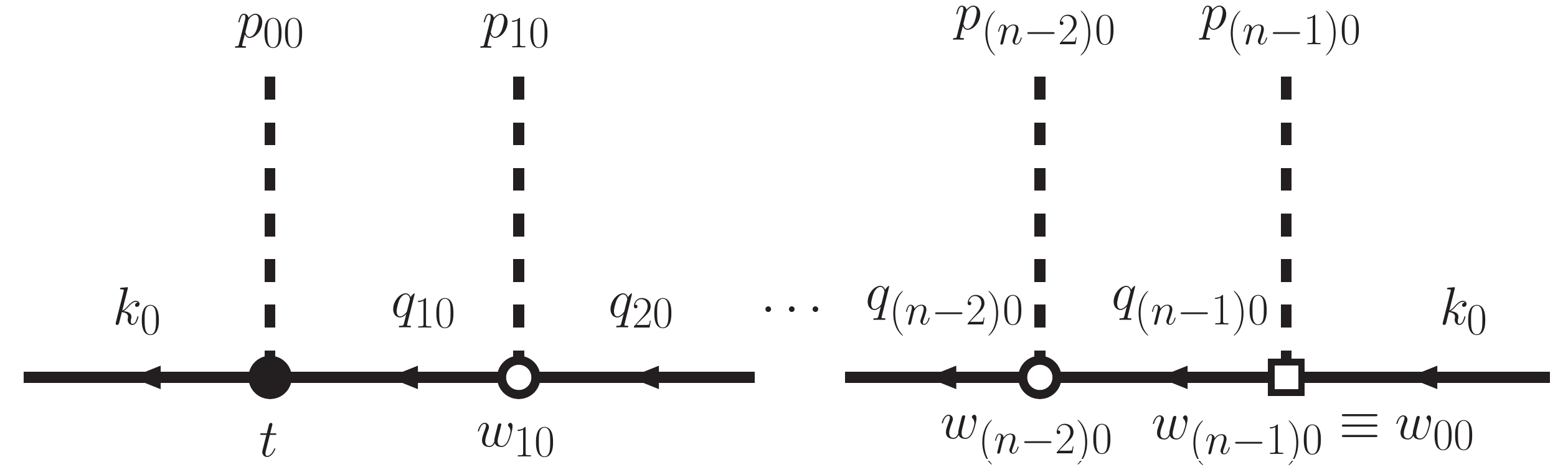}
\caption{A cut through an arbitrary diagram around the continuous fermion line contributing to $\Sigma_{\rm eff}$. The scalar lines may be interconnected or connected to separated closed fermion loops not shown.}
    \label{fig:lineDiagram}
\end{figure}

Consider first a generic sub-diagram of the type shown in Fig.~\ref{fig:lineDiagram}. This is the continuous fermion line going through an arbitrary fermionic self-energy graph. It contains $n$ vertices and $n-1$ fermion propagators, and the final time on the fermionic line is the special external time $t$. The scalar lines may either be interconnected, or they may be connected to closed fermion loops, not shown in the diagram. To be able to associate correct phases with the vertices, we first have to introduce the notion of the ordering of the arguments in the two-time propagator (\ref{eq:effphase}) into the mixed representation. This can be done by associating propagators with a unique {\em direction of flow}. We start by defining the flow in the two-time representation according to the flow of time in time-ordered propagators $G^{11}(u,v)$; that is, from the vertex $v$ to the vertex $u$. This choice induces a natural definition for the flow direction in the mixed representation as being {\em along the four-momentum of the positive energy states.} For fermions this corresponds choosing the direction along the fermion number flow.
Given this definition, Eq.~(\ref{eq:effphase}) implies that a fermion propagator with an energy $q_{k0}$, connected to a vertex $w_{j0}$ gives rise to a phase factor that depends on the direction of the flow according to:
\beqa
e^{-iq_{k0}w_{j0}} \, \qquad {\rm ``incoming"} \phantom{\,.}\nonumber \\
e^{ ic_{qk} q_{k0}w_{j0} + iq_{k0}(1-c_{qk})t}\, \qquad {\rm ``outgoing"}\,. 
\label{eq:fermionphases}
\eeqa
For a neutral scalar line we do not have a similar natural orientation \footnote{This is so because we are considering a neutral scalar field here. For a charged scalar field the charge conservation could be used to define a natural flow orientation.} and we have to use the general expression:
\beq
e^{is_j c_{pk} p_{k0}w_{j0} + ip_{k0}\frac{1}{2}(1+s_j)(1-c_{pk})t}\,,
\label{eq:scalarphase}
\eeq
where the index $s_j=1$ for an outgoing and $s_j=-1$ for an incoming scalar line. The definition (\ref{eq:scalarphase}) is clearly consistent with the fermion phases in Eq.~(\ref{eq:fermionphases}). (Note the implicit rule that for $s_j=-1$ we always have $c_{pk} = 1$ in the associated propagator.)

With these definitions it is easy to show that the time integration in each of the internal vertices in the diagram in Fig.~\ref{fig:lineDiagram} ($j$ runs from 1 to $n-2$)  gives rise to a delta-function:
\begin{equation}
\int {\rm d} w_{0j}\, 
  e^{i w_{j0} ( c_{qj} q_{j0} + s_j c_{pj} p_{j0} - q_{(j+1)0} )} e^{i\phi_{j} t}  
= 2\pi\, \delta(c_{qj} q_{j0} + s_jc_{pj} p_{j0} - q_{(j+1)0}  ) e^{i \phi_j t} \,,
\label{eq:deltaphase}
\end{equation}
where the extra phase factor is
\begin{equation}
\phi_j = (1-c_{qj})q_{j0} + \frac{1}{2}(1+s_j)(1-c_{pj})p_{j0}\,.
\end{equation}
Clearly the extra phase vanishes if $c_{qj}=1$ (normal fermion propagator) and either $s_j=-1$ (incoming scalar field) or $s_j=1$ and $c_{pj}=1$ (normal outgoing scalar propagator). Taking into account the special vertices at the ends of the fermion line, and integrating over the times at internal vertices as well as over the specific time $w_0$, the generic expression for the self-energy contribution from diagram~\ref{fig:lineDiagram} becomes:
\begin{eqnarray}
\Sigma^{<,>}_{{\rm eff}}(k_0,{\bf k},t) 
  &=& \int {\rm d} w_0 e^{ i k_0 (t-w_0)}\Sigma(t,w_0,{\bf k})Ê
\nonumber \\
  &\propto& 
  \Big( \prod_{j=1}^{n} 
   2\pi \delta(c_{qj} q_{j0} + s_jc_{pj} p_{j0} - q_{(j+1)0}) \Big) 
   e^{i \phi_{\rm line} t} \times C_{\rm loops}\,,
\label{effsigma2}
\end{eqnarray}
with the understanding that $q_{n0}\equiv k_0$, and the global phase factor is
\beq
\phi_{\rm line} =  k_0 - q_{10} + s_0 c_{p0} p_{00} + \frac{1}{2}(1+s_0)(1-c_{p0})p_{00}  + \sum_{j=1}^{n-1}\phi_j \,.
\eeq
In the second line of Eq.~(\ref{effsigma2}) we suppressed all momentum integrations and the effective propagators associated with the internal lines. They can easily be inserted back afterwards. The explicitly shown delta functions and phases come from the vertices in the diagram~\ref{fig:lineDiagram} and the factor $C_{\rm loops}$ contains the contributions from all possible closed internal loops in the full diagram. By a recursive use of the delta-functions in the internal vertices, one can show that $\phi_{\rm line}$ can be associated with a sum of scalar momenta:
\begin{eqnarray}
\phi_{\rm line} &=&  
\frac{1}{2}(1+s_0)p_{00} - \frac{1}{2}(1-s_0)c_{p0} p_{00} \nonumber \\
&+&k_0 - q_{20} + s_1 c_{p1} p_{10} + \frac{1}{2}(1+s_1)(1-c_{p1})p_{10}  + \sum_{j=2}^{n-1}\phi_j
\nonumber \\
&=& ... \;\;=  \sum_{j=1}^{n-1} 
\big[ \sfrac{1}{2}(1+s_j)p_{j0} - \sfrac{1}{2}(1-s_j)c_{pj} p_{j0} \big]
\nonumber \\
&=&  \sum_{j=0}^{n-1} s_jp_{j0} \,,
\label{eq:totalphase1}
\end{eqnarray}
where in the last step we used the fact that for an incoming particle with $s_j=-1$ we always have $c_{pj}=1$, and that for an outgoing state with $s_j=1$ the $c_{pj}$-term vanishes in Eq.~(\ref{eq:totalphase1}). The final sum in Eq.~(\ref{eq:totalphase1}) would vanish if all scalar lines in the graph were interconnected (no closed fermion loops in the graph), because then each energy $p_{i0}$ would appear twice in the sum, once with $s_j=+1$ and once with $s_j=-1$. However, as some of the lines may be connected to loops, the sum
does not vanish in general. Physically this means that coherence information can be transported between separate fermion lines by the scalar fields.

\begin{figure}
\centering
\includegraphics[width=0.3 \textwidth]{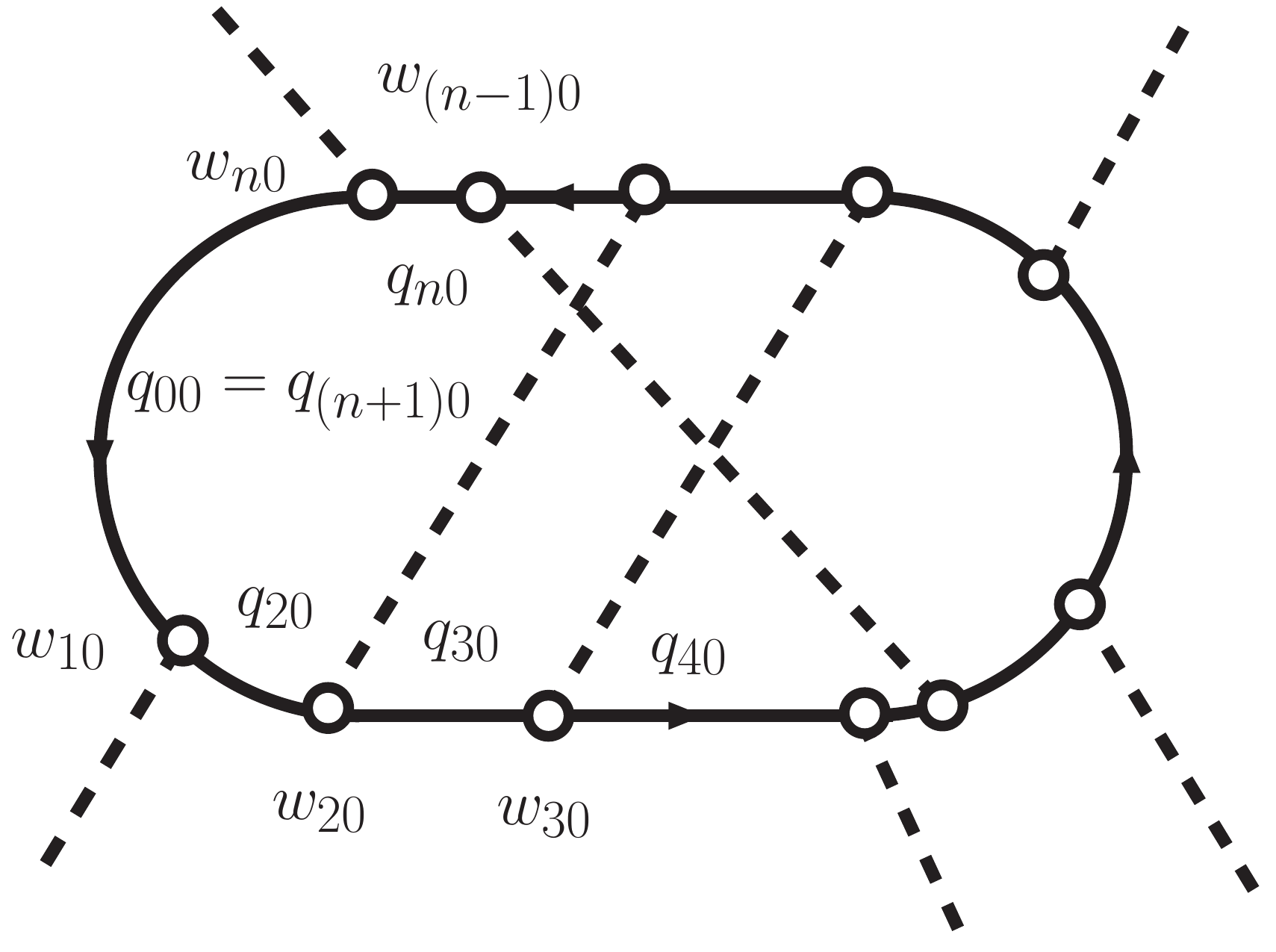}
\caption{A generic closed fermion loop diagram with all internal vertices. The scalar lines can again be interconnected, or connected to some other loops or to a continuous line going through a fermion self-energy diagram.}
    \label{fig:floopDiagram}
\end{figure}
Now consider a closed internal fermion loop depicted in Fig.~\ref{fig:floopDiagram}. For a closed loop the direction of flow could be chosen arbitrarily, but we follow our earlier definition of going along the fermion number flow. The calculation proceeds similarly to the case with the open fermion line, the sole difference being that now all the propagators and vertices are internal ones. After a straightforward calculation one finds that for a loop with $n_i$ vertices:
\beq
  C^i_{\rm loop} \propto \prod_{j=1}^{n_i} 
    2\pi \delta(c^q_j q_{j0} + s_jc_{pj} p_{j0} - q_{(j+1)0}) 
    e^{i\phi^i_{\rm loop}t}
\eeq
where
\beqa
\phi^i_{\rm loop} &=& \sum_{j=1}^{n_i}\phi_j 
\nonumber \\
  &=& q_{11} - q_{20} 
  + \left[\sfrac{1}{2}(1+s_1)p_{10} - \sfrac{1}{2}(1-s_1)c_{p1} p_{j0}\right]
  + \sum_{j=2}^{n_i}\phi_j 
\nonumber\\
  &=& q_{10} - c_{qn}q_{n0} - s_nc_{pn}p_{n0} 
  + \sum_{j=1}^{n_i}
  \big[ \sfrac{1}{2}(1+s_j)p_{j0} - \sfrac{1}{2}(1-s_j)c_{pj} p_{j0} \big]  
\nonumber\\
  &=&  
  \sum_{j=1}^{n_i}s_jp_{j0} \,, 
\eeqa
where in the last step we used the fact that due to cyclicity $n+1$'th and the first fermion propagators are the same: $c_{nq}q_{n0} + s_nc_{pn}p_{n0}=q_{(n+1)0}=q_{10}$. Combining the phase factors from the open fermion line and all closed fermion loops we find that the total extra phase proportional to the external time $t$, counting all $2N$ vertices in a diagram with a total of $N$ internal scalar lines is:
\beq
\phi_{\rm TOT} = \phi_{\rm line} + \sum_{i} \phi_{\rm loop}^i = \sum_{l=1}^{2N} s_{l}p_{l0} = 0\,.
\label{eq:totalphasezero}
\eeq
The total phase $\phi_{\rm TOT}$ vanishes because each scalar propagator appears twice in the sum, both as an outgoing ($s_l=+1$) and an incoming ($s_l=-1$) one with the same energies, and these contributions cancel pairwise.
\begin{figure}
  \centering
  \begin{minipage}[b]{6 cm}
    \includegraphics[width=1\textwidth]{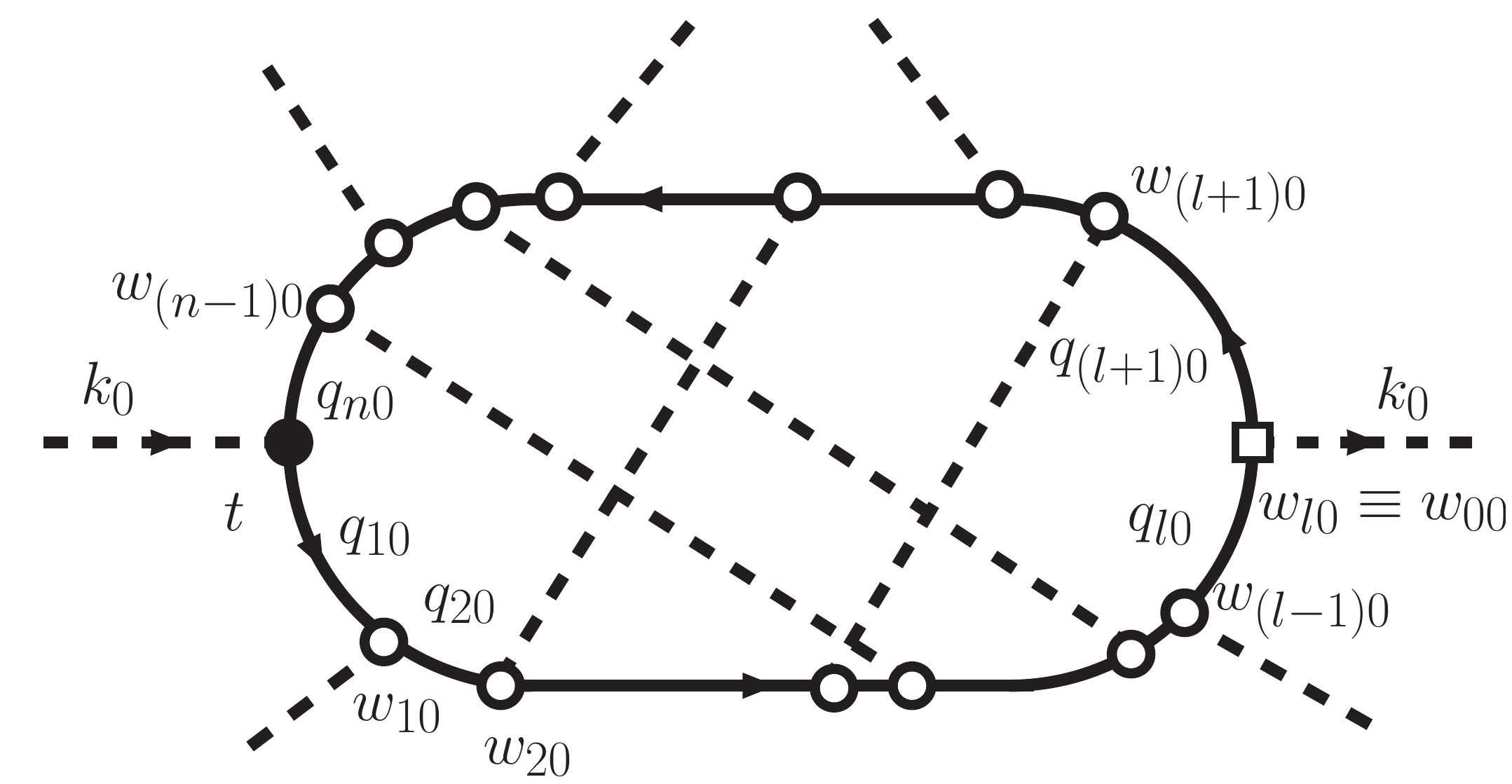}  
  \end{minipage} \hskip 1cm
  \begin{minipage}[b]{6 cm}
    \includegraphics[width=1\textwidth]{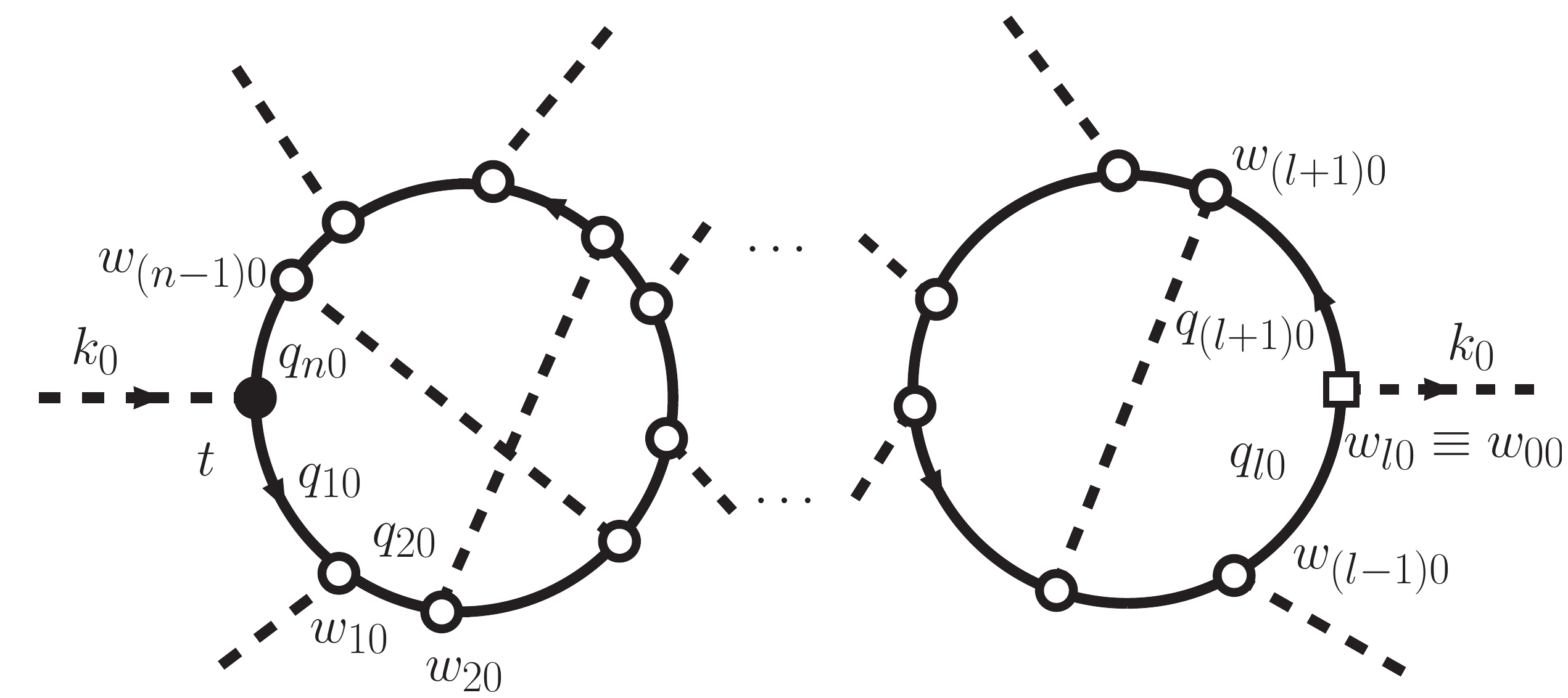}
  \end{minipage}
  \caption{Generic diagrams contributing to the scalar self-energy function $\Pi_{\rm eff}$ where the special times $t$ and $w_0$ are connected with the same (left) and with different (right) closed fermion loops.}
  \label{fig:sloopDiagrams}
\end{figure}
Physically the vanishing of $\phi_{\rm TOT}$ in Eq.~(\ref{effsigma2}) shows that while the energy is not conserved in the internal vertices in connection with the coherence propagators, the overall energy in the complete diagram is conserved.

Showing the vanishing of the total phase factor for an arbitrary scalar self-energy function $\Pi_{\rm eff}$ proceeds similarly to the fermionic case. In addition to the closed internal fermion loop one now has to evaluate the phases coming from the two sub-diagrams shown in Fig.~\ref{fig:sloopDiagrams}, which express the two possible ways of connecting the special vertices $t$ and $w_0$ into a most general diagram for  $\Pi_{\rm eff}(t,w_0,{\bf k})$. It is by now quite straightforward to show that the extra phase, in either of these cases becomes 
\begin{equation}
\phi_{\rm figs.5} = \sum_{j=1(\neq l)}^{n-1}\phi_j 
        + (1-c_{ql})q_{l0} + q_{n0} - q_{10} + k_0  = ... =  
  \sum_{j=1}^{n-1}s_jp_{j0} \,.
\end{equation}
Combining this result with the all possible phases from the internal closed loops gives the total phase factor for a $\Pi_{\rm eff}$ with $N$ internal scalar lines:
\beq
\phi^{\rm scalar}_{\rm TOT} = \phi_{\rm figs.5} + \sum_{i} \phi_{\rm loop}^i = \sum_{l=1}^{2N} s_{l}p_{l0} = 0\,,
\label{eq:totalphasezeroscal}
\eeq
where the final sum over the scalar energies vanishes by the same argument as in the fermionic case in Eq.~(\ref{eq:totalphasezero}). This result completes our proof that the local extra phases coming from the coherence propagators cancel in arbitrary self-energy diagrams. As a result, we can neglect all such phases in the actual calculations and use the local momentum space vertex Feynman rule:
\beq
  \phi\bar\psi\psi:\;\; y \; (2\pi)^4 \delta(q_0' - c_q q_{0} - s_p c_p p_0) \delta^3({\bf q}' - {\bf q} - s_p{\bf p} )\,.
\label{eq:frule0}
\eeq
The only difference from the usual rule then is the appearance of extra sign factors in the energy delta function in association with outgoing coherence propagators, leading to a local energy non-conservation within the loop.
\begin{figure}
\centering
\includegraphics[width=0.8 \textwidth]{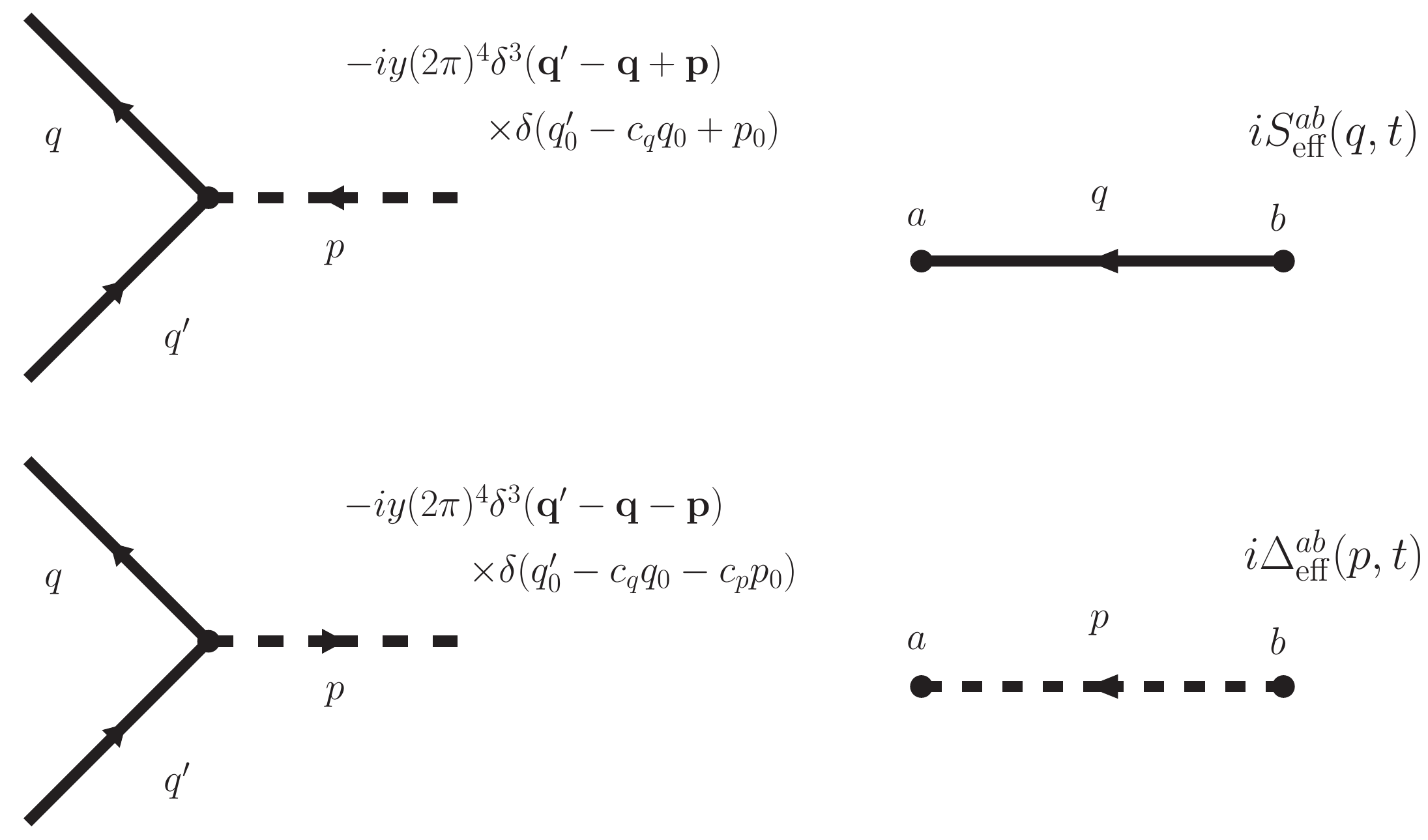}
\caption{The Feynman rules including coherence with an explicit orientation of the scalar lines following from Eqs.~(\ref{eq:Geffsnew} - \ref{eq:sigmaekaeff}) and Eq.~(\ref{eq:frule0}). The sign functions are $c_{p,q}=-1$ for the coherence parts of the associated propagators and $c_{p,q}=1$ otherwise.}
\label{fig:FeynmanRules}
\end{figure}
The complete set of momentum space Feynman rules for computing the fermionic and scalar self-energy functions including coherence propagators in the Yukawa theory with interaction Eq.~(\ref{interaction}) are shown in figure  \ref{fig:FeynmanRules}.  The arrows in the propagator lines indicate the direction of the flow corresponding to the 4-momentum of the positive energy state.

\section{Examples}
\label{sect:self-energies}
\begin{figure}
\centering
\includegraphics[width=0.7\textwidth]{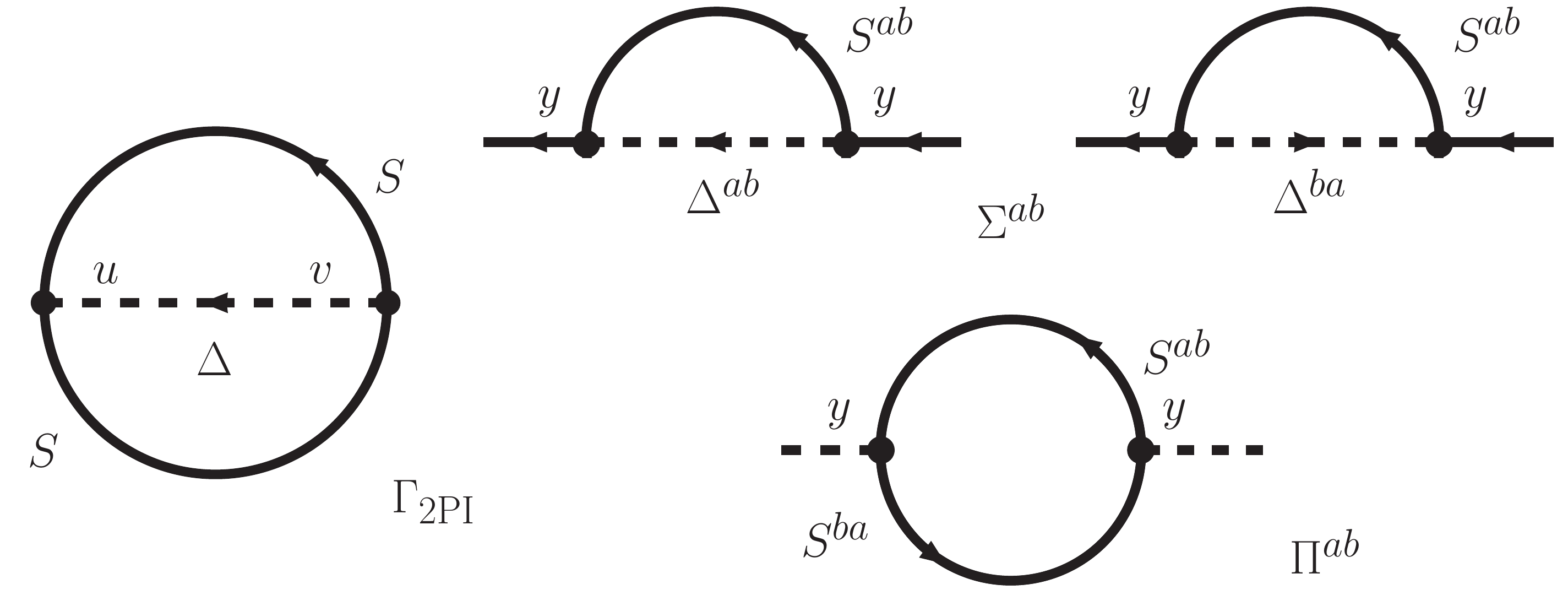}
    \caption{On left: the diagram contributing to the 2PI effective action at the lowest order for interaction (\ref{interaction}). On top right: the two diagrams
    contributing to the fermion self-energy $\Sigma^{ab}_{\rm eff}(k,t)$ and on bottom right: the single diagram contributing to the scalar self-energy $\Pi^{ab}_{\rm eff}(k,t)$.}
    \label{fig:2piS}
\end{figure}

We shall now compute some examples of self-energies and collision integrals in the Yukawa theory  described by the Lagrangian (\ref{interaction}), starting with with the simplest one-loop self-energy diagrams shown in the upper right panel of Fig.~\ref{fig:2piS}. These diagrams can be obtained for example from the two-particle irreducible (2PI) effective action vacuum diagram shown on the left in Fig.~\ref{fig:2piS}: 
\begin{equation} 
\Gamma_{\rm 2PI} = -\frac{y^2}{2} \int_C {\rm d}^4u\, {\rm d}^4v\, {\rm Tr}\left[S(u,v)\,S(v,u)\right]\Delta(u,v) \,,
\label{gamma2pI}
\end{equation}
where 
the integration is over the Keldysh path~\cite{HKR1}. For example the fermion self-energy now follows by a direct functional differentiation:
\begin{equation}
   \Sigma^{ab}(u,v) =  -iab \frac{\delta \Gamma_2[S]}{\delta S^{ba}(v,u)} = \frac{i y^2}{2} S^{ab}(u,v) \left[\Delta^{ab}(u,v) + \Delta^{ba}(v,u)\right] \,.
\label{sigma-ab2}
\end{equation}   
This much of the calculation is straightforward even without the Feynman rules developed in the last section. However, to proceed further would be tedious, since the two-time representation of Eq.~(\ref{sigma-ab2}) should be integrated over $w_0$ according to Eq.~(\ref{effsigma}) using the two-time effective propagators given by Eqs.~(\ref{eff_prop}) and (\ref{two-time_sca}), while taking great care of the different phase factors in the number of coherence- and mass-shell propagators.  

With the Feynman rules of Fig.~\ref{fig:FeynmanRules} at hand none of the extensive labour discussed above is needed and we can directly write down the final mixed representation expression for $\Sigma_{\rm eff}(k,t)$:
\begin{eqnarray} 
i\Sigma^{ab}_{\rm eff}(k,t) =  
\frac{y^2}{2} \sum_{s_b=\pm 1} \int \frac{{\rm d}^4q}{(2\pi)^4} \frac{{\rm d}^4p}{(2\pi)^4}  
iS^{ab}_{c_q}(q,t) \big[ i\Delta^{ab}_{c_p}(p,t) \delta_{s,1} 
                 + i\Delta^{ba}_{\rm eff}(p,t) \delta_{s,-1} \big] \times 
\nonumber \\
\times (2\pi )^4 \delta(k_0 - c_qq_0 - \hat c_p[s_b]p_0) 
             \delta^3({\bf k} - {\bf q} - s{\bf p})
\,.
\label{wignerself}
\end{eqnarray} 
Three observations are in place here: first we used a sum over the index $s^b$ together with the Kronecker delta functions associated with the scalar propagators to account automatically for the direction of the four momentum flow.  Second, we defined $G^{ab}_{c_{k}}$ to denote effective propagators Eqs.~(\ref{eq:Geffsnew} - \ref{eq:sigmaekaeff}) for which the coherence parts $G^{<,>}_{c,{\rm eff}}$ are combined with the correct $c_{p,k}$-factors in the vertex $\delta$-functions. Finally we took care of the implicit $s^b$-dependence of $c_p$-factor in the scalar line by introducing the notation
\beq
\hat c_p[s^b] = c_p \delta_{s^b,1} - \delta_{s^b,-1} \,
\label{csignautomate}
\eeq
inside the vertex delta-function. These notations will be useful later on. Note that the implicit dependence in $c$ signing out the coherent parts of the propagators still remains as explained above.

Expanding the sum in $s$ and separating the coherence solutions, we find that for example the self-energy functions $i\Sigma^< = i\Sigma^{12}$ and $i\Sigma^> = i\Sigma^{21}$ become:
\begin{eqnarray} 
i\Sigma^{<,>}_{\rm eff}(k,t) &=&  
 y^2 \int \frac{{\rm d}^4q}{(2\pi)^4} \frac{{\rm d}^4p}{(2\pi)^4}  
 (2\pi )^3 \delta^3({\bf k} - {\bf q} - {\bf p}) \times
\nonumber \\
&&\times 
\Big[ \;
 (2\pi ) \delta(k_0 - q_0 - p_0) iS_m^{<,>}(q,t) i\Delta_m^{<,>}(p,t)  
\nonumber \\
&& \phantom{i}
+ (2\pi ) \delta(k_0 + q_0 - p_0) iS_{c,{\rm eff}}^{<,>}(q,t) i\Delta_m^{<,>}(p,t)  
\nonumber \\
&& \phantom{i}
+ (2\pi ) \delta(k_0 - q_0 + p_0) iS_m^{<,>}(q,t) \Delta_{c,{\rm eff}}(p,t)  
\nonumber \\
&& \phantom{i}
+(2\pi ) \delta(k_0 + q_0 + p_0) iS_{c,{\rm eff}}^{<,>}(q,t) 
\Delta_{c,{\rm eff}}(p,t) 
\, \Big] \,, 
\label{wignerself1}
\end{eqnarray} 
where we used the fact $\Delta_m^>(-p,t) = \Delta_m^<(p,t)$ and $\Delta_{c,{\rm eff}}(p_0,-{\bf p},t) = \Delta_{c,{\rm eff}}(p_0,{\bf p},t)$ (because of the isotropy and the identity $\Delta_{c,{\rm eff}}^> = \Delta_{c,{\rm eff}}^< \equiv \Delta_{c,{\rm eff}}$). Equation (\ref{wignerself1}) shows explicitly how different types of energy conservation are associated with the coherence propagators; the delta-function associated with the $S_m\Delta_m$-term has the normal signature. Using the isotropy again we can combine  the coherence and mass terms together under the same delta function $\delta^4(k-q-p)$ which can then be used to integrate over the momentum $p$, eventually giving just:
\begin{equation} 
i\Sigma^{<,>}_{\rm eff}(k,t) =   
 y^2 \int \frac{{\rm d}^4q}{(2\pi)^4}   
 i\tilde S^{<,>}_{\rm eff}(q,t) i\tilde \Delta^{<,>}_{\rm eff}(k-q,t)  \,,
\label{wignerself2}
\end{equation} 
where we have further defined 
\begin{eqnarray}
i\tilde S^{<,>}_{\rm eff\pm} &=&  2\pi \big( {\cal S}^{<,>}_{m\pm} + {\cal S}^{<,>}_{c\mp} \big) 
   \,\delta(k_0 \mp \omega_{\bf k})  
 \nonumber \\[2mm]
i\tilde\Delta^{<,>}_{\rm eff\pm} &=&  \frac{\pi}{\omega_{\bf k}}  
 \big( \pm f^{<,>}_{m\pm} + f^{<,>}_{c\mp}\big) 
 \,\delta(k_0 \mp \omega_{\bf k})  \, .
\label{eq:Seffsnew2}
\end{eqnarray}
Note that these functions differ from the effective propagators in Eq.~(\ref{eq:Geffsnew}) in that here the coherence-shell functions appear in ``wrong" energy shells. 

Equation (\ref{wignerself2}) is remarkably simple; it can be obtained  from the standard expression for $\Sigma^{<,>}$ by a direct substitution $S^{<,>} \rightarrow \tilde S^{<,>}_{\rm eff}$ and $\Delta^{<,>} \rightarrow \tilde \Delta^{<,>}_{\rm eff}$. Unfortunately such a simple rule does not generalize to arbitrary diagrams, as can be seen already from the one-loop scalar self-energy function $\Pi_{\rm eff}$. Indeed, a direct evaluation of the scalar self-energy diagram shown in Fig.~\ref{fig:2piS} gives
\begin{eqnarray} 
i\Pi^{ab}_{\rm eff}(k,t) =  
-\frac{y^2}{2} \int \frac{{\rm d}^4q_1}{(2\pi)^4} \frac{{\rm d}^4q_2}{(2\pi)^4}  
{\rm Tr} [ iS^{ab}_{c^q_1}(q_1,t) iS^{ba}_{c^q_2}(q_2,t) ] \times 
\nonumber \\
\times (2\pi )^4 \delta(q_{20} - c^q_1q_{10} - k_0) 
             \delta^3({\bf q}_{20} - {\bf q}_{10} - {\bf k})
\,,
\label{wignerself3}
\end{eqnarray} 
from which it is now easy to see that the self-energy functions $\Pi^{<,>}_{\rm eff}$ become:
\begin{equation} 
i\Pi^{<,>}_{\rm eff}(k,t) =  
- \frac{y^2}{2} \int \frac{{\rm d}^4q}{(2\pi)^4} 
{\rm Tr} [ i\tilde S^{<,>}_{\rm eff}(q,t) iS^{{>,<}}_{\rm eff}(k-q,t) ] \,.
\label{wignerself3b}
\end{equation} 
That is, the substitution of $S^{<,>} \rightarrow \tilde S^{<,>}_{\rm eff}$ is made only for the fermion line which is flowing {\em out} from the special vertex $t$, while for the fermion propagator flowing {\em into} the vertex the substitution is $S^{<,>} \rightarrow S^{<,>}_{\rm eff}$. These simplifications generalize for propagators directly connected to the special $t$-vertex in arbitrary diagrams. In all other cases the assignment of energy signs in the coherence propagators is more complicated and can only be worked out by a use of the full Feynman rules of section~\ref{sect:effective-feynman}.
\subsection{A two-loop example}

\begin{figure}
\centering
\includegraphics[width=0.5 \textwidth]{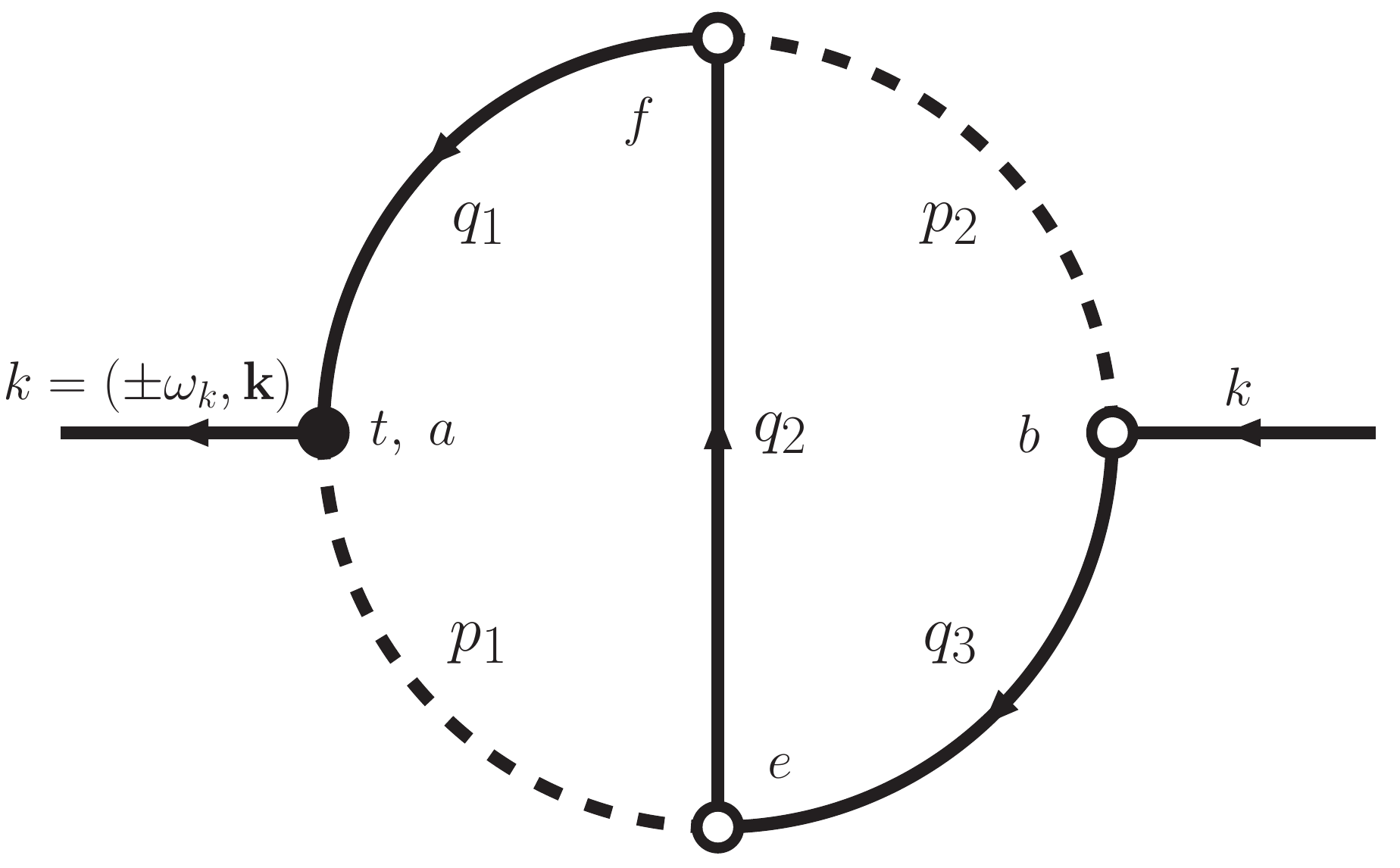}
    \caption{The two-loop 2PI-fermion self-energy diagram. Each scalar propagator goes in both directions, so the diagram represents four independent terms. The special vertex $a$ corresponding to the external time $t$ is again marked by a black dot, while the internal vertices are shown by open dots. Full explanation in text.}
    \label{fig:2-loop}
\end{figure}

Let us next consider a more complicated 2-loop example. The contribution from the diagram shown in Fig.~\ref{fig:2-loop} to $\Sigma^{ab}$ is:
\begin{eqnarray}
i\Sigma^{ab}_{\rm eff}(k,t) &=& 
\frac{y^4}{4} \sum_{fe}\sum_{s^e_1,s^f_2,s^b_2} \delta_{s_2^f,-s^b_2} \int 
\prod_{i=1}^3 \frac{{\rm d}^4q_i}{(2\pi)^4}\,
\prod_{j=1}^2 \frac{{\rm d}^4p_j}{(2\pi)^4}\;
\,[ iS^{af}_{c_{q1}}(q_1)iS^{fe}_{c_{q2}}(q_2)iS^{eb}_{c_{q3}}(q_3)] \times\,
\nonumber \\
&\times&  
(2\pi)^{12} \,
\delta^3({\bf q}_2 - {\bf q}_1 - s^f_2{\bf p}_2)
\delta^3({\bf q}_3 - {\bf q}_2 - s^e_1{\bf p}_1)
\delta^3({\bf k}   - {\bf q}_3 - s^b_2{\bf p}_2)
\nonumber \\
&\times&  
\delta(q_{20} - c_{q1}q_{10} - \hat c^{f}_{p2}p_{20})
\delta(q_{30} - c_{q2}q_{20} - \hat c^{e}_{p1}p_{10})
\delta(k_{0}  - c_{q3}q_{30} - \hat c^{b}_{p2}p_{20})
\nonumber \\
&\times&
\big[ i\Delta^{ae}_{c_{p1}}(p_1) \delta_{s^e_1,1}
      + i\Delta^{ea}_{\rm eff}(p_1) \delta_{s^e_1,-1}  \big] \, 
\big[   i\Delta^{bf}_{c_{p2}}(p_2) \delta_{s^f_2,1} 
      + i\Delta^{fb}_{c_{p2}}(p_2) \delta_{s^b_2,1} \big] 
 \,,
\label{eq:twoloopbig}
\end{eqnarray}
where we have suppressed the $t$-arguments in propagators for clarity. We also continued using the tagging on propagators $G^{ab}_{c_k}$ introduced in the 1-loop example above, as well as the Kronecker delta notation and $s^g_k$-sums to indicate all possible directings of the scalar vertices. We also used the shorthand $\hat c^b_p \equiv \hat c_{p}[s_p^b]$ for the notation Eq.~(\ref{csignautomate}) in the vertex delta functions. These notations unambiguously indicate the scalar momentum flow in the internal vertices and whether the coherence solution is to get a nontrivial sign or not.  

Now, because the momentum $p_{10}$ appears only in one of the delta-functions, we can absorb the index $c^{e}_{p1}$ to a scalar propagator by effecting a transformation $p_{10}\rightarrow c_{p1}[s^e]p_{10}$, which amounts to the substitution $\Delta^{ae}_{c_{p1}}(p_1)\rightarrow \tilde\Delta^{ae}_{\rm eff}(p_1)$. 
The same argument applies to the energy variable $q_{10}$ corresponding to the fermion propagator flowing into the special $t$-vertex, which we can deal with by a change $q_{10}\rightarrow c_{q1}q_{10}$, causing $S^{af}_{c_{q1}}(q_1)\rightarrow \tilde S^{af}_{\rm eff}(q_1)$. Beyond these simplifications associated with the special vertex $t$ no significant reduction is possible. Being careful  in particular with the scalar flow direction assignments on vertices $e$ and $f$ one eventually finds: 
\begin{eqnarray}
i\Sigma^{ab}_{\rm eff}(k,t) &=& \frac{y^4}{2} \sum_{fe} \int 
\prod_{i=1}^3 \frac{{\rm d}^4q_i}{(2\pi)^4}\,
\prod_{j=1}^2 \frac{{\rm d}^4p_j}{(2\pi)^4}\;
\, [ i\tilde S^{af}_{\rm eff}(q_{1}) i\tilde \Delta^{ae}_{\rm eff}(p_{1}) ]  \,
\nonumber \\
&\times& 
 iS^{fe}_{c_{q2}}(q_2)iS^{eb}_{c_{q3}}(q_3)\big[ i\Delta^{bf}_{c_{p2}}(p_2) + i\Delta^{fb}_{c_{p2}}(-c_{p2}p_2) \big] \,
\nonumber \\
&\times&  
(2\pi)^{12} \delta^3({\bf q}_2 - {\bf q}_1 - {\bf p}_2)
\delta^3({\bf q}_3 - {\bf q}_2 - {\bf p}_1)
\delta^3({\bf k}   - {\bf q}_3 + {\bf p}_2) \,,
\nonumber \\
&\times&  
\delta(q_{20} - q_{10} - c_{p2}p_{20})
\delta(q_{30} - c_{q2}q_{20} - p_{10})
\delta(k_{0}  - c_{q3}q_{30} + p_{20})\,.
\label{eq:twoloopself}
\end{eqnarray}
The effective Wightman functions $\tilde S^{<,>}_{\rm eff}$ and $\tilde \Delta^{<,>}_{\rm eff}$ were defined in Eqs.~(\ref{eq:Seffsnew2}). The special diagonal propagators $\tilde G^{11}_{\rm eff}$ and $\tilde G^{22}_{\rm eff}$ ($G$ again denotes $S$ or $\Delta$) are given by the analogous relations to Eq.~(\ref{eq:sigmaekaeff}) with $G^<_{\rm eff}$ replaced by $\tilde G^<_{\rm eff}$ everywhere, and the functions $\tilde G^-_{\rm eff}$ again correspond to functions $\tilde G^<_{\rm eff}$ from which the vacuum parts have been subtracted out. 

Particular self-energy functions, and particular sub-contributions to these self-energies are obtained by assigning special values for the time-contour indices $a$, $b$, $e$ and $f$ in Eq.~(\ref{eq:twoloopself}). Let us consider one example here for illustration. Choosing $aefb=1112$ we get a two-loop correction to the  self-energy function $\Sigma^{<}=\Sigma^{12}$:
\begin{eqnarray}
i\Sigma^{<}_{\rm eff}(k,t)_{1112} &=& \frac{y^2}{2} \int 
\frac{{\rm d}^4q_3}{(2\pi)^4}\frac{{\rm d}^4p_2}{(2\pi)^4}
i\Lambda^{111}(q_3,p_2,t) iS^{12}_{c_{q3}}(q_3)
\big[ i\Delta^{12}_{c_{p2}}(p_2) + i\Delta^{21}_{c_{p2}}(-c_{p2}p_2) \big] \,
\nonumber \\
&& \times (2\pi)^{4} \delta(k_{0}  - c_{q3}q_{30} - c_{p2}p_{20})
\delta^3({\bf k}   - {\bf q}_3 - {\bf p}_2) \,,
\label{eq:selfenergy1112}
\end{eqnarray}
where the function $\Lambda^{111}$ is given by
\begin{eqnarray}
i\Lambda^{111}(q_3,p_2,t) &\equiv& y^2 \int 
\frac{{\rm d}^4q_2}{(2\pi)^4}
\frac{{\rm d}^4p_1}{(2\pi)^4}
[ i\tilde S^{11}_{\rm eff}(q_2+p_2) i\tilde \Delta^{11}_{\rm eff}(p_{1}) ]  
 iS^{11}_{c_{q2}}(q_2)
 \nonumber\\
&&\times  
(2\pi)^{4}
\delta(q_{30} - c_{q2}q_{20} - p_{10})
\delta^3({\bf q}_3 - {\bf q}_2 - {\bf p}_1) \,.
\label{eq:vertexfun111}
\end{eqnarray}
Note that the self-energy function (\ref{eq:selfenergy1112}) becomes equivalent with the 1-loop result shown in Eq.~(\ref{wignerself}) when $i\Lambda^{111} \rightarrow 1$. In this sense $\Lambda^{111}$ can be interpreted as a one-loop vertex function correction to the 1-loop self-energy. On practical side, observe that the coherence parts of $S^{12}_{c_{q3}}(q_3)$ in Eq.~(\ref{eq:selfenergy1112}) and $S^{11}_{c_{q2}}(q_2)$ in Eq.~(\ref{eq:vertexfun111}) are associated with a different phase space integrals than the non-coherent parts, due to the $c_{qi}$-dependence of the remaining energy delta functions. This dependence is passed onto the arguments of the other propagators when one performs the integral over the remaining delta function. Of course there are other contributions to $\Sigma^<$ and also $\Sigma^>$ is needed to compute the corresponding collision integrals in the quantum Boltzmann equations. Nevertheless, we believe that these examples amply display the feasibility of the use of our Feynman rules for practical diagrammatic calculations in coherent perturbation expansions. In particular, a two-loop calculation similar to the one presented above, but generalized to the multiflavour case~\cite{FHKR}, will be relevant in the context of resonant leptogenesis \cite{resonant_lepto}.

\subsection{Direction independent Feynman rules}

\begin{figure}
\centering
\includegraphics[width=0.8 \textwidth]{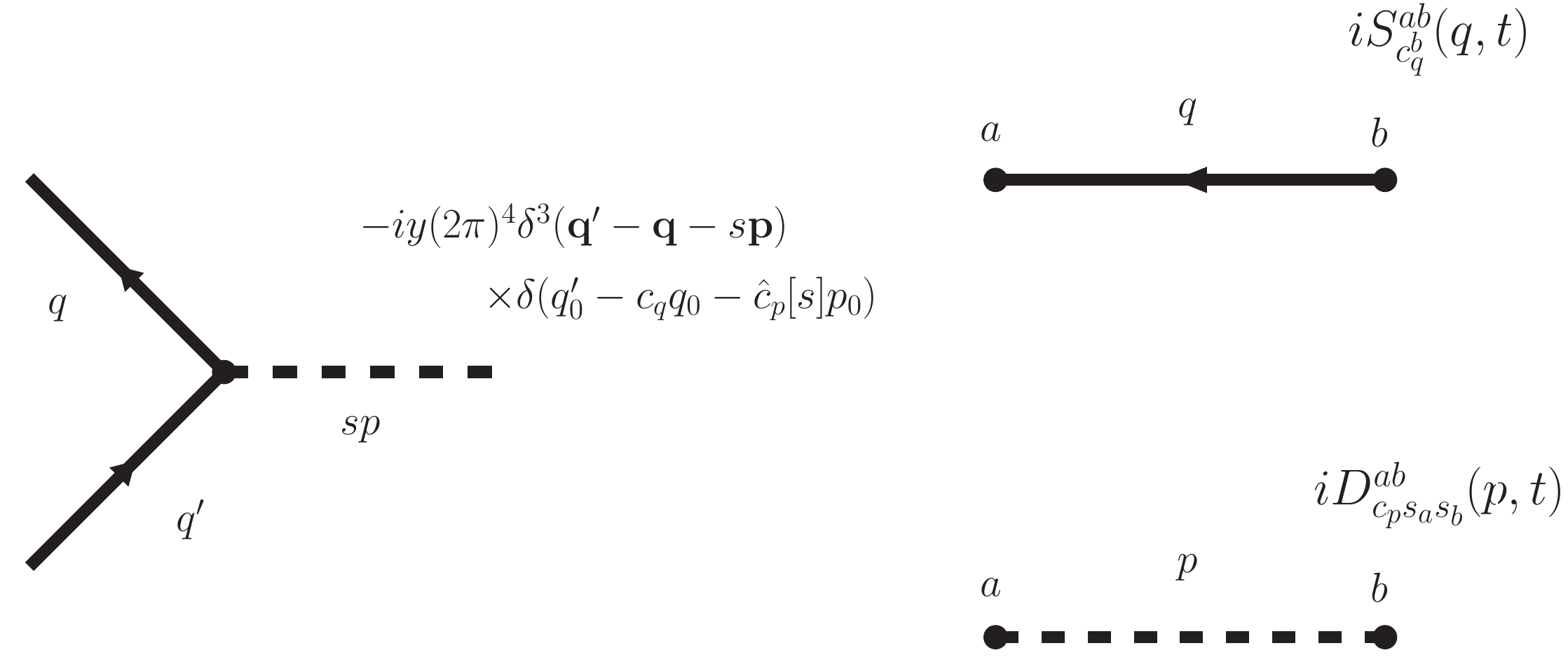}
\caption{The final direction independent rules including the coherence.}
\label{fig:FeynmanRules2}
\end{figure}

The result  (\ref{eq:twoloopbig}) for the first time displays fully internal scalar and fermion propagators. In particular, it shows how the internal scalar propagators combine to a form which is insensitive to the choice of flow directions of the scalar lines. We can use these findings to reformulate the direction dependent Feynman rules of Fig.~\ref{fig:FeynmanRules} in an even more useful form. The complete rules can now be stated as follows:

\begin{itemize}

\item{} Draw the diagram just as in the vacuum field theory and associate the usual vacuum symmetry factor with it.

\item{} Give each vertex an index $s^b_p$ associated with the direction of the scalar field, and use the vertex Feynman rule %
\beq
  \phi\bar\psi\psi:\;\; -i y \; (2\pi)^4 \delta(q_0' - c_q q_{0} - \hat c_p[s^b_p] p_0) \delta^3({\bf q}' - {\bf q} - s^b_p{\bf p} )\,,
\label{eq:frule1}
\eeq
where $\hat c_p[s^b_p] \equiv c_p\delta_{s^b_p,1}-\delta_{s^b_p,-1}$, where $c_p=-1$ for the coherence parts of the propagators, and $c_p=1$ otherwise.

\item{} For each fermion line in the diagram, substitute a {\em sign-tagged}  propagator $S^{ab}_{c_q}(q,t)$, where the index $c_q$ explicitly keeps track of the coherence sign in the effective fermion propagator $S^{ab}_{\rm eff}(q,t)$ defined in Eqs.~(\ref{eq:Geffsnew}-\ref{eq:sigmaekaeff}).

\item{} For each scalar line substitute a propagator
\beq
iD^{ab}_{c_p,s^a,s^b}(p,t) \equiv \frac{1}{2} 
\left[ i\Delta^{ab}_{c_p}(p,t) \delta_{s^b,1} 
      + i\Delta^{ba}_{c_{p}}(p,t) \delta_{s^a,1}\right] \delta_{s^a, -s^b} \,,
\label{eq:unsignedscalarprop}
\eeq
where $\Delta^{cd}_{c_p}(p,t)$ is again the sign-tagged propagator corresponding to the effective scalar propagator $\Delta^{cd}_{\rm eff}(p,t)$ defined in Eqs.~(\ref{eq:Geffsnew}-\ref{eq:sigmaekaeff}).

\item{} Sum over all indices $s^b_p$ associated with all vertices.

\item{} All other Feynman rules associated with the momentum integrations and negative signs associated with closed fermion loops are as usual.

\end{itemize}
Special simplifications of the rules apply for propagators connected to the external vertex $t$; see the discussion below Eq.~(\ref{wignerself3b}).
These direction independent rules are presented graphically in the figure~\ref{fig:FeynmanRules2}.  

\subsection{Explicit 1-loop collision integrals for fermions}

We now compute the final expressions for the fermionic collision integrals (\ref{coll_types1}) following from our above examples for $\Sigma_{\rm eff}$. To keep our expressions simple we only consider the 1-loop diagram, and that the scalar field is in thermal equilibrium with (see eg. \cite{PSW}):
\begin{equation}
 i \Delta^{<,>}_{{\rm eq}} (p) = 
 2\pi\,{\rm sgn}(p_0) f^{<,>\phi}_{{\rm eq},{\rm sgn}(p_0)}({\bf p}) \delta(p^2-m_\phi^2)\,,
\label{effectiveD1}
\end{equation}
where $f^{<\phi}_{{\rm eq} \pm}({\bf p}) 
= 1/(e^{\pm \beta \,\omega^\phi_{\bf p}} - 1)$
and $f^{>\phi}_{{\rm eq} \pm}({\bf p}) = 1 + f^{<\phi}_{{\rm eq} \pm}({\bf p})$ with $\beta = 1/T_\phi$. Using the equilibrium propagator (\ref{effectiveD1}) in Eq.(\ref{wignerself2}), going back to the form with both $q$- and $p$-integrals and integrating over $q_0$ and $p_0$ using the on-shell delta functions in the propagators, we find
\begin{eqnarray}
i\Sigma^{<,>}_{\rm eff}(\pm\omega_{\bf k},{\bf k};t) 
&=&  y^2\sum_{h,s_{\bf p},s_{\bf q}}\int \frac{{\rm d}^3 {\bf q}}{(2\pi)^3} 
              \frac{{\rm d}^3 {\bf p}}{2\omega_{\bf p}(2\pi)^3}
           (2\pi)^4 \delta^3({\bf k} - {\bf q} - {\bf p}) 
\nonumber\\
&& \times \,  
   \delta(\pm\omega_{\bf k} - s_{\bf q} \omega_{\bf q}  - s_{\bf p} \omega^\phi_{\bf p})  
    f^{<,>\phi}_{{\rm eq}, s_{\bf p}}({\bf p}) 
    [{\cal  S}^{<,>}_{mh,s_{\bf q}}({\bf q})+ {\cal  S}^{<,>}_{mh,-s_{\bf q}}({\bf q})]\,. \phantom{i} 
\label{effectiveself}
\end{eqnarray} 
The sign-factors $s_{\bf q,p}$ appear in Eq.~(\ref{effectiveself}) as a result of projections on positive and negative energy shells after integrations over $q_0$ and $p_0$; they should not be confused with the orientation signs encountered in our Feynman rules. According to the standard 3-particle kinematics the momentum delta functions have roots only if $m_\phi \geq 2|m(t)|$ and only for the signatures, $\delta(\omega_{\bf k} + \omega_q  - \omega^\phi_p)$ and $\delta(-\omega_{\bf k} - \omega_q  + \omega^\phi_p)$, so that each on-shell self-energy function has only one contribution coming from the sum in Eq.~(\ref{effectiveself}). Employing the decompositions (\ref{4spectral2Cov}) we finally get:
\begin{eqnarray}
i\Sigma^{<,>}_{\rm eff}(\pm\omega_{\bf k},{\bf k},t) 
&=& y^2 \int \frac{{\rm d}^3 {\bf q}}{2\omega_{\bf q}(2\pi)^3} 
             \frac{{\rm d}^3 {\bf p}}{2\omega^\phi_{\bf p}(2\pi)^3} 
             (2\pi)^4
             \delta(\omega_{\bf k} + \omega_{\bf q}  - \omega^\phi_{\bf p})
             \delta^3({\bf k} - {\bf q} - {\bf p}) \;
\nonumber \\ 
&& \times  f^{<,>\phi}_{{\rm eq} \pm}({\bf p})
  \sum_{h'} \left( 
        {\cal K}_{mh'\mp}({\bf q}) f^{<,>}_{mh'\mp}({\bf q};t) 
     +  {\cal K}_{ch'\pm}({\bf q}) f^{<,>}_{ch'\pm}({\bf q};t)
           \right)  \,. 
\label{Y-self1}
\end{eqnarray}
Given Eq.~(\ref{Y-self1}) it is a simple matter to calculate the scalar self-energy functions $\Sigma^{<,>}_{jh\pm}$ which appear in Eqs.~(\ref{coll_types1})\footnote{Note that the signature of the $\omega_{\bf q}$ and ${\bf q}$ in the energy- and momentum conservation delta functions are different. This is because one of the fermions labelled by ${\bf k}$ and ${\bf q}$ must be an antiparticle corresponding to a negative energy state in our language. That is if, say, the ${\bf k}$-state is a positive energy particle, then the ${\bf q}$ state must have a negative frequency. However, since in Eq.~(\ref{Y-self1}) we wrote the energies in terms of positive physical frequencies, which makes the momentum ${\bf q}$ to appear with a ``wrong" sign. The solution is of course that here the {\em physical} 3-momentum of the antiparticle is just $-{\bf q}$. The same argument applies to the case where the ${\bf k}$-state is an antiparticle. We could reinstate the normal signatures by a change $\bf k\rightarrow \pm {\bf k}$. $\bf p\rightarrow \pm {\bf p}$ and $\bf q\rightarrow \mp {\bf q}$, and the corresponding Feynman-St{\"u}ckelberg reinterpretation of the $f$-factors. However, there is no practical advantage of making this change and it is easier to keep using the unphysical momenta instead.}.
Multiplying (\ref{Y-self1}) by the projectors ${\cal K}_{jh\pm}$ and taking the traces we find:
\begin{eqnarray}
   \Sigma^{<,>}_{jh\pm} &=&  
        \int \frac{{\rm d}^3 {\bf q}}{2\omega_q(2\pi)^3} 
             \frac{{\rm d}^3 {\bf p}}{2\omega^\phi_p(2\pi)^3} 
             \delta(\omega_{\bf k} + \omega_{\bf q}  - \omega^\phi_{\bf p})
             \delta^3({\bf k} - {\bf q} - {\bf p}) \;f^{<,>\phi}_{{\rm eq} \pm}({\bf p})
\nonumber \\  
  && \times    \sum_{h'}
   \big( {\cal M}^2_{mj\pm}(h',h,{\bf q},{\bf k})  f^{<,>}_{mh'\mp}({\bf q};t) + {\cal M}^2_{cj\pm}(h',h,{\bf q},{\bf k}) f^{<,>}_{ch'\pm}({\bf q};t)\big) \,,
\label{Y-self2}
\end{eqnarray}
where the generalized squared matrix elements are given by the traces:
\begin{eqnarray}
 {\cal M}^2_{mj\pm}(h',h,{\bf q},{\bf k}) 
  &=& y^2{\rm Tr}\big[{\cal K}_{mh'\mp}({\bf q})
                          {\cal K}_{jh\pm}({\bf k}) \big] 
\nonumber\\
 {\cal M}^2_{cj\pm}(h',h,{\bf q},{\bf k}) 
  &=& y^2{\rm Tr}\big[{\cal K}_{ch'\pm}({\bf q})
                          {\cal K}_{jh\pm}({\bf k}) \big] \,,
\label{Y-matrixelements}
\end{eqnarray}
with $j=m,c$. Note that these matrix elements are not in general real (see Eqs.~(\ref{Y-matrixelements2}-\ref{Y-matrixelements1}) below). Given the self-energy functions $\Sigma^{<,>}_{jh\pm}$ we can now combine all terms appearing in the {\em r.h.s.}~of the Eqs.~(\ref{coll_types1}) to form the complete collision integrals ${\cal C}_{jh\pm}$. The mass-shell collision integrals are in this way found to be: 
\begin{eqnarray}
{\cal C}_{mh\pm}[f_\alpha]
  &=&  \Re \, \frac{1}{2\omega_{\bf k}}\sum_{h'} 
        \int \frac{{\rm d}^3 {\bf q}}{2\omega_q(2\pi)^3} 
             \frac{{\rm d}^3 {\bf p}}{2\omega^\phi_p(2\pi)^3} 
             (2\pi)^4\hat \delta^4(k - q - p) \times 
\nonumber\\
&&
\times \Big[ \,\, 
      {\cal M}^2_{mm\pm} 
         f^{<\phi}_{{\rm eq} \pm}({\bf p}) f^<_{mh'\mp}({\bf q})     
   		 f^>_{mh\pm}({\bf k})  
\nonumber\\
&&\phantom{l}
   + {\cal M}^2_{cm\pm}
         f^{<\phi}_{{\rm eq} \pm}({\bf p}) f^<_{ch'\pm}({\bf q})  
         f^>_{mh\pm}({\bf k})  
\nonumber\\
&&\phantom{l}
   + {\cal M}^2_{mc\mp}
         f^{<\phi}_{{\rm eq} \mp}({\bf p}) f^<_{mh'\pm}({\bf q}) 
         f^>_{ch\mp}({\bf k})
\nonumber\\
&& \phantom{l}
  + {\cal M}^2_{cc\mp} 
         f^{<\phi}_{{\rm eq} \mp}({\bf p}) f^<_{ch'\mp}({\bf q})   
         f^>_{ch\mp}({\bf k}) \;
     \Big] - \big[>\leftrightarrow < \big] \,, \phantom{hi}
\label{Y-collisionintegral1}
\end{eqnarray}
where we defined a shorthand notation $\hat \delta^4(k-q-p) \equiv \delta(\omega_{\bf k} + \omega_{\bf q}  - \omega^\phi_{\bf p}) \delta^3({\bf k} - {\bf q} - {\bf p})$.
As usual, we see that the distribution products in the Boltzmann equations are independent of the form of the interactions, the details of which are entirely encoded in the matrix elements. The first line in equation (\ref{Y-collisionintegral1}), which contains only the mass-shell distribution functions, corresponds to the usual collision integral in the standard Boltzmann equations. Indeed, rewriting the distribution functions in terms of the usual particle and antiparticle numbers, one finds for example the familiar form:

\begin{equation}
f^{>\phi}_{{\rm eq}+}({\bf p}) f^{>}_{h'-}(-{\bf q}) f^{<}_{h+}({\bf k})   
= (1 + n^{\phi}_{\rm eq}(p)) \bar n_{h'}(q) n_h(k) \,.  
\label{3bodycollisioterm3a}
\end{equation}
One could similarly rewrite all the other products of distributions appearing in Eq.~(\ref{Y-collisionintegral1}) using the particle and antiparticle distribution functions. However, since $f_\alpha$'s are the quantities which naturally appear in all our loop calculations and in our equations of motion, we will stick to  this notation in what follows. The second line in Eq.~(\ref{Y-collisionintegral1}) comes from an internal coherence line in our one-loop diagram, corresponding to a coherence modulation of a scattering rate off a state with momentum $\bf q$. The third line represents an on-shell contribution to the collisional coupling between the mass and coherence shells in the mass-shell equation, and the last term gives the coherence modulation of this collisional coupling term. The coherence collision integral ${\cal C}_{ch\pm}$ is somewhat more complicated:
\begin{eqnarray}
{\cal C}_{ch\pm}[f_\alpha]
 	&=&  \frac{1}{2\omega_{\bf k}} \sum_{h',s} 
        \int \frac{{\rm d}^3 {\bf q}}{2\omega_{\bf q}(2\pi)^3} 
             \frac{{\rm d}^3 {\bf p}}{2\omega^\phi_{\bf p}(2\pi)^3} 
             (2\pi)^4\hat \delta^4(k - q - p) \times 
\nonumber \\ 
    && 
    \times \;  
 \Big[ \; 
  {\cal M}^2_{mm\pm} \; 
       f^{<\phi}_{{\rm eq}\pm}({\bf p}) f^<_{mh'\mp}({\bf q}) 
                                        f^>_{ch\pm}({\bf k})
\nonumber \\ && \phantom{a} 
+ {\cal M}^2_{cm\pm} \; 
       f^{<\phi}_{{\rm eq}\pm}({\bf p}) f^<_{ch'\pm}({\bf q}) 
                                        f^>_{ch\pm}({\bf k})  
\nonumber \\ && \phantom{a} 
 + \xi_{\bf k} {\cal M}^{2\;*}_{mc\pm} \; 
       f^{<\phi}_{{\rm eq}\pm}({\bf p}) f^<_{mh'\mp}({\bf q})
                                        f^>_{mh\pm}({\bf k})
\nonumber \\ && \phantom{a} 
 +  \xi_{\bf k} {\cal M}^{2 \;*}_{cc\pm} \; 
       f^{<\phi}_{{\rm eq}\pm}({\bf p}) f^<_{ch'\mp}({\bf q}) 
                                        f^>_{mh\pm}({\bf k}) 
 \nonumber \\ 
    && \phantom{ll} + (\pm \leftrightarrow \mp)^*
 \Big]  - \big[>\leftrightarrow < \big]\,.
\label{Y-collisionintegral2}
\end{eqnarray}
The interpretation of the various terms in this equation should be obvious now.
Finally, for completeness, we give the explicit expressions of the matrix elements ${\cal M}_{ab\pm}$ following from our 1-loop example. A straightforward evaluation of the traces in Eq.~(\ref{Y-matrixelements}) gives:
\begin{eqnarray}
 {\cal M}^2_{mm\pm} &=& {\cal M}^2_{0} \nonumber\\
 {\cal M}^2_{cm\pm} &=& \pm \frac{m_R}{\omega_{\bf q}}{\cal M}^2_{0} \pm {\cal M}^2_{R} - i {\cal M}^2_{I} \nonumber\\
 {\cal M}^2_{mc\pm} &=& \mp \frac{m_R}{\omega_{\bf k}}{\cal M}^2_{0} \pm {\cal M}^2_{R} - i {\cal M}^2_{I} \nonumber\\
 {\cal M}^2_{cc\pm} &=& (1-\frac{m_R^2}{\omega_{\bf k}\omega_{\bf q}}){\cal M}^2_{0} 
 +(\frac{m_R}{\omega_{\bf k}}-\frac{m_R}{\omega_{\bf q}}) {\cal M}^2_{R} \pm i (\frac{m_R}{\omega_{\bf q}}+\frac{m_R}{\omega_{\bf k}}) {\cal M}^2_{I} \,,
\label{Y-matrixelements2}
\end{eqnarray}
where we have defined:
\begin{eqnarray}
  {\cal M}^2_{0}&\equiv&  y^2 \Big[ (1 + h'h\; {\bf \hat q} \cdot {\bf \hat k}) (q \cdot k - m_R^2 + m_I^2) 
 + h'h|{\bf q}||{\bf k}|(1-({\bf \hat q} \cdot {\bf \hat k})^2)\Big]   \nonumber\\
 {\cal M}^2_{R} &\equiv&  y^2 \Big[ (1 + h'h\; {\bf \hat q} \cdot {\bf \hat k})\; m_R(\omega_{\bf q}-\omega_{\bf k}) \Big]\nonumber\\
 {\cal M}^2_{I} &\equiv&  y^2 \Big[ (1 + h'h\; {\bf \hat q} \cdot {\bf \hat k})\; m_I(h'|{\bf q}|- h|{\bf k}|)\Big]\,,
\label{Y-matrixelements1}
\end{eqnarray}
where ${\bf \hat q} \cdot {\bf \hat k} = \cos\theta$ is the angle between fermion momenta and the product of fermion four momenta is $q \cdot k = \frac{1}{2}m_{\phi}^2 - |m(t)|^2$. Note in particular that the squared matrix elements ${\cal M}^2_{mm\pm}$ are real and equivalent to the standard expression for a scalar field decaying to fermions after one makes the usual Feynman-St\"uckelberg interpretation of the negative energy states as the positive energy antiparticles.

\section{Applications}
\label{sec:applications}

In this section we consider numerical examples in the case where the fermion mass changes abruptly but continuously from zero to a finite value. To be specific, we model the change by a kink profile: 
\beqa
|m(t)| &=& \frac{m_\infty }{2}\big(1+\tanh \frac{t-t_0}{\tau} \big) 
\nonumber \\
\theta(t) &=& \frac{\Delta \theta}{2}\big(1-\tanh\frac{t-t_0}{\tau} \big)\,,
\label{kink}
\eeqa
and we assume that the fermion is interacting with a scalar $\phi$ and another fermion field $q$, through a non-diagonal Yukawa-interaction term
\beq 
{\cal L}_{\rm int} = - y\; \bar \psi_R \phi \, q_L + h.c. \,.
\label{interaction2}
\eeq
For simplicity we also assume that the scalar $\phi$ is thermal and fermion $q$ is in kinetic equilibrium with an effective chemical potential $\mu_q(t)$. With these assumptions the self energies $\Sigma_{\rm eff}$ reduce to the usual thermal expressions, where $\Sigma^{<,>}$ are related by the Kubo-Martin-Schwinger relation: $\Sigma^> = e^{\beta(k_0 - \mu_q)}\Sigma^<$, and 
\beq 
\Sigma^<(k) = 
     \big( \Sigma^<_0 \,\gamma^0 - \Sigma^<_3 \,
     \, \hat {\bf k}\cdot {\gamma} \big) \, P_R \,, 
\label{self3}
\eeq
where $\hat {\bf k} = {\bf k}/|{\bf k}|$ and $i\Sigma^<_{0,3}$ are real-valued functions, whose expressions at the mass shells are (remember that $\omega_{\bf k}=\omega_{\bf k}(t)$ depends on time through the mass function) 
\beqa 
i\Sigma^<_0(\pm\omega_{\bf k},|{\bf k}|) 
   &=& \frac{y^2 T^2}{8\pi |{\bf k}|} |I_1(\pm\omega_{\bf k},|{\bf k}|)| 
\\ 
v_{{\bf k}}i\Sigma^<_3(\pm\omega_{\bf k},|{\bf k}|) 
   &=& \pm\frac{y^2 T^2}{8 \pi |{\bf k}|} \, 
             \left(|I_1(\pm\omega_{\bf k},|{\bf k}|)| 
         - \frac{|\alpha |\,|m|^2}{\omega_{\bf k}^2} 
             |I_0(\pm\omega_{\bf k},|{\bf k}|)|\right) \,, 
\label{Sigma_0,3}
\eeqa    
where we defined $v_{{\bf k}} \equiv |{\bf k}|/\omega_{\bf k}$ and  	
\beq 
I_n(k_0,|{\bf k}|) 
  = \theta(\lambda) \int_{\alpha - \delta}^{\alpha + \delta} {\rm d}y \: y^n 
       \frac{1}{(e^{y - \mu_q/T} + 1) (e^{k_0/T - y} - 1 )} \,,
\label{I_n}
\eeq  
with 
\beqa
\alpha &=& \frac{|m|^2+m_q^2-m_\phi^2}{2|m|^2} \frac{k_0}{T}
\nonumber \\
\delta  &=& \frac{\lambda^{1/2}(|m|^2,m_q^2,m_\phi^2)}{2|m|^2}
               \frac{|\bf{k}|}{T} \,, 
\eeqa
where $\lambda(a,b,c) = (a+b-c)^2-4bc$ is the usual kinematic phase space function.
\begin{figure}
\centering
\includegraphics[width=1 \textwidth]{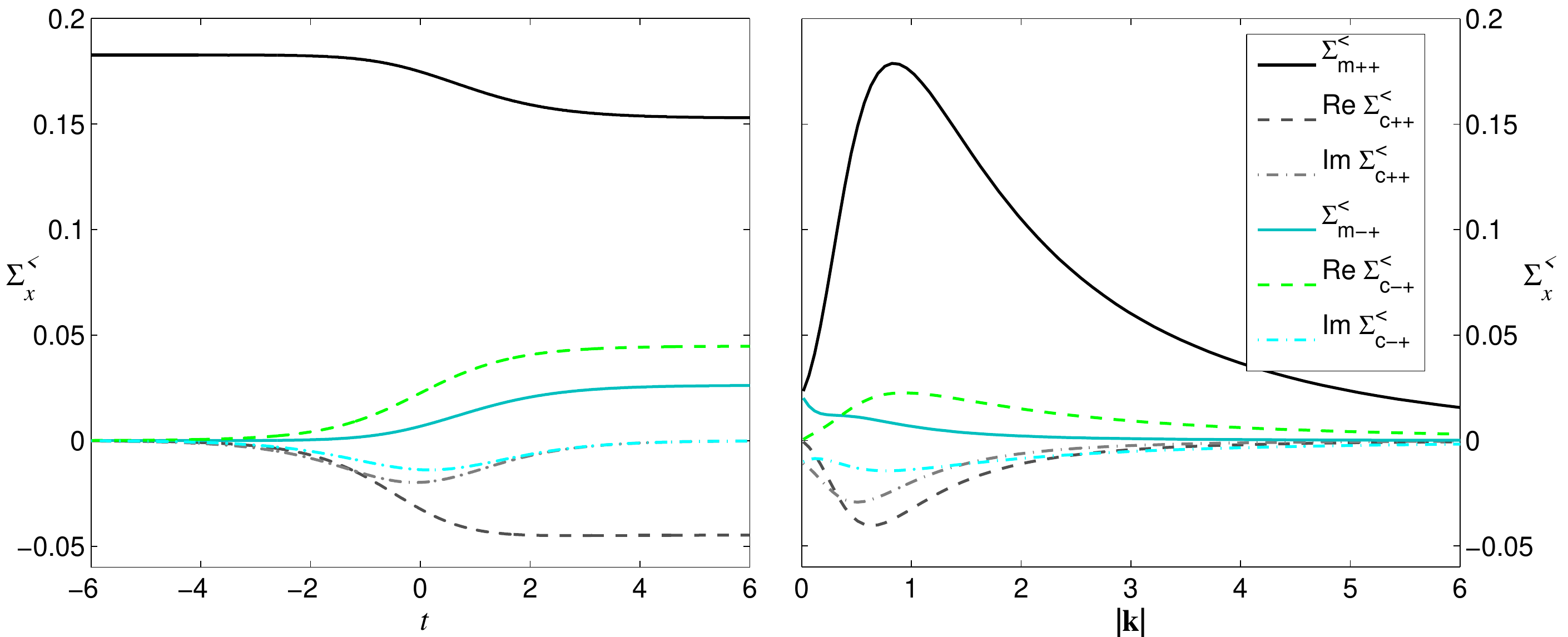}
\caption{Shown are the various self-energy components $\Sigma^<_{jh+}$ (for positive energies only) defined in Eqs.~(\ref{eq:projsigmasN1}) as a function of $tm_\infty$ for a fixed $|{\bf k}|/m_\infty\equiv 1$ (left panel) and as a function of $|{\bf k}|/m_\infty$ for a fixed $tm_\infty \equiv 0$ (right panel).   
For other parameters we used $t_0=0 $, $\tau= 2/m_\infty$, $\Delta \theta = -1.0$, $T = 10 m_\infty$, $m_q = 10 m_\infty$, $m_{\phi}= 5 m_\infty$, $\mu_q=0$ and $y=0.3.$}
\label{fig:sigma}
\end{figure}   
Note that both $\Sigma_0$ and $\Sigma_3$ have finite limit when $|{\bf k}|\rightarrow 0$ despite the apparent singularity, because the integration region $\propto \delta \propto |{\bf k}|$. The chemical potential $\mu_q$ is calculated from the conservation of the fermionic charge in the specified interaction.  For the projected self-energy functions $\Sigma^{<,>}_{jh\pm}$ in the collision integrals (\ref{coll_types1}) we find the expressions:
\begin{eqnarray}
\Sigma^{<,>}_{mh\pm} &=& 
\frac{1}{2}(1 \pm h v_{{\bf k}})\big( i\Sigma^{<,>}_0(\pm\omega_{\bf k})  - h i
\Sigma^{<,>}_3(\pm\omega_{\bf k}) \big)
\nonumber \\
\Sigma^{<,>}_{ch\pm} &=&
-\frac{1}{2\omega_{\bf k}}( h m_R v_{{\bf k}} \mp i m_I) 
\big( i\Sigma^{<,>}_0(\pm\omega_{\bf k})  - h i\Sigma^{<,>}_3(\pm\omega_{\bf k}) \big) \,.
\label{eq:projsigmasN1}
\end{eqnarray}

Let us note that we used the self-energy function (\ref{self3}) also in the numerical examples in our earlier work~\cite{HKR2}. However, at that time the relevance of the resummation of the coherence oscillations was not realized, and so the coherence-shell self-energy was erroneously computed in ref.~\cite{HKR2} by naively projecting $\Sigma^{<,>}$ to the $k_0=0$ shell. This shows that the effects of resummation are nontrivial even when the primary self-energy is thermal. Typical behaviour of the self-energies is displayed in figure \ref{fig:sigma}. On the left panel we show the time dependence of the self-energies $\Sigma^<_{jh+}$ over the wall region, and on the right panel their $|{\bf k}|$-dependence at a fixed time $tm_\infty =1$ during the transition. In all figures in this section the time is measured in units $m_\infty^{-1}$ and the momenta in units $m_\infty$, where $m_\infty$ is the absolute value of the mass of the fermion at the infinite future $m_\infty\equiv |m(t=\infty)|$. Note that at very early times, where the $\psi$-field is massless, only $\Sigma^<_{m++}$ is nonzero as expected for a right chiral interaction term in association with a positive helicity state. Once the field becomes massive all self-energies evolve differently as a function of the increasing mass. In particular the imaginary parts of $\Sigma^<_{ch+}$ are very small and nonzero only inside the wall region, confined to $|tm_\infty | \lsim 2$. In the right panel we again see that the right helicity mass-shell self-energy is by far the largest, while the other components rapidly become vanishing when the momentum increases and one again effectively approaches the massless limit.
It should also be observed that while the mass-shell functions are strictly positive definite, $\Sigma^<_{ch+}$'s which appear as cross-interaction terms between the mass- and coherence-shells in the collision terms (\ref{coll_types1}) can have either sign.

\begin{figure}
\centering
\includegraphics[width=1.0\textwidth]{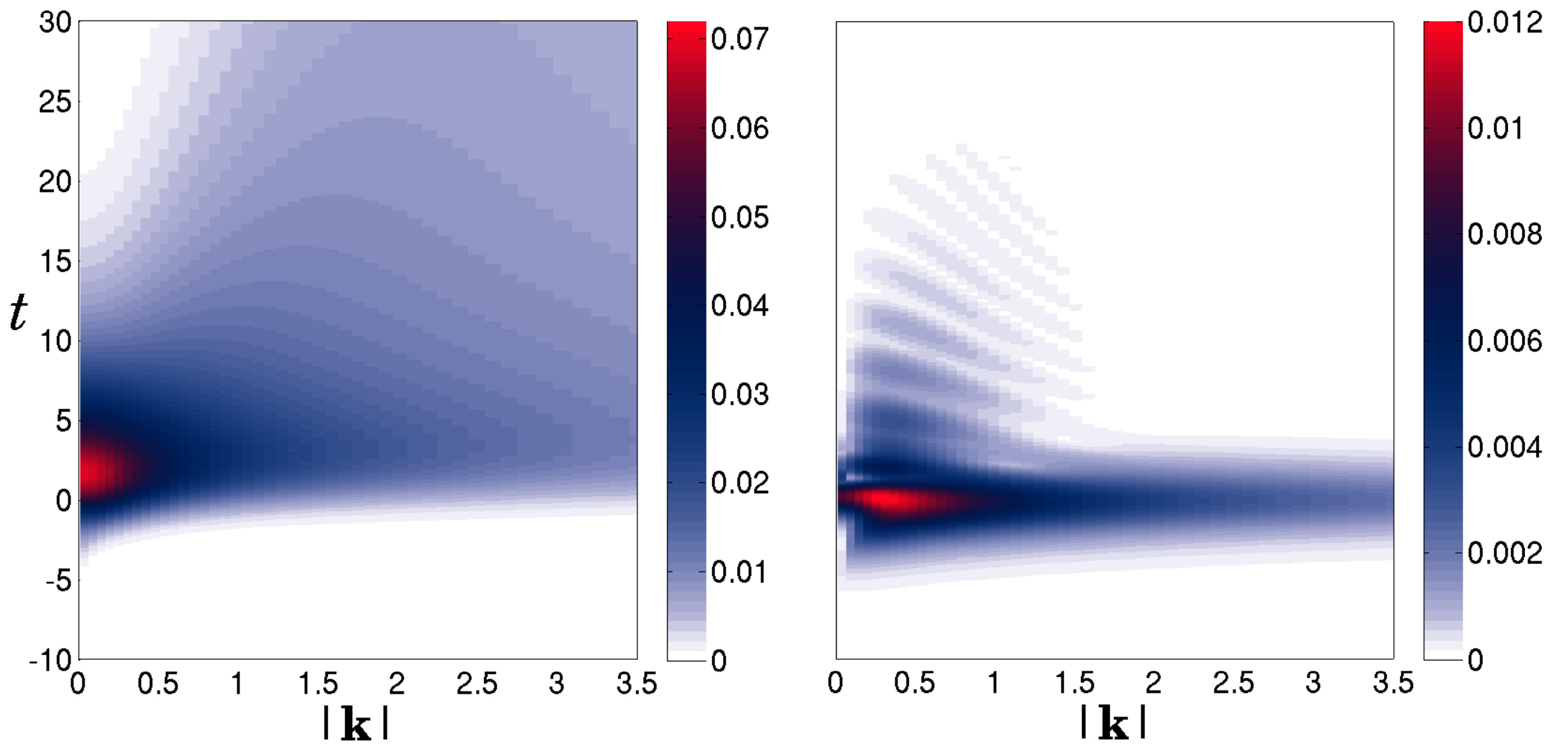}
\vskip-0.3truecm
\caption{Shown is the excess total particle number density 
$(n_{{\bf k}} + \bar n_{{\bf k}}) - (n_{{{\bf k}}\,\rm eq} +  \bar n_{{{\bf k}}\,\rm eq})$ (left) and the magnitude of total particle - antiparticle coherence density $\tilde f_{c}({\bf k})$ (right).} 
\label{fig:results1}
\end{figure}   

\begin{figure}
\centering
\vskip-0.4truecm
\includegraphics[width=1.0 \textwidth]{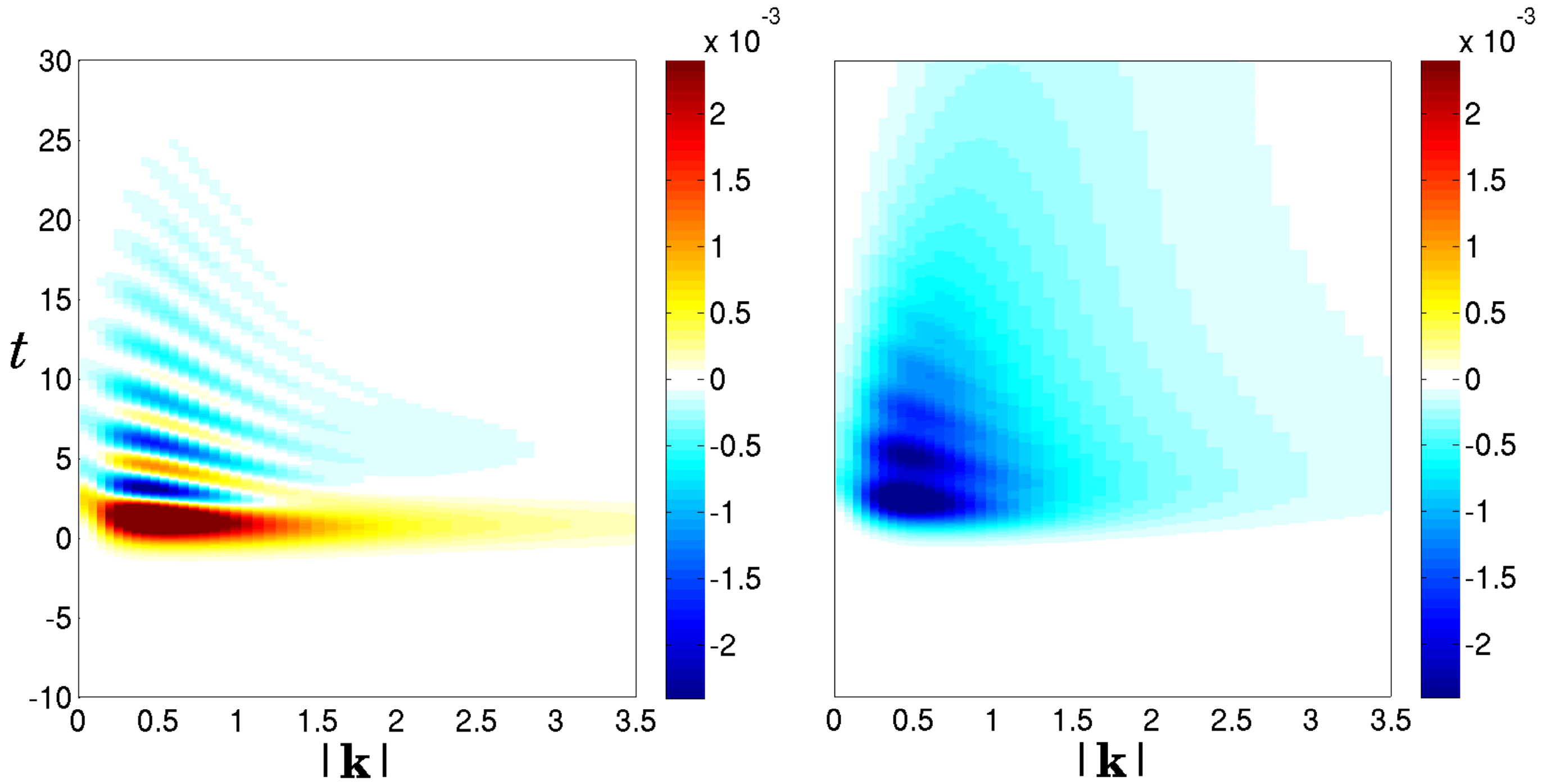}
\vskip-0.3truecm
\caption{Shown is the total right chiral number density $j_{\rm R}^0({\bf k})$ (left) and the total charge density (net particle - antiparticle asymmetry) $j^0({\bf k})$ (right). We used the same parameters as in Fig.~\ref{fig:results1}.}
\label{fig:results2}
\end{figure}

We have solved numerically the quantum Boltzmann equations (\ref{eom_on-shell1}-\ref{eom_on-shell2}) with collision integrals (\ref{coll_types1}) where the projected functions $\Sigma_{jh\pm}$ are given by Eq.~(\ref{eq:projsigmasN1}) for the mass profile defined in Eq.~(\ref{kink}). Our results are shown in Figs.~\ref{fig:results1} and~\ref{fig:results2}. In the left panel of Fig.~\ref{fig:results1} we show the evolution of the $|{\bf k}|$-dependent excess total particle number density $(n_{{\bf k}} + \bar n_{{\bf k}}) - (n_{{{\bf k}}\,\rm eq} +  \bar n_{{{\bf k}}\,\rm eq})$, where $n_{{\bf k}} = \sum_h n_{{\bf k}h}$ (similarly for $\bar n$), and $n_{{\bf k}h}$ and $\bar n_{{\bf k}h}$ were defined through the Feynman-St\"uckelberg interpretation in Eq.~(\ref{Fey-Stuck}). In the right panel of Fig.~\ref{fig:results1} we show the total particle-antiparticle coherence density defined as $\tilde f_c({{\bf k}}) \equiv \sfrac{1}{2}\sum_{h\pm} |\tilde f_{ch\pm}({{\bf k}})|$, where the individual coherence functions $\tilde f_{ch\pm}$ are canonically normalized according to Eq.~(\ref{eq:canonicalnorm}). The $|{\bf k}|$-dependent baseline for the colour coding in figures was defined by the adiabatic thermal Fermi-Dirac distribution $n_{{\bf k}h\,{\rm eq}} = \bar n_{{\bf k}h\,{\rm eq}} = 1/(\exp(\beta\omega_{\bf k}) + 1) $, which ranges from $2$ at the low end to $\approx 1$ at the high end of the plotted phase space area. Finally, we used parameters $t_0=0 $, $\tau= 2/m_\infty$ and $\Delta \theta = -1.0$ for the kink, and $T = 10 m_\infty$, $m_q = 10 m_\infty$ and $m_{\phi}= 5 m_\infty$ for the self-energy. For the Yukawa coupling we used $y=0.3$. As a boundary condition, all fields including the coherently evolving fermion were are assumed to be in thermal equilibrium without chemical potentials in the distant past. 

Both quantities shown in Figs.~\ref{fig:results1} vanish in the equilibrium conditions, when no coherence evolution takes place. However, creation of extra excitations and coherence is clearly seen in the plots. As expected, most of the excitations are created in non relativistic modes near the transition time. As time goes on, the asymmetry diffuses towards the higher momenta in the phase space.  The exact quantitative details are of course dependent on the kink and the interaction parameters. The creation of the extra modes seen in the left panel, is associated with narrow bands of damped oscillations in the coherence distribution $\tilde f_{c}$ shown in the right panel. Just like the difference from equilibrium, the coherence function is restricted in time to near the phase transition time and in momentum to non- or near relativistic region. The canonically normalized coherence functions are roughly by a factor of five smaller than the deviations from equilibrium in the mass-shell functions.

In Fig.~\ref{fig:results2} we show some current densities derived from the same data we used to create Figs.~\ref{fig:results1}. In the left panel we show the total right chiral number density $j_{\rm R}^0 =\langle \bar \psi \gamma^0P_{\rm R}\psi\rangle = \sum_h j_{h \rm R}^0$. Using the cQPA correlation function we can write $j_{h \rm R}^0 = {\rm Tr}[\bar {\cal S}_h^< P_{\rm R}]$ in terms of the on-shell distribution functions as follows:
\beq
j_{h \rm R}^0({\bf k}) =  
\frac{1}{2}  \Big[(1 + h v_{{\bf k}})n_{{\bf k}h} 
                - (1 - h v_{{\bf k}})\bar n_{{\bf k}h}\Big] 
- h v_{{\bf k}}\frac{m_R}{\omega_{\bf k}}\Re f^<_{ch+}
+ \frac{m_I}{\omega_{\bf k}}\Im f^<_{ch+} \,, 
\eeq
where the vacuum contribution has been subtracted off. The result again displays the oscillatory behaviour driven by nonzero coherence distributions $f_{c\pm}$ in near relativistic modes right after the transition. The right chirality of our chosen interaction term (\ref{interaction2}) leads to a temporal charge separation between fields $\psi$ and $q$ after the transition. This effect produces the effective chemical potential for the field $q$ mentioned above, and a net particle-antiparticle number density asymmetry 
$j^0 = \langle \bar \psi \gamma^0\psi \rangle = \sum_h j_{h}^0= \sum_h{\rm Tr}[\bar {\cal S}_h^<]$ for the field $\psi$. With the vacuum part again subtracted off we find:
\beq
j_{h}^0({\bf k}) = n_{{\bf k}h} - \bar n_{{\bf k}h}\,. 
\eeq
We show the net asymmetry in the right panel of Fig.~\ref{fig:results2}. Of course the total, {\em combined} fermion number of $\psi$ and $q$ fields is conserved, because no net fermion number production takes place in the absence of sphaleron processes.  

\subsection{Error analysis}

In the cQPA scheme the collision terms are computed to the zeroth order of the gradient expansion in time derivatives of the mass function $m(t)$. The effects of the neglected gradient corrections are twofold: first, the gradient corrections to the constraint equations would broaden the singular phase space structure of the 2-point functions $S^{<>}$~\cite{Joni}. As a result {\em e.g.}~the self-energy functions in Eq.~(\ref{effectiveC3}) (or in expressions  (\ref{eq:projsigmasN1})) would not be evaluated at exactly on shell, but rather convolved with distributions which are peaked but nonsingular functions at the on-shell momenta. However if, as is generally the case, the self-energy functions are smooth in the scale set by the width of the distribution functions, the correction from this effect should be very small.

Second, and more importantly, there are explicit gradient corrections to the collision term involving the derivatives $\partial_t S$ and $\partial_t \Sigma$, which (after performing the zeroth order resummation) are proportional to mass gradients. It is not easy to estimate the precise numerical error coming from neglecting these gradients. However, a naive estimate for the relative error in the collision terms is given by dimensionless gradient expansion parameter:
\beq
\epsilon \equiv |\frac{m^\prime}{m\omega}|\,.
\eeq
Note that $\epsilon$ depends both on time and on the momenta involved. In the example of this section the bulk asymmetry is generated at momentum $|{\bf k}| \approx 0.5\,m_\infty$ (see Fig.~\ref{fig:results2}), and for these modes, in the middle of the kink profile (\ref{kink}) at $t=0$, where the gradients are largest, we find $\epsilon \approx 0.8$. However, the error rapidly gets smaller as one moves away from the wall and it dies exponentially when $|\Delta t| > \tau$. Thus $\epsilon$ alone does not provide a direct estimate of the error on the final results. Now, the quantities proportional to particle-antiparticle asymmetries are the most sensitive to variation of collision terms. In these cases the true error should be bounded by the product of $\epsilon$ with the wall width and $\Gamma$: $\delta^\Gamma_{\rm rel} \sim \epsilon \tau \Gamma$. In the current example we then find $\delta^\Gamma_{\rm rel} \sim 0.1$, suggesting our results may be accurate to within 10 per cent. For the ``bulk"-quantities (not proportional to asymmetry) the collision term is not dominating the evolution and consequently we should expect a much smaller induced error in these quantities.  To test these arguments we redid our calculations by adding an arbitrary correction to all self-energies of the form
\beq
\Sigma^{<,>}_{a;h\pm} \rightarrow \Sigma^{<,>}_{a;h\pm} (1 \pm \epsilon) \,,
\eeq
where $a=m,c$. Numerical results verify the analytical error estimates given above: the chiral number densities and charge densities shown in figure 12 and the chemical potential for the $q$-species shown in figure 13 were seen to change by the expected 10\%, while in the excess of the total particle number density and the total particle antiparticle coherence solutions of Fig.~11 the changes were at level $< 1 \%$. We believe that these evaluations provide a generous upper limit for the true errors, because in reality not all gradient corrections work coherently in the same direction.

Finally, we wish to emphasize that these gradient corrections described above affect {\em only} to the evaluation of the collision term; the flow term and thus the free theory evolution is exact in this approximation. Indeed, the set of distribution functions $f_\alpha$ form a complete reparametrization of the components of zeroth moment integral of $S^<$, so that even in the region of large gradients where their interpretation as singular phase space distribution functions would break, they would still present the free theory evolution exactly.

\subsection{A toy model for coherent baryogenesis}

We can construct a very simple toy model of baryogenesis using the above results. Let us assume that the field $q$ in our example corresponds to a standard model quark, while $\psi$ is some new field with the complex time-varying mass $m$. We do not provide any details of the phase transition characteristics, except of noting that the order parameter field giving rise to $\psi$ mass need not be related to the field $\phi$ in the interaction (\ref{interaction2}). 
\begin{figure}
\centering
\includegraphics[width=0.8 \textwidth]{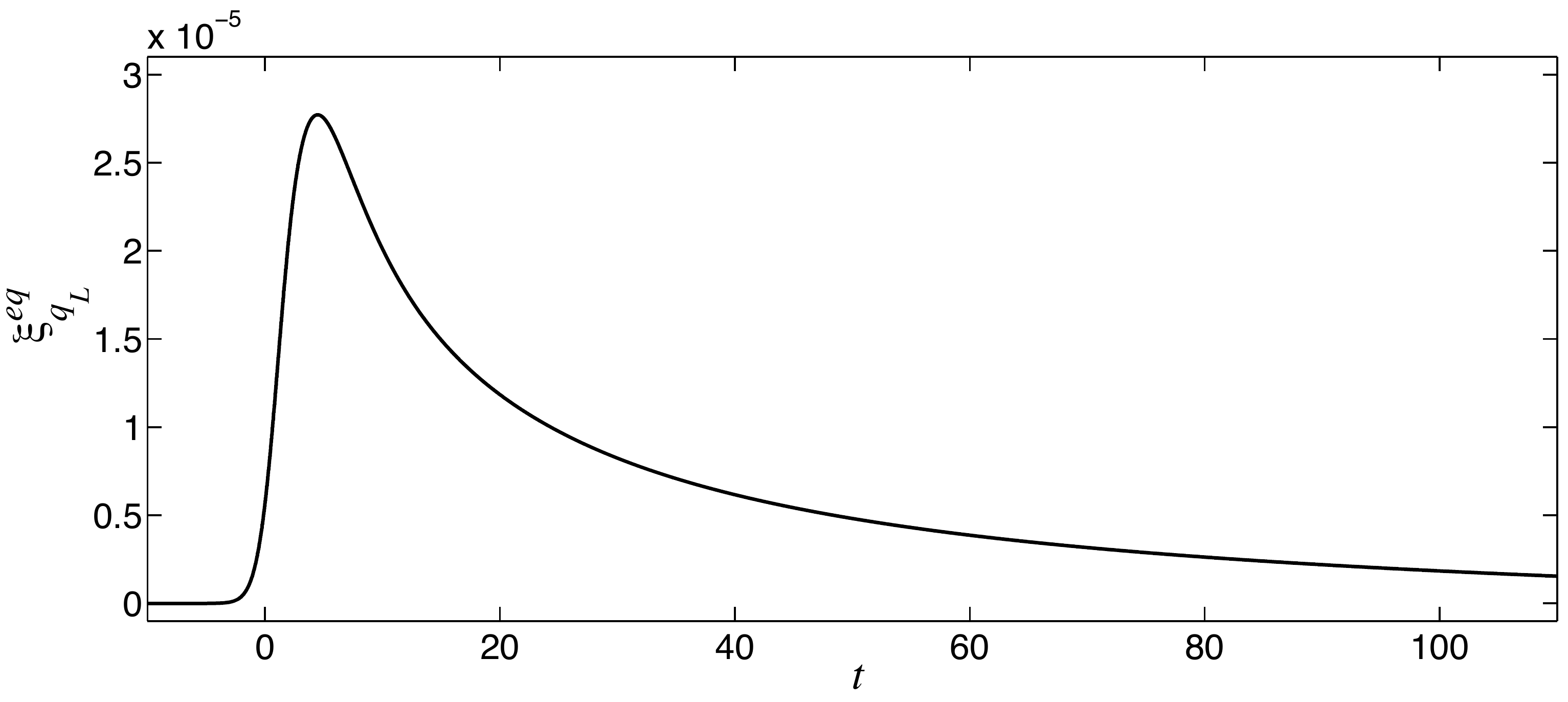}
\caption{Left-chiral quark chemical potential $\xi_{q_{L}} \approx \sfrac{1}{2}\sfrac{\mu_{q_L}}{T}$ as a function of time. All components of the quark field are assumed to be in kinetic equilibrium at all times.}
\label{fig:results3}
\end{figure}   
While not needed for a qualitative picture, such details would of course be necessary in a more serious modeling effort. First, if $q$ is a quark field, it would be held in kinetic equilibrium by the strong gauge interactions. We can assume that only one of the quarks (say the top quark) or any number of them are coupled to $\psi$ through interaction (\ref{interaction2}); variations of this type can be modeled by changing the strength of the coupling $y$. In addition, the strong chiral anomaly equilibrates the total left- and right chiral quark chemical potentials~\cite{ClassForceomat}, whereby the effective chemical potential sourcing the baryon number production is reduced by a factor of two:
\beq
\xi^{\rm eq}_{q_L} \approx \frac{1}{2}\frac{\mu_{q_{L}}}{T} \,. 
\eeq
Given $\xi^{\rm eq}_{q_L}$ the total baryon asymmetry produced by the Electroweak anomaly can be computed from~\cite{ClassForceomat}:
\begin{equation} 
\frac{\partial n_B}{\partial t} = \frac32 \Gamma_{\rm sph} 
\left(\xi^{\rm eq}_{q_L} - A \frac{n_B}{T^3} \right),
\label{Beqn}
\end{equation}
where $\Gamma_{\rm sph} \equiv \kappa_{\rm sph} \alpha_W^5 T^4$ is the
Chern-Simons number diffusion rate across the energy barrier
which separates $N$-vacua of the SU$(2)$ gauge theory, where
$\kappa_{\rm sph} = 20\pm 2$~\cite{MooreSp}.  The second term
describes sphaleron-induced relaxation of the baryon asymmetry in the
symmetric phase:
\beq
A \frac{n_B}{T^2} \equiv \mu_{CS} = 3 \sum_i \mu_{q_i} + \sum_i \mu_{l_i}\,,
\label{Adef}
\eeq
where $\mu_{CS}$ is the Chern-Simons- and $\mu_l$ the leptonic chemical potential and constant $A$ is a model dependent quantity of order unity. We shall use a value $A = 15/2$ corresponding to the Standard Model particle content with flavour equilibrated chemical potentials apart from the right chiral leptons (see {\eg}~ref.~\cite{ClassForceomat}), but the actual value is not relevant. It is now easy to integrate the equation (\ref{Beqn}) for the total baryon number produced:
\beq
n_B =  \frac{3}{2} \Gamma_{\rm sph} \int_{-\infty}^{\infty} {\rm d}t\,
\xi^{\rm eq}_{q_L} e^{-k_B t}\,,
\label{Btotal}
\eeq
where
\beq
k_B \equiv \frac{3 A}{2}\, \frac{\Gamma_{\rm sph}}{T^3}\,.
\label{keibii}
\eeq
Now using $\eta_B \equiv n_B/n_\gamma \approx 7n_B/s$ and $\eta_{10} \equiv 10^{10}\eta_B$, we find:
\beq
\eta_{10} \approx 1800 \big(\frac{T}{m_\infty}\big) \, \int_{-\infty}^{\infty} {\rm d}\hat t\,
\xi^{\rm eq}_{q_L}(\hat t) e^{-\hat k_B \hat t}\,, 
\label{Btotal2}
\eeq
where $\hat t \equiv tm_\infty$ and we took $g_{*S} = 110$ for the number of entropy degrees of freedom. The back-scattering term $\hat k_B \equiv k_B/m_\infty \approx 1.24\times 10^{-6} T/m_\infty$ is too small to affect our results as suggested above. For the results displayed in figures \ref{fig:results1}-\ref{fig:results2} we chose the parameters to optimize the display of the qualitative features of coherence solutions in the plots. We redid the numerical analysis for the baryogenesis model taking $t_0=0 $, $\tau= 3m_\infty$, $\Delta \theta = \pi/2$ for the kink, and $T = 5m_\infty$, $m_q = 0$, $m_{\phi}= 5 m_\infty$ and $y=0.8$ for the interactions. The resulting chemical potential $\xi_{q_L}(\hat t)$ is shown in Fig.~\ref{fig:results3}, and the baryon asymmetry given by Eq.~(\ref{Btotal2}) is
\beq
\eta_{10} \approx 7.5 \,.
\eeq
Let us point out that this model is not realistic in the sense that a single fermion complex mass term breaks the CPT-symmetry. The model should then be thought of being a part of a more complex multi-field mixing scenario where the single-field treatment can be used to approximate diagonal mass eigenstates in the non-degenerate limit. Another obvious concern is the speed of transition giving rise to the $\psi$-mass. We do not try to speculate on this issue any further, except note that perhaps such a ``quench" could be effected through a coupling to another scalar field along the lines discussed in ref.~\cite{Enqvist:2010fd}.

\section{Conclusions and outlook}
\label{sec:discussion}

In this work we have reformulated and extended our recently introduced quantum kinetic theory for coherent, interacting fermionic and scalar fields in spatially homogeneous and isotropic systems. Our formalism is based on the coherent quasiparticle approximation (cQPA), where nonlocal coherence information is encoded in new spectral solutions at off-shell momenta. We have used the Schwinger-Keldysh formalism of non-equilibrium quantum field theory with pertinent approximations to derive the cQPA propagators in the spatially homogeneous background. In particular, we have now introduced the familiar 4-dimensional Dirac notation for fermions, which allows us to use the standard Dirac algebra in the reduction of the matrix elements. The Dirac-structure of the coherence propagator living at the shell $k_0 = 0$ in particular is expressible as products of the usual mass-shell projectors onto positive and negative energies, or alternatively as direct products of positive and negative energy eigenspinors. The coherence propagators are always rapidly oscillating functions of time.

We have shown that in the cQPA the Kadanoff-Baym equations for the 2-point functions in the Wigner representation reduce to a closed set of extended quantum Boltzmann equations (qBE) for the cQPA mass- and coherence-shell distribution functions.  We observe that the collision integrals in these qBE's need to be resummed to all orders in gradients because of the rapid oscillations of the coherence-shell functions. We have performed this resummation and derived a set of generalized momentum space Feynman rules for the theory, including  effective 4-dimensional propagators endowed with the coherence, as well as an interaction vertex rule for a Yukawa theory. As a result of the resummation the vertex rule displays a local energy non-conservation in association with the coherence propagators, however in a way that never violates the energy conservation at the level of complete self-energy diagrams.

We have given several examples of the diagrammatic calculations in our formalism, including 1-loop fermionic and scalar self-energies and a 2-loop fermionic self-energy in a Yukawa theory. We have also applied the formalism to construct a simple toy model for baryogenesis, where a coherent fermionic field is interacting with a thermal background scalar field and a Standard Model quark. In this model the CP-violation resulting from the C-breaking complex mass term of the $\psi$-field and the chirality breaking decay interaction leads to a nonzero chemical potential for the left chiral quarks, which then acts as a seed asymmetry biasing the Chern-Simons diffusion rate to produce baryons via the Electroweak anomaly. The final baryon asymmetry created via this mechanism can be roughly of the correct magnitude to explain the observed baryon excess in the universe.

In this work we have considered only the case of a single fermionic and scalar field. A generalization of the current analysis to multiple coherently mixing fermion and scalar fields, including full flavour coherent propagators, flavoured quantum Boltzmann equations and a full set of flavour dependent Feynman rules will be presented in ref.~\cite{FHKR}. The natural applications of the formalism with multiple mixing fields include for example the neutrino flavour oscillations in the early universe and the resonant leptogenesis \cite{resonant_lepto}, where the quantum coherence between the neutrino flavours may play an important role in the dynamics. For a realistic application to electroweak baryogenesis the formalism of this work needs to be generalized to the case of a stationary planar symmetry, where the coherence lives at the shell $k_z = 0$. Based on our earlier work~\cite{HKR1}, we expect this case to be highly analogous, although technically somewhat more complicated than the spatially homogeneous formalism presented here. Also that formalism need to be developed for many mixing fields. 
However, despite the fact that most realistic applications of the cQPA formalism appear to need a generalization to mixing fields, most novel aspects related to the new coherence shells and resummation procedures necessary to define the appropriate Feynman rules should be qualitatively very similar to the techniques developed here.

\section*{Acknowledgments}
The work of MH was supported by the Gottfried Wilhelm Leibniz programme of the Deutsche Forschungsgemeinschaft. 
The work of PMR was supported by the Magnus Ehrnrooth foundation.

%
%

\end{document}